\documentclass{aa}
\usepackage{natbib}
\usepackage{pdflscape} % or {lscape}
\usepackage{wasysym}
\usepackage[varg]{txfonts}
\usepackage{graphicx}
\usepackage{hyperref}
\bibpunct{(}{)}{;}{a}{}{,} % to follow the A&A style
\usepackage{longtable}

\usepackage{afterpage}

\usepackage[usenames,dvipsnames]{color}
\definecolor{mygreen}{rgb}{0,0.5,0}
\definecolor{myorange}{rgb}{0.5,0.5,0}
\definecolor{myred}{rgb}{0.5,0,0}

\def\ms{\hbox{\,m\,s$^{-1}$}}         %m.s -1
\def\m2s2{\hbox{\,m$^{2}$\,s$^{-2}$}} %m2.s -2
\def\kms{\hbox{\,km\,s$^{-1}$}}       %km.s -1
\def\vsini{\hbox{$v$\,sin\,$i$\,}}      %vsini
\def\logrhk{$\log$(R$^{\prime}_{HK}$)}

\begin{document}

\title{Radial-Velocity Fitting Challenge
\thanks{Based on observations collected at the La Silla Parana Observatory,
ESO (Chile), with the HARPS spectrograph at the 3.6-m telescope.}}

\subtitle{II. First results of the analysis of the data set}

 %\and F. Borsa, M. Damasso, R. Diaz, P. C. Gregory, N. C. Hara, A. Hatzes, V. Rajpaul, M. Tuomi
 %H. R. A. Jones
\author{X. Dumusque\inst{1,2}
	    \thanks{Society in Science -- Branco Weiss Fellow (url: \url{http://www.society-in-science.org})}    
	    \and F. Borsa\inst{3}
	    \and M. Damasso\inst{4}
	    \and R. D\'iaz\inst{1}
	    \and P. C. Gregory\inst{5}
	    \and N.C. Hara\inst{6}
	    \and A. Hatzes\inst{7}
	    \and V. Rajpaul\inst{8}
	    \and M. Tuomi\inst{9}
	    \and S. Aigrain\inst{8}
            \and G. Anglada-Escud\'e\inst{9,10}
            \and A.S. Bonomo\inst{4}
            \and G. Bou\'e\inst{6}
            \and F. Dauvergne\inst{6}
            \and G. Frustagli\inst{3}
            \and P. Giacobbe\inst{4}
            \and R. D. Haywood\inst{2}
            \and H. R. A. Jones\inst{9}
            \and M. Pinamonti\inst{11,12}
            \and E. Poretti\inst{3}
            \and M. Rainer\inst{3}
            \and D. S\'egransan\inst{1}
            \and A. Sozzetti\inst{4}
            \and S. Udry\inst{1}
	    }
	    
%	    M. Damasso\inst{3}
%	    \and A. Sozzetti\inst{3}
%	    \and R. Haywood\inst{2,4}
%	    \and A. S. Bonomo\inst{3}
%	    \and M. Pinamonti\inst{5}
%	    \and P. Giacobbe\inst{3}}

\institute{Observatoire de Gen\`eve, Universit\'e de Gen\`eve, 51 ch. des Maillettes, CH-1290 Versoix, Switzerland \email{xavier.dumusque@unige.ch} 
	      \and Harvard-Smithsonian Center for Astrophysics, 60 Garden Street, Cambridge, Massachusetts 02138, USA
	      \and INAF $-$ Osservatorio Astronomico di Brera, Via E. Bianchi 46, 23807 Merate (LC), Italy
	      \and INAF $-$ Osservatorio Astrofisico di Torino, via Osservatorio 20, 10025, Pino Torinese, Italy
	      \and Physics and Astronomy Department, University of British Columbia, 6224 Agricultural Rd., Vancouver, BC V6T 1Z1, Canada
	      \and ASD, IMCCE-CNRS UMR8028, Observatoire de Paris, UPMC, 77 Av. Denfert-Rochereau, 75014 Paris, France
	      \and Th\"uringer Landessternwarte Tautenburg, Sternwarte 5, 07778 Tautenburg, Germany
	      \and Sub-department of Astrophysics, Department of Physics, University of Oxford, Oxford OX1 3RH, UK
	      \and University of Hertfordshire, Centre for Astrophysics Research, Science and Technology Research Institute, College Lane, AL10 9AB, Hatfield, UK
	      \and School of Physics and Astronomy, Queen Mary University of London, 327 Mile End Rd., E1 4NS, London, UK
	      \and Dipartimento di Fisica, Universita degli Studi di Trieste, via G. B.Tiepolo 11, I-34143 Trieste, Italy
	      \and INAF $-$ Osservatorio Astronomico di Trieste, via G. B. Tiepolo 11, I-34143, Trieste, Italy
	      }

\date{Received XXX; accepted XXX}

\abstract
% Context, Aims, Methods, Results, Conclu (not mandatory)
{Radial-velocity (RV) signals arising from stellar photospheric phenomena are the main limitation for precise RV measurements
Those signals induce RV variations an order of magnitude larger than the signal created by the orbit of Earth-twins, thus preventing their detection.}
{Different methods have been developed to mitigate the impact of stellar RV signals. The goal of this paper is to compare the efficiency of these different methods to recover extremely low-mass planets despite stellar RV signals. However, because observed RV variations at the meter-per-second precision level or below is a combination of signals induced by unresolved orbiting planets, by the star, and by the instrument, performing such a comparison using real data is extremely challenging.}
{To circumvent this problem, we generated simulated RV measurements including realistic stellar and planetary signals. Different teams analyzed blindly those simulated RV measurements, using their own method to recover planetary signals despite stellar RV signals. By comparing the results obtained by the different teams with the planetary and stellar parameters used to generate the simulated RVs, it is therefore possible to compare the efficiency of these different methods.}
{The most efficient methods to recover planetary signals {\bf take into account the different activity indicators,} use red-noise models to account for stellar RV signals and a Bayesian framework to provide model comparison in a robust statistical approach. Using the most efficient methodology, planets can be found down to $K/N= K_{\mathrm{pl}}/\mathrm{RV}_{\mathrm{rms}}\times\sqrt{N_{\mathrm{obs}}}=5$ with a threshold of $K/N=7.5$ at the level of 80-90\% recovery rate found for a number of methods. These recovery rates drop dramatically for $K/N$ smaller than this threshold. In addition, for the best teams, no false positives with $K/N > 7.5$ were detected, while a non-negligible fraction of them appear for smaller $K/N$. A limit of $K/N = 7.5$ seems therefore a safe threshold to attest the veracity of planetary signals for RV measurements with similar properties to those of the different RV fitting challenge systems.}
{}


\maketitle
\titlerunning{The Radial-Velocity fitting challenge}
\authorrunning{X. Dumusque}

\section{Introduction} \label{sect:1}

The radial-velocity (RV) technique is an indirect method that measures with Doppler spectroscopy the stellar wobble induced by a planet orbiting its host star. The technique is sensitive not only to possible companions, but also to signals induced by the host star. Now that the \ms\, precision level has been reached by the best spectrographs, it is clear that solar-like stars introduce signals at a similar level. Those stellar signals, often referred to as \emph{stellar jitter}, currently prevent the RV technique from detecting and measuring the mass of Earth-twins orbiting solar-type stars, i.e., Earth analogues orbiting in the habitable zone of GK dwarfs, because such planets induce signals an order of magnitude smaller. It is therefore extremely important to investigate new approaches to mitigate the impact of stellar signals if we want the RV technique to be efficient at characterizing the Earth-twins that will be found by TESS \citep[][]{Ricker-2014} and PLATO \citep[][]{Rauer-2014}.

At the \ms\,precision level, RV measurements are affected by stellar signals, that depend on the spectral type of the observed star { \citep[][]{Dumusque-2011a,Isaacson-2010,Wright-2005}}. For GK dwarfs, those stellar signals can be decomposed, to our current knowledge, in four different components:
\begin{itemize}
\item solar-type oscillations \citep{Dumusque-2011a,Arentoft-2008,OToole-2008,Kjeldsen-2005}, 
\item granulation phenomena \citep{Dumusque-2011a,Del-Moro-2004a,Del-Moro-2004b,Lindegren-2003,Dravins-1982},
\item short-term activity signals on the stellar rotation period timescale \citep[][]{Haywood-2016,Borgniet-2015,Robertson-2015a,Robertson-2014,Dumusque-2014b,Boisse-2012b,Saar-2009,Meunier-2010a,Saar-1997b},
\item and long-term activity signals on the magnetic cycle period timescale \citep[][]{Lanza-2016,Diaz-2016,Meunier-2013,Lovis-2011b,Dumusque-2011c,Makarov-2010}.
\end{itemize}
{For more details about these signals and their origins, readers are referred to Section 2 in \citet{Dumusque-2016a} and references therein.}

Stellar signals creates RV variations that are larger than the signal induced by small-mass exoplanets, such as Earth-twins. There is several examples in the literature, where by analyzing the same RV measurements different teams detected different planetary configurations. This is the case of the famous planetary system GJ581, for which the number of planet detected is ranging between 3 and 6 \citep[][]{Hatzes-2016,Anglada-Escude-2015, Robertson-2014, Baluev-2013, Vogt-2012, Gregory-2011, Vogt-2010b, Mayor-2009b}, of HD40307, for which 4 to 6 planets have been announced \citep[][]{Diaz-2016,Tuomi-2013a}, and GJ667C, for which 3 to 7 planets have been detected \citep[][]{Feroz-2014,Anglada-Escude-2012a,Gregory-2012}. All those systems are affected by stellar signals, and therefore depending on the model used to analyze the data, different teams arrives to different conclusions. This shows that optimal models do not exist at the moment to analyze RV measurements affected by stellar signals and this pushes the community towards finding an optimal solution. The RV fitting challenge is one of the efforts pursued today in this direction. The development of the HARPS-N solar telescope \citep[][]{Dumusque-2015b} is another one that should deliver the optimal data set for characterizing and understanding stellar signals in detail. 

In principle, the nature of RV stellar and planetary signals is different. RV signal induced by a planet is periodic over time, while stellar signals are in the best case semi-periodic. In addition, a planet induces a pure Doppler shift of the observed stellar spectrum, while stellar signals change the shape of the spectral lines. Therefore, it should be possible to find techniques to differentiate between planetary and stellar signals.

Stellar oscillations are often averaged out in RV surveys by fixing an exposure time to 15 minutes. To obtain the best RV precision, it is also possible to observe the same star several times per night, with measurements spread out during the night, to sample better the signature of granulation and supergranulation \citep[][]{Dumusque-2011a}. It has been shown that this simple approach reduces the observed daily RV rms of measurements, however it does not fully average out this signal \citep[][]{Meunier-2015,Dumusque-2011a}, and more optimal techniques need to be investigated. For short-term activity, which is by far the most difficult stellar signal to deal with due to the non-periodic, stochastic, long-term signals arising from the evolution and decay of active regions, several correction techniques have been investigated:
\begin{itemize}
\item fitting sine waves at the rotation period of the star and harmonics \citep[][]{Boisse-2011},
\item using red-noise models to fit the data \citep[e.g.][]{Feroz-2014, Gregory-2011, Tuomi-2013a},
\item using the FF${}^\prime$ method if contemporaneous photometry exists \citep[][]{Dumusque-2015b, Haywood-2014,Aigrain-2012},
\item modeling activity-induced signals in RVs with Gaussian process regression, whose covariance properties are shared either with the star's photometric variations \citep[][]{Haywood-2014,Grunblatt-2015} or a combination of several spectroscopic indicators \citep[][]{Rajpaul-2015}, or determined from the RVs themselves \citep[][]{Faria-2016a},
\item using linear correlations between the different observables, i.e., RV, bisector span (BIS SPAN) and full width at half maximum (FWHM) of the cross correlation function \citep[CCF,][]{Baranne-1996,Pepe-2002a}, photometry \citep[][]{Robertson-2015a,Robertson-2014,Boisse-2009,Queloz-2001}, and magnetic field strength \citep[][]{Hebrard-2014},
\item checking for season per season phase incoherence of signals \citep[][]{Santos-2014,Dumusque-2014a,Dumusque-2012},
\item avoiding the impact of activity by using wavelength dependence criteria for RV signal \citep[e.g. in HD40307 and HD69830,][]{Tuomi-2013a, Anglada-Escude-2012}.
\end{itemize}
Finally, long-term activity seems to correlate well with the calcium chromospheric activity index, which provides a promising approach to mitigation of this source of stellar RV noise \citep[][]{Lanza-2016, Diaz-2016,Meunier-2013,Dumusque-2012}.

The goal of this paper is to test the efficiency of different approaches to retrieve low eccentricity planetary signals despite stellar signals. To do so, we present the results of a RV fitting challenge, where several teams analyzed blindly the same set of real and simulated RV measurements affected by planetary and stellar signals. Each team used their own method to recover planetary signals despite stellar signals. At the \ms\,precision level reached by the best spectrographs, RV measurements are affected by unresolved planets, but also stellar and instrumental signals. Without knowing which part of the RV variations is due to planets and which is due to the star or the instrument, it is extremely difficult to test which method is the most efficient at finding low-mass planets despite stellar signals. For such an exercise, it is crucial to use simulated RV measurements so that a comparison can be performed between the results of the different analysis and what was initially injected into the data. The set of simulated and real RV measurements used for this RV fitting challenge is described in detailed in \citet{Dumusque-2016a}. { As said in this paper, most of the planets injected in the data have very low eccentricities, which is common is observed multi-planetary systems.} Those RVs correspond to typical quiet { solar-like stars} targeted by high-precision RV surveys. Therefore, the conclusions of this paper are relevant for most high-precision RV surveys.

In Sections \ref{sect:2} and \ref{sect:3}, we describe the methods used by the different teams to recover planetary signals despite stellar signals; Section \ref{sect:2} focuses on methods relying on a Bayesian framework, while Section \ref{sect:3} on other methods. For those sections, the number assigned to each team does not have any particular meaning. In Section \ref{sect:4}, we discuss the results of the different teams and compare the efficiency of their method to recover low-mass planetary signals despite stellar signals. We conclude in Section \ref{sect:5}.

\section{Methods to deal with stellar signals using a Bayesian Framework} \label{sect:2}

In total eight different teams have analyzed the RV fitting challenge data set, using different approaches. This section is dedicated to the description of these different methods.
The first five teams used a Bayesian framework and model comparison to find the most favorable solution for each system. Teams 1 through 4 used red-noise models to account for stellar signals, while team 5 used a white noise model. Team 6, 7 and 8 used \emph{pre-whitening},{compressed sensing} and/or filtering in the frequency domain. Table \ref{tab:0} summarizes the different techniques used to deal with stellar signals, and the different stellar and orbital parameters reported by each team.
\begin{table*}
\begin{footnotesize}
\begin{center}
\caption{Techniques to deal with stellar signals used by the different teams, as well as planetary and stellar parameters reported. P$_{rot}$ corresponds to the stellar rotation period, $P$, $K$ $T_0$, ecc and $\omega$ to the planetary period, semi-amplitude, transit time, eccentricity and argument of periastron, respectively.} \label{tab:0}
\begin{tabular}{ccccccccc}
\hline\hline
 & Team & Techniques & $P_{rot}$ & $P$ & $K$ & $T_0$ & ecc & $\omega$\\
\hline
1 & Torino       & Bayesian framework with Gaussian process to account for red noise & Yes & Yes & Yes & Yes & Sometimes & Sometimes\\
2 & Oxford      & Bayesian framework with Gaussian process to account for red noise & No  & Yes & Yes & Yes & Yes & Yes\\
3 & M. Tuomi  & Bayesian framework with Moving Average to account for red noise        & Yes & Yes & Yes & No  & No & No\\
4 &  P. Gregory   & Bayesian framework with apodized Keplerians to account for red noise  & No  & Yes & Yes & Yes & Yes & Yes\\
5 & Geneva    & Bayesian framework with white noise                                                      & No  & Yes & Yes & No  & No & No\\
6 & A. Hatzes & Pre-whitening                    				  				     & Yes & Yes & Yes & No  & No & No\\
7 & Brera        & Filtering in frequency space     				  				     & No  & Yes & Yes & Yes & Yes & Yes\\
8 & IMCCE     & Compressed sensing and filtering in frequency space (preliminary results) & Yes & Yes & Yes & No & No & No\\
\hline
\end{tabular}
\end{center}
\end{footnotesize}
\end{table*}
%

%#######################################################################
% GROUP 1
%#######################################################################

\subsection{Team 1: Torino team - Bayesian framework with Gaussian process regression to account for stellar signals} \label{sect:2-0}

The Torino team is composed of, in order of contribution, M. Damasso, A. Sozzetti, R. D. Haywood, A.S. Bonomo, M. Pinamonti, and P. Giacobbe. Their activities are in the framework of the {\it Global Architecture of Planetary Systems} (GAPS) project (e.g., Poretti et al. 2015). This team analyzed the 14 valid systems of the RV fitting challenge\footnote{as explained in \citet{Dumusque-2016a}, system 6 was not considered due to a problem when generating the RV measurements.}. For each system, team 1 reported the period, semi-amplitude, time of periastron passage and sometimes eccentricity and argument of periastron of the detected planets, as well as their best estimate of the stellar rotation period. For most of the planetary system, team 1 analyzed only the most significant signals with small $p$-values, as the team had not enough time and computational power to explore more complex solutions obtained by adding smaller significance signals. For the same reason, as a first approach team 1 favored circular over more complex eccentric orbits.

\subsubsection{General framework}

The Torino team used Gaussian processes \citep[GP,][]{Rasmussen-2006} to model, in a non-parametric way, stellar activity effects in RV data. GPs are a powerful tool for mitigating the contribution of stellar activity in RV measurements, especially when contemporaneous stellar activity indicators are available (e.g. light curves, spectroscopic activity indexes).

When simulating short-term activity signals for the RV fitting challenge data set, \cite{Dumusque-2016a} only considered slow rotators, i.e., \vsini$<4$\kms. In this case, plages are expected to be responsible for the majority of the short-term activity RV variation \citep[see introduction, and for more details][]{Haywood-2016,Dumusque-2014b, Meunier-2010a}. The calcium activity index \logrhk, which is a measure of the emission in the core of the \ion{Ca}{II} H$\&$K spectral lines, was provided within the RV fitting challenge data set, and appeared to be the best proxy to trace out short-term activity contribution to the RVs.

The real strength of a GP regression approach is that it is non-parametric and does not assume any physical model about active regions or the physical processes at play. Here, the only assumption made by Team 1 is that short-term activity variations in RV and \logrhk\,share the same covariance properties. This assumption is reasonable as the short-term activity signal in RV and \logrhk\,is induced by stellar rotation and active region evolution.

Team 1 first trained a GP on the time series of the activity index \logrhk, and then injected the resultant covariance function into another GP that is part of a RV model. This GP absorbs correlated noise due to the stellar activity through a global fit (Keplerian signals + GP). This approach is inspired in particular by the works of \citet{Haywood-2014} and \cite{Grunblatt-2015}. For the Kepler-78 system discussed in this last paper, the same global fit was able to recover a periodicity in the RV consistent with the stellar rotation period found via photometry.

For the sake of a homogeneous analysis, all the systems were processed generally using the same recipe, following a sequence of few steps, as described in the appendix (see Section \ref{app:2-1}). In all cases, the team used only one covariance function and analyzed the full data sets, i.e., without rejecting outliers, without binning the data to get a single point per night, and without dividing the analysis into sub-sets. The analysis did not use the information provided by the bisector span \citep[BIS SPAN,][]{Queloz-2001} of the CCF.

Because an optimal way of correctly identifying the number of planetary signals, and their orbital properties, based on the available RV data is through the evaluation of the Bayesian statistical evidence $\mathcal{Z}$ for each model, the Torino group carried out this task by testing a limited number of models and calculating $\mathcal{Z}$, which is notoriously complicated to assess using only one analytical procedure among those proposed in literature.

{The details about the method used by team 1 to analyze the data of the RV fitting challenge can be found in the appendix of the paper (Section \ref{app:2-1}). In the next subsection we illustrate the method using as example system 2.}

\subsubsection{Example for system 2}

\begin{figure*}
\begin{center}
\includegraphics[width=9cm]{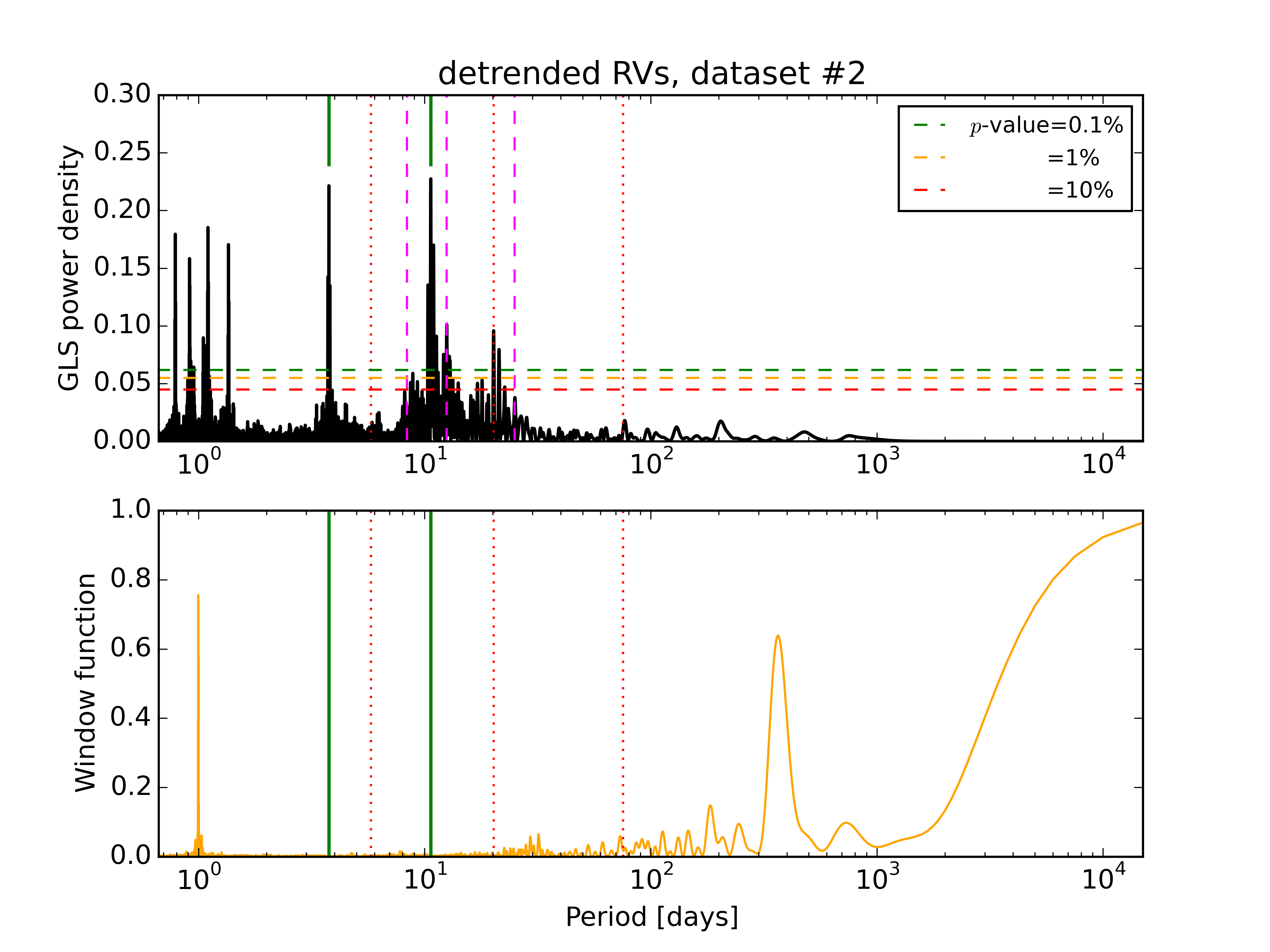}
\includegraphics[width=9cm]{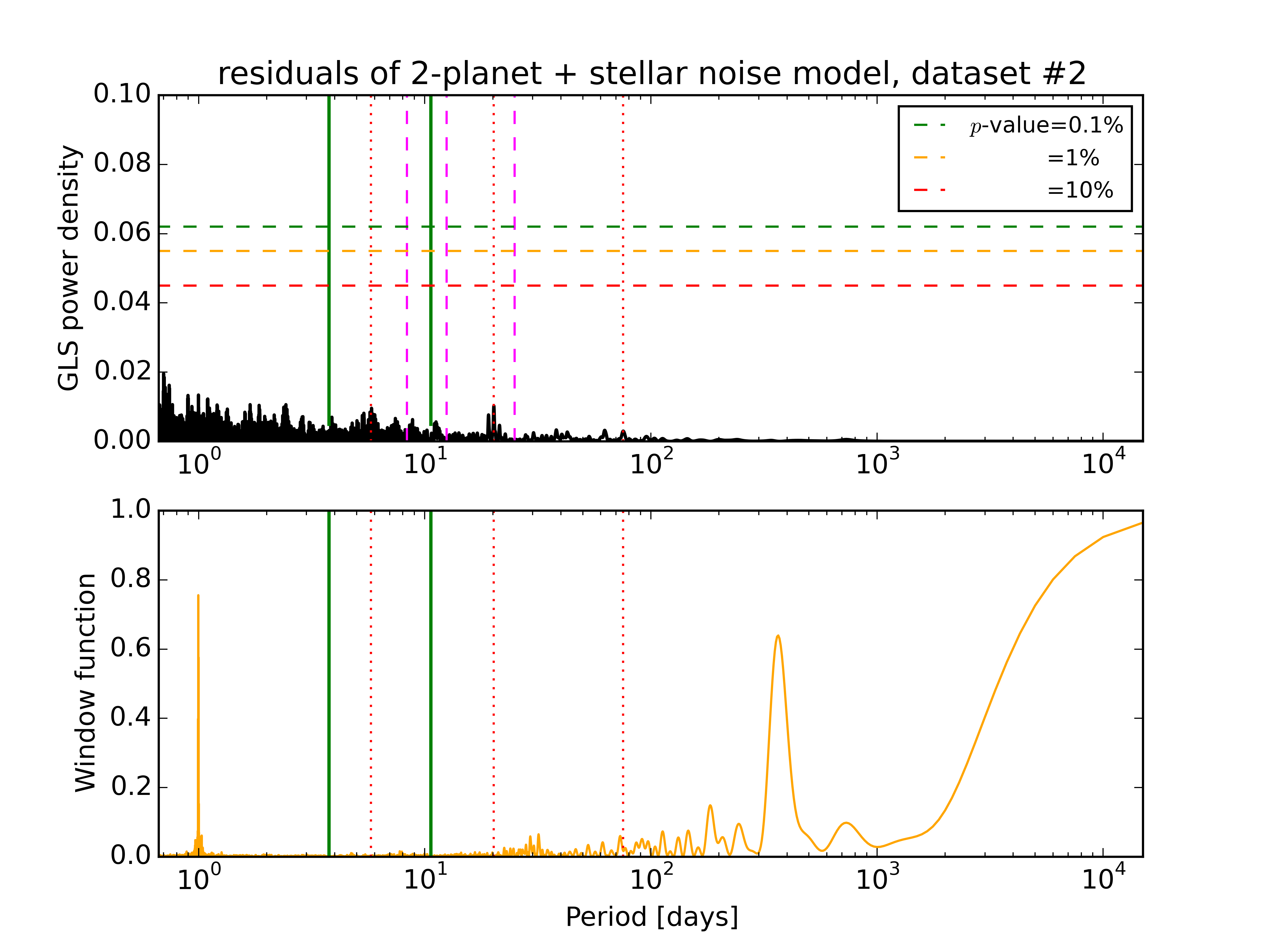}
\includegraphics[width=9cm]{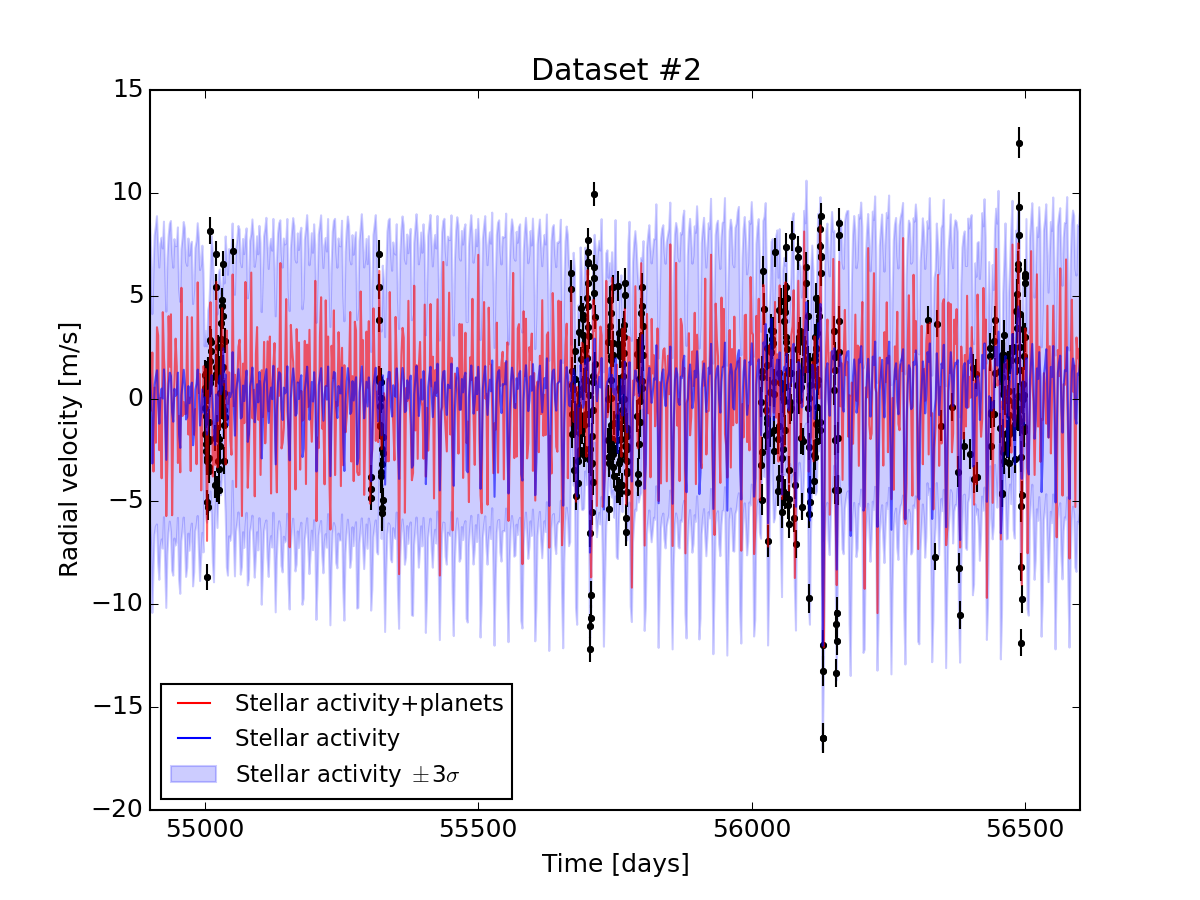}
\includegraphics[width=9cm]{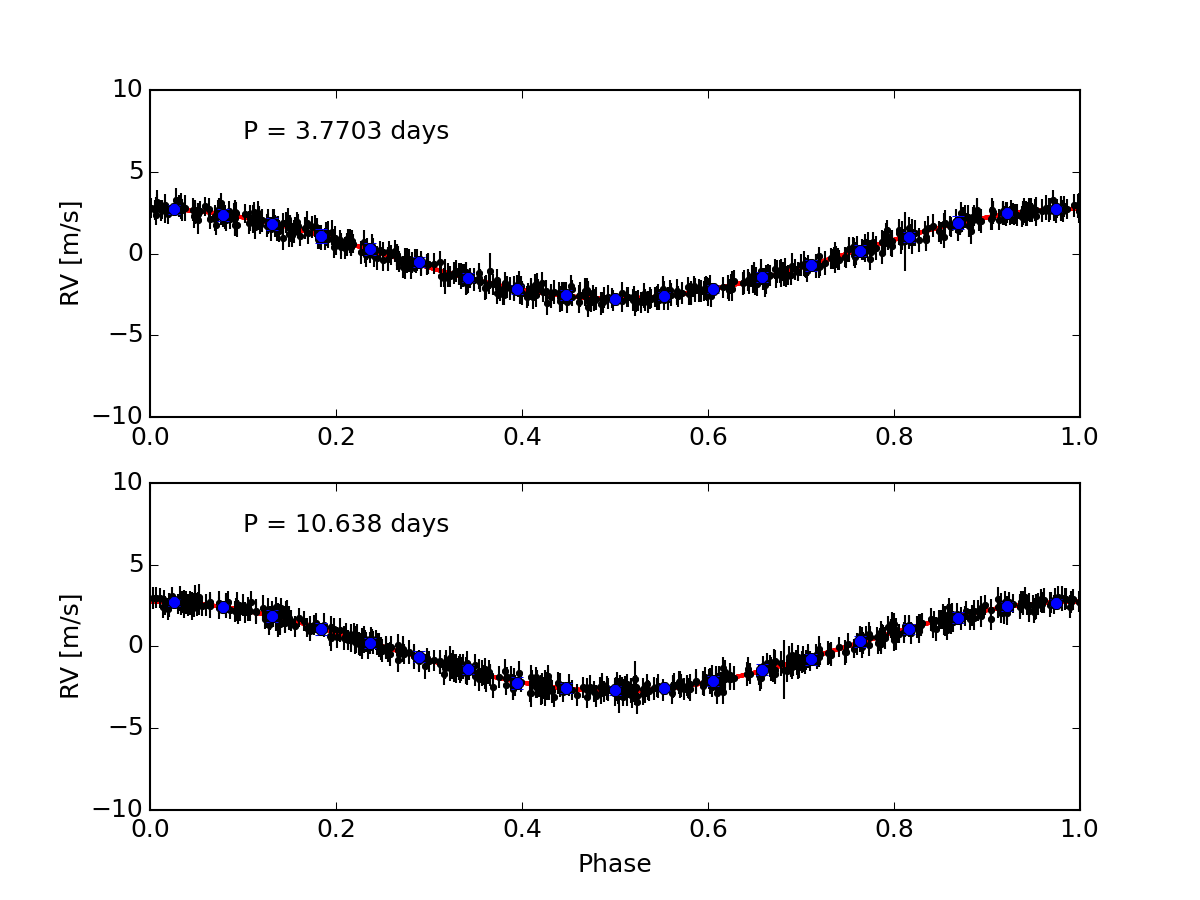}
\caption{\emph{Top left panel:} GLS periodogram of the challenge RV dataset number 2 analysed by the team 1, and the corresponding window function. The original RVs data were first corrected through a linear regression with \logrhk and then through a sinusoidal fit to remove a 2727-day periodicity likely related to a long-term stellar activity cycle. Vertical lines indicate \textit{i)} the orbital frequencies of the planetary candidates identified by the team (solid green lines), which turned out to be real; \textit{ii)} the missed planets (dotted red lines); \textit{iii)} the simulated stellar rotation period and its first two harmonics (dashed magenta lines). \emph{Top right panel:} GLS periodogram of the RV residuals after the best-fit global model including the two planets detected was removed from the detrended dataset. A bootstrap analysis performed on these RV residuals shows no evidence of significant signals left.
\emph{Bottom left panel:} RVs detrended with the best-fit global model superposed (solid red line) and the mean contribution due to the stellar activity predicted by the GP (blue solid line). The shaded area spans the 3$\sigma$ region centered around the best-fit noise model. 
\emph{Bottom right panel:} RV curves folded at the periods of the two candidate Keplerian solutions found by the team, with the superposed red continuous line representing the best-fit orbital model. Data in each plot refers to a single planet, and are obtained from the original RVs by subtracting the offset, the GP stellar activity noise model, and the Keplerian of the other planet.}
\label{fig:Torino-1}
\end{center}
\end{figure*}

{In Fig. \ref{fig:Torino-1}, we show the results obtained by team 1 for system 2. The original RVs data show a very significant correlation with \logrhk\,(Spearman’s rank correlation coefficient $\rho$ = 0.88), thus they were first corrected using a linear regression with \logrhk. The GLS periodogram of the RV residuals shows a peak at $\sim$ 2727 days, that corresponds to a peak value observed in the GLS periodogram of the \logrhk\,time series at exactly the same frequency, probably due to a long-term stellar activity cycle. We removed this signal from the RV residuals and \logrhk\,time series by fitting a sinusoid with the same periodicity. Team 1 correctly recovered the rotation period of the star ($P_{\rm rot}$=25 days) through a GP analysis of the detrended \logrhk\,time series. They recovered two out of the 5 planets injected in system 2. The GLS periodogram of the corrected RVs shows a series of significant frequencies with p-value<0.1$\%$, the most prominent ones corresponding to the orbital period of the two recovered planets and to the first harmonic of the stellar rotation period. Despite the dominant frequency related to the stellar rotation appearing at the first harmonic $P_{\rm rot}/2$, the GP noise model converged towards the stellar rotation period. There is also a couple of peaks with p-value<0.1$\%$, one of them clearly corresponding to the $\sim$20 days period planet that Team 1 did not report. To prevent announcing false positives, Team 1 assumed this peak to be produced by differential rotation of the star, because of a period close to stellar rotation.}

%\begin{table}
%\caption{Orbital parameters of the two planet candidates identified in the dataset 2 by the Torino team, and estimates of the hyper parameters of the covariance function used to model the short-term stellar activity noise in the radial velocities (see Eq. \ref{eq:2-0}). }
%\label{table:torino_ex2}
%\begin{tabular}{cc}
%\hline
%parameter             & value \\ \hline
%& \\
%$A$ [m s$^{-1}$] & 6.9$^{+1.3}_{-0.6}$  \\ [3pt]
%$l$ [days] & 0.032$^{+0.018}_{-0.012}$  \\  [3pt]
%$L$ & 0.29$^{+0.044}_{-0.035}$   \\[3pt]
%$\theta$ [days] & 25.01$\pm0.013$   \\ [3pt]
%$\alpha$ & 0.019$^{+0.01}_{-0.008}$  \\  [3pt]
%$RV_{\rm offset}$ [m s$^{-1}$] & 9.4e-02$\pm1.7$   \\ [3pt]\hline
%& \\
%\textbf{Planet 1} & \\ [5pt]
%$K_{\rm 1}$ [m s$^{-1}$] & 2.7$\pm$0.15   \\[3pt]
%$Period_{\rm 1}$ [days] & 3.7703$\pm$0.00027   \\[3pt]
%$T_{\rm periastron, 1}$ [days] & 55002.43$\pm0.07$\\   [3pt] 
%& \\
%\textbf{Planet 2} & \\ [5pt]
%$K_{\rm 2}$ [m s$^{-1}$] & 2.7$\pm$0.2  \\  [3pt]
%$Period_{\rm 2}$ [days] & 10.638$\pm0.004$ \\ [3pt]  
%$T_{\rm periastron, 2}$ [days]  & 55010.2$\pm0.4$   \\ 
%& \\ \hline
%\end{tabular}
%\end{table}

%#######################################################################
% GROUP 2
%#######################################################################

\subsection{Team 2: Oxford team - Bayesian framework with GP regression to account for stellar signals} \label{sect:2-1}

Team 2 is composed of, in order of contribution, V. Rajpaul and S. Aigrain. This team analyzed the first 5 systems of the RV fitting challenge. For each system, team 2 reported the period, semi-amplitude, time of periastron passage, eccentricity and argument of periastron of the detected planets. Team 2 did not report stellar rotation periods.

\subsubsection{General framework}

Team 2 analyzed the data of the RV fitting challenge using, like team 1, a GP regression to account for red noise induced by stellar signals. However their approach is slightly different and is described in details in \citet{Rajpaul-2015}. Rather than using only the calcium activity index \logrhk\,as a proxy for activity and first training the GP on the activity indicator and then using the best estimate of the GP hyper-parameter to fit the RVs, team 2 modeled all time series (\logrhk, FWHM, BIS SPAN, RV) simultaneously. Team 2 treated each time series as a linear combination of a single unobserved GP and its derivative, adding a polynomial function to fit long-term trends. In addition, and only for the RV data, team 2 added one or several Keplerians. This approach is statistically more robust as it does not require an iterative fitting process, however, it is more computationally demanding.

At the outset, team 2 wanted to marginalize fully over all the hyper-parameters and parameters of their model, using an MCMC sampler. However, because this increased computational times by orders of magnitude, team 2 found it was not able to produce reasonable results within the time allocated for the RV fitting challenge. Team 2 therefore found the maximum \emph{a posteriori} (MAP) values for all model hyper-parameters and parameters. Then, with the GP covariance hyper-parameters fixed at their MAP values (type-II maximum likelihood approximation), team 2 used a nested sampling algorithm to explore the parameter space of interest, i.e., the planet parameters. Note that team 2 used uniform priors for all parameters, except for eccentricity for which a log-uniform prior was used. As a result of the nested sampler, team 2 obtained posterior distributions for all parameters, as well as model evidences (log\,$\mathcal{Z}$). Team 2 compared models with up to $N=5$ planets and added planets until the evidence for a model with an additional planet was smaller than the model without this extra complexity, i.e., log\,$\mathcal{Z}$ | ($N+1$) $<$ log\,$\mathcal{Z}$ | ($N$).

Team 2 notes that given the type-II maximum likelihood approximation, computed evidence values are perhaps not reliable. Team 2 also note that the code used to pre-process the data before GP regression included a polynomial-subtraction component that removed long-term trends from the time series. In retrospect, this ended up removing all planetary signals with periods longer than a couple of months -- a simple though costly error.

\subsubsection{Example for system 2}

\begin{figure*}
\begin{center}
\includegraphics[width=16cm]{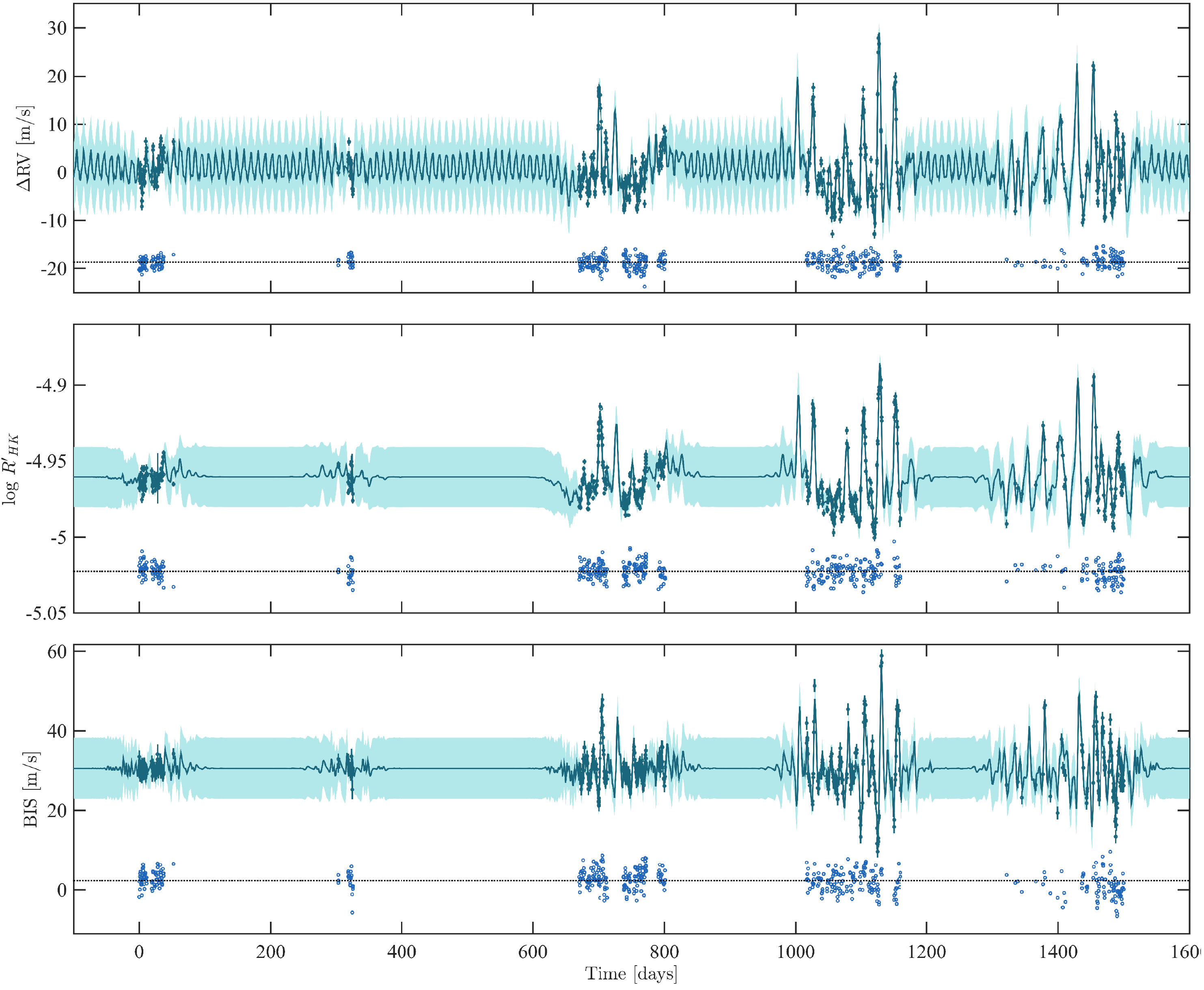}
\caption{GP model MAP fit to system 2 for which team 2 could recover accurately two of the four injected planetary signals. As in other datasets, the crude pre-processing pipeline used by team 2 wrongly removed signals longer than a couple of months, therefore removing the signal of the 75-day planet present in the data. All plotted time series were fitted simultaneously, using a single set of GP hyperparameters. The green dots indicate the raw time series; the solid lines are model posterior means, and the shaded regions denote $\pm \sigma$ posterior uncertainty. The blue dots indicate model residuals. A zoom on the fourth epoch of observation can be seen in Fig. \ref{fig:oxford2}.}
\label{fig:oxford1}
\end{center}
\end{figure*}

\begin{figure*}
\begin{center}
\includegraphics[width=16cm]{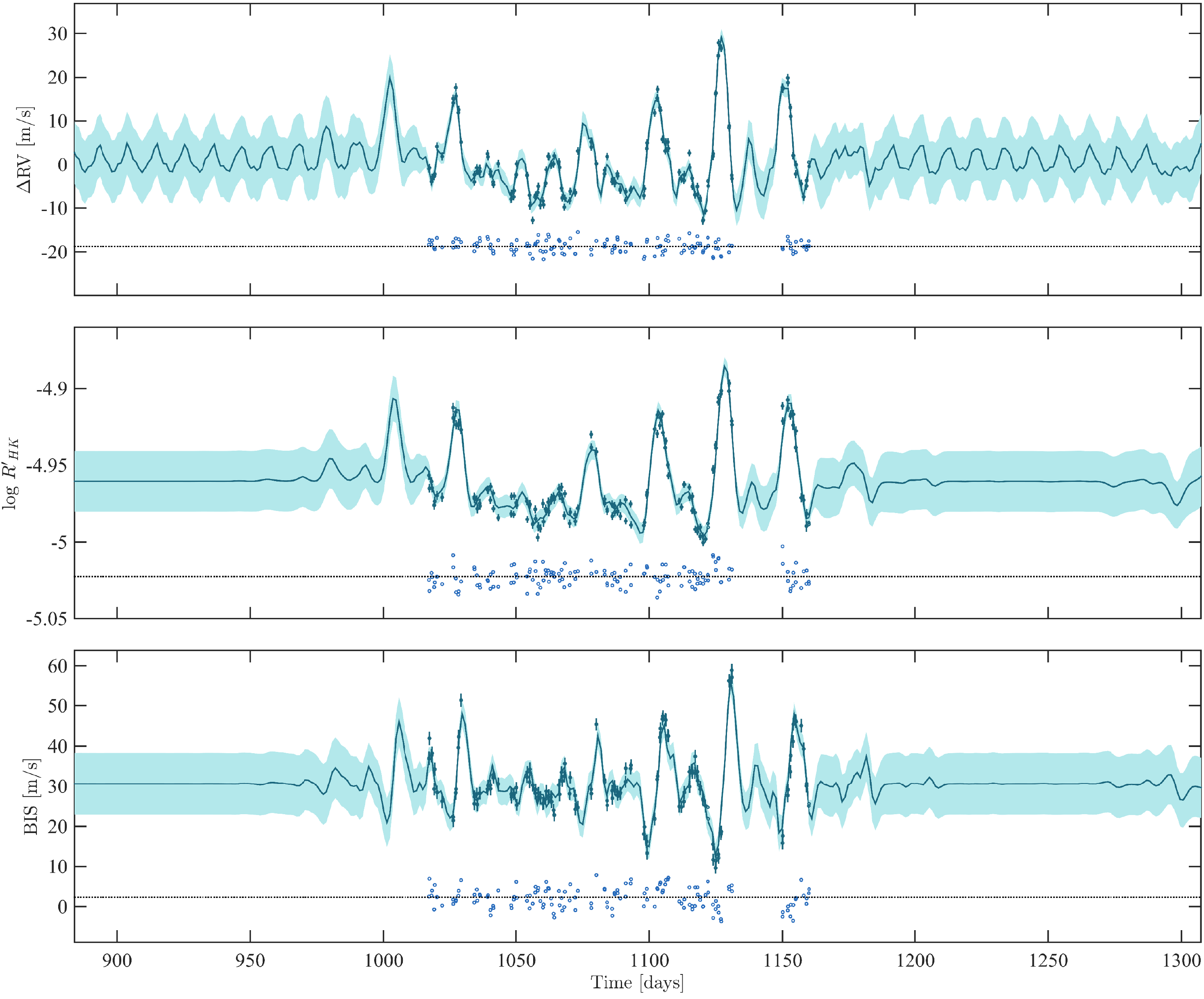}
\caption{Same legend as Fig. \ref{fig:oxford1}, just zoomed in on the fourth epoch of observation}
\label{fig:oxford2}
\end{center}
\end{figure*}

\begin{figure}
\begin{center}
\includegraphics[width=8cm]{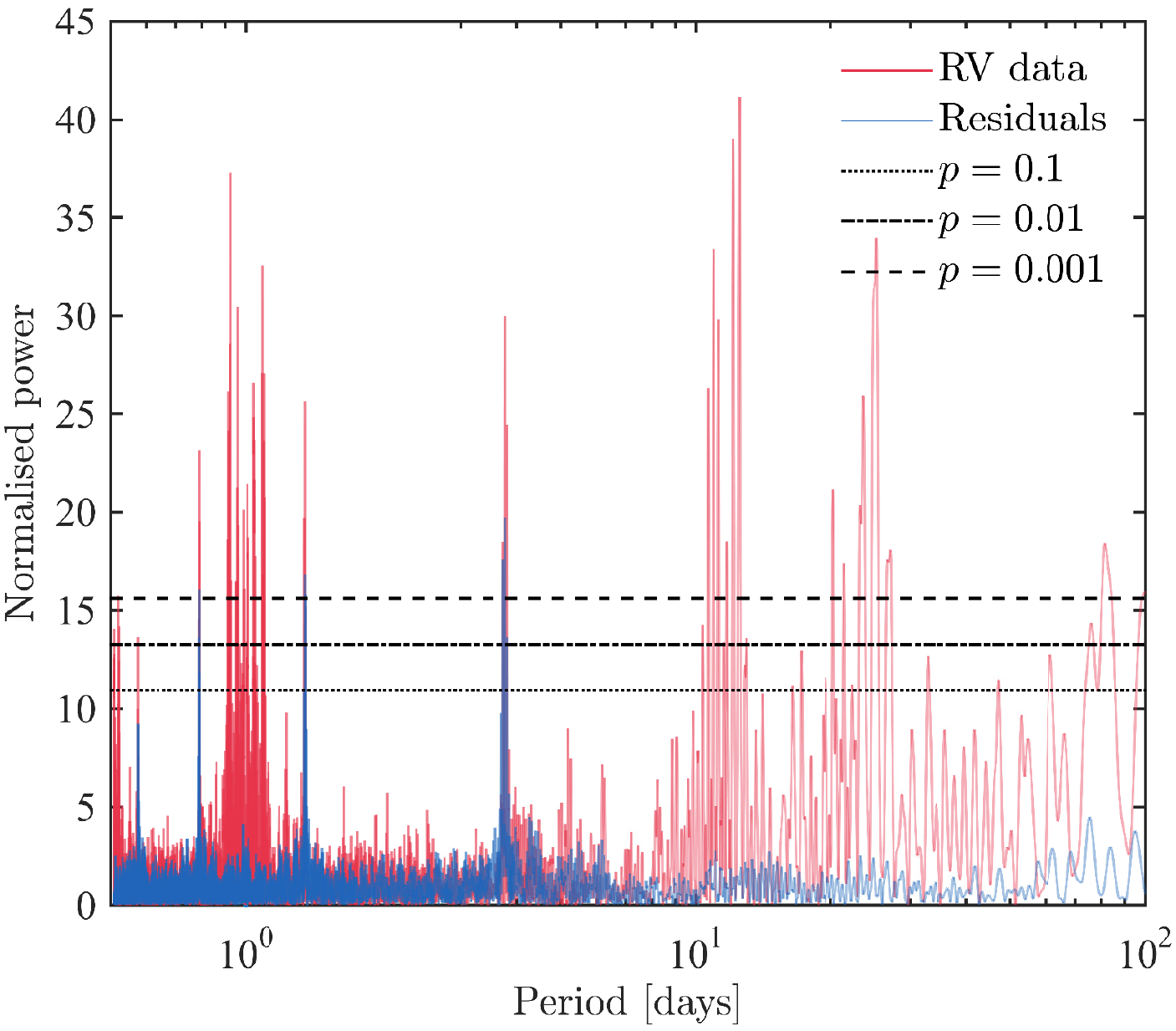}
\caption{Periodograms of the raw RVs (red) and of the residual RVs (blue) after removing the best GP plus 2-planet model to account for stellar signals and the 3.77 and 10.64-day planets present in the time series. The horizontal lines show from top to bottom the 0.1, 1 and 10\% p-values used as a first guess to estimate signal significance. The non-white residuals suggest an imperfect model.}
\label{fig:oxford3}
\end{center}
\end{figure}

In Figs.  \ref{fig:oxford1} and \ref{fig:oxford2}, we show the best-fit obtained by team 2 on system 2 of the RV fitting challenge. In Fig. \ref{fig:oxford3} we compare the periodogram of the raw RVs with the residual RVs after removing the same best-fit. We see that the GP fitted simultaneously to the RV, \logrhk\,and BIS SPAN allows to strongly mitigate the variations induced by stellar signals. For more examples on the technique used by team 2, readers are referred to \citet{Rajpaul-2015}.

%#######################################################################
% GROUP 3
%#######################################################################

\subsection{Team 3: M. Tuomi and G. Anglada-Escud\'e - Bayesian framework with first order Moving Average to account for stellar signals} \label{sect:2-2}

Team 3 is composed of, in order of contribution, M. Tuomi, G. Anglada-Escud\'e and H. R. A. Jones. This team analyzed the 14 systems of the RV fitting challenge. For each system, team 3 reported the period and semi-amplitude of the detected planets, as well as their best estimate of the stellar rotation period.

\subsubsection{General framework}

Team 3 also analyzed the data of the RV fitting challenge using a correlated noise model to account for stellar activity. However, in their case, team 3 used a first order moving average component with exponential smoothing. Team 1 used a GP and trained it on the \logrhk\,data to obtain the best hyper-parameters that are then used to fit the RVs. This implies therefore an iterative fitting process, which can be dangerous. Team 2 overcomed this problem by fitting simultaneously all the time series with a GP. However a strong assumption is made during the process: the covariance of short-term activity should be the same in the RVs than in the activity observables (\logrhk, BIS SPAN and FWHM). Using a first order moving average, team 3 avoided this assumption, which could imply significantly different results if this assumption turns out not to be valid. In addition, team 3 also considered in their RV model linear correlations with the different activity observables, therefore fitting everything at once implying a robust statistical approach.

{The details about the method used by team 3 to analyze the data of the RV fitting challenge can be found in the appendix of the paper (Section \ref{app:2-3}). In the next subsection we illustrate the method using as example system 2.}

\subsubsection{Example for system 2}

{\bf Here we present the results of the analysis of system 2 performed by team 3. 

\vspace{0.2cm}
{{\bf{Analysis of the activity indicators of system 2}}}
\vspace{0.2cm}

\noindent The team first analyzed the activity indicators by calculating the
likelihood-ratio periodograms (see Fig. \ref{fig:periodogram_tuomi}). This analysis indicates a strong signal
at a period of 12.5 days in the BIS SPAN time  series. The time-series for both
FWHM and \logrhk\,activity indices show a very strong signal at 24.9 days, twice
the period found in the BIS SPAN value, suggesting that this is the rotation period. The fact that BIS SPAN
shows a signal at half the rotation is the expected period for a spots showing
only one half of the rotation \citep[][]{Dumusque-2014b}. Signals in the activity indices were also searched when adding a first order moving average
component to the model but they didn't change the periods found much in this
case. Fig. \ref{fig:periodogram_tuomi} shows that these signals (12.4 and 24.9 days) are detected
well below the usual 1\% (even 0.1\%) p-value thresholds (shown as horizontal
lines).
\begin{figure*}
\center
\includegraphics[width=6cm]{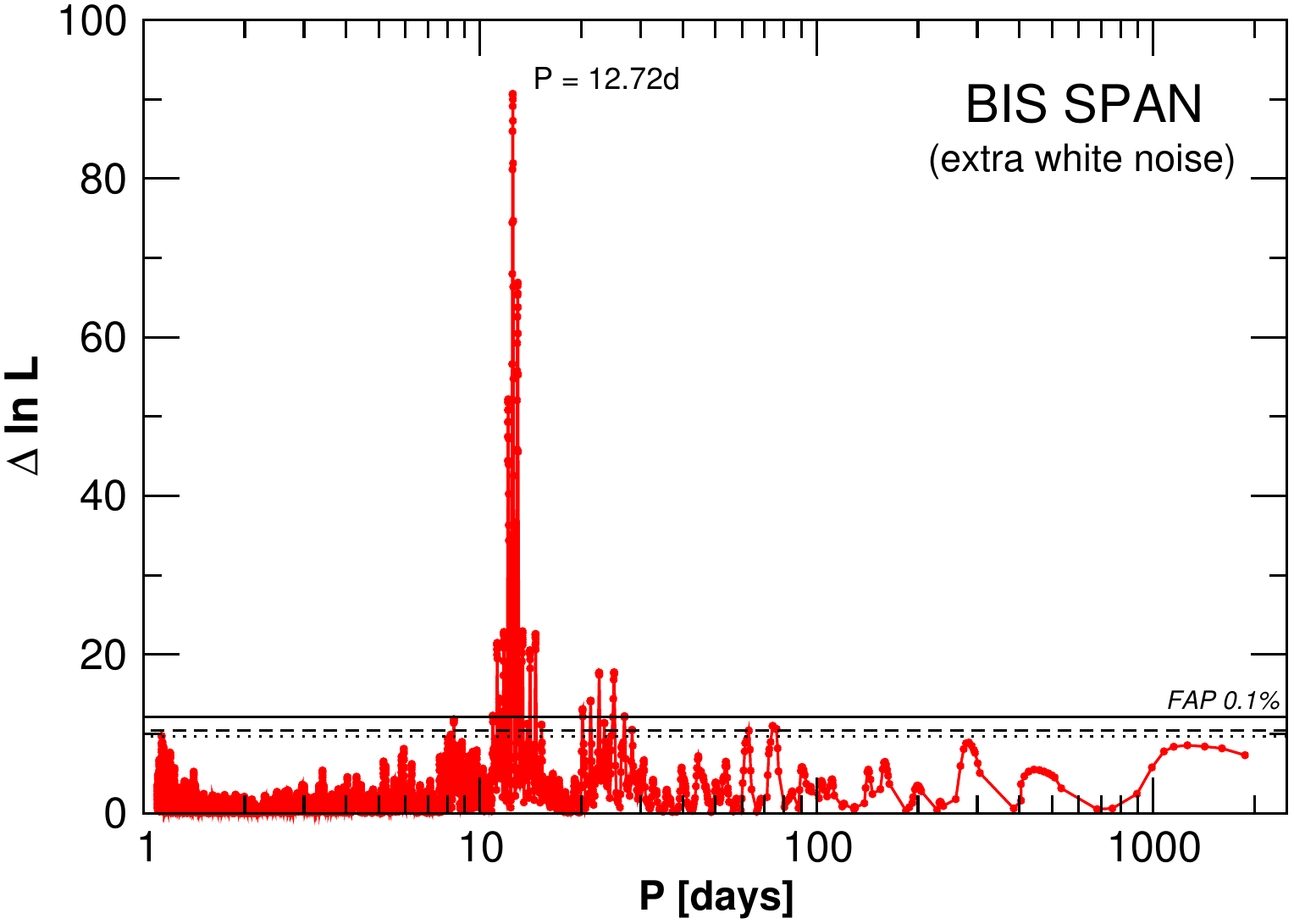}
\includegraphics[width=6cm]{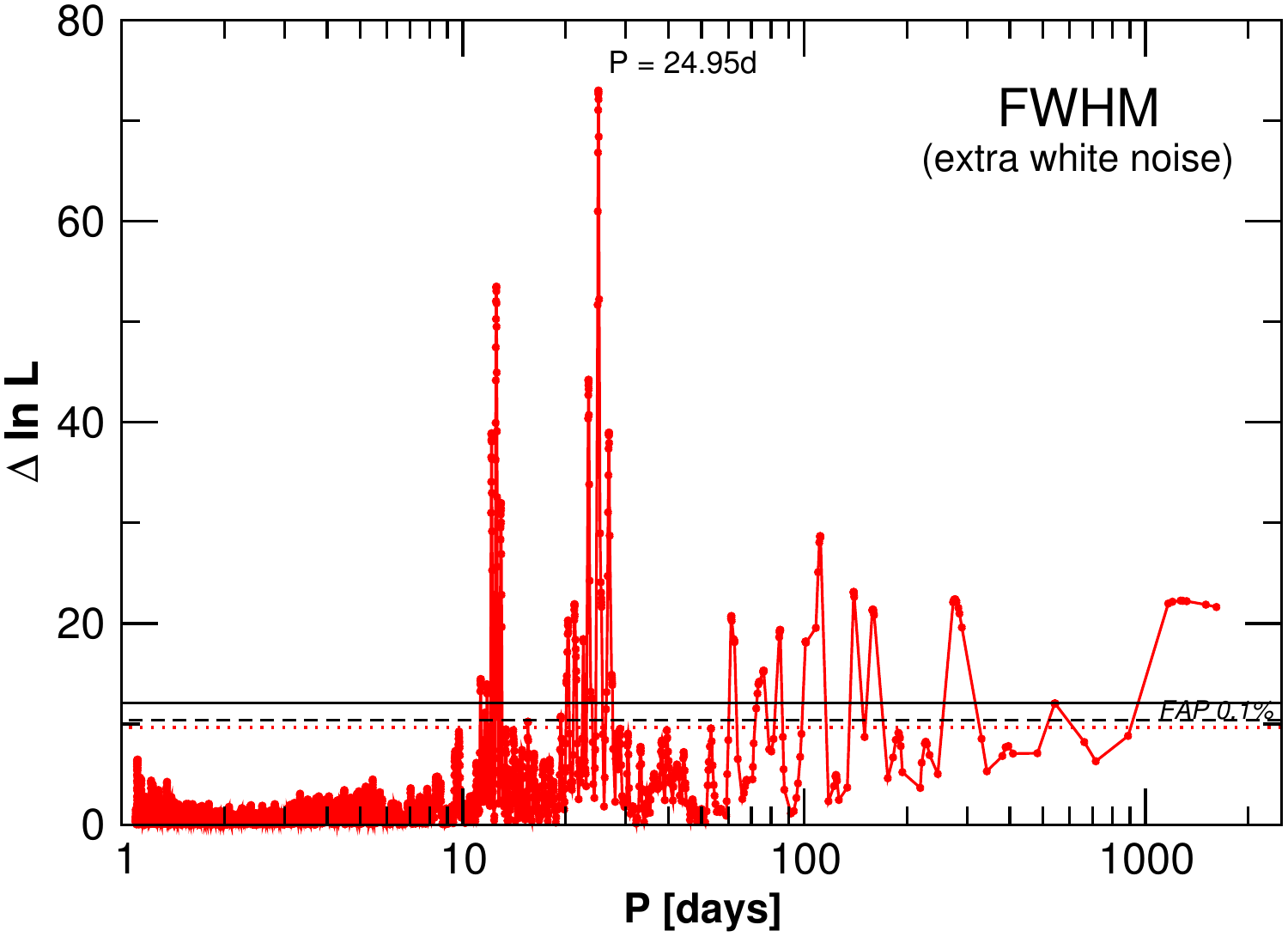}
\includegraphics[width=6cm]{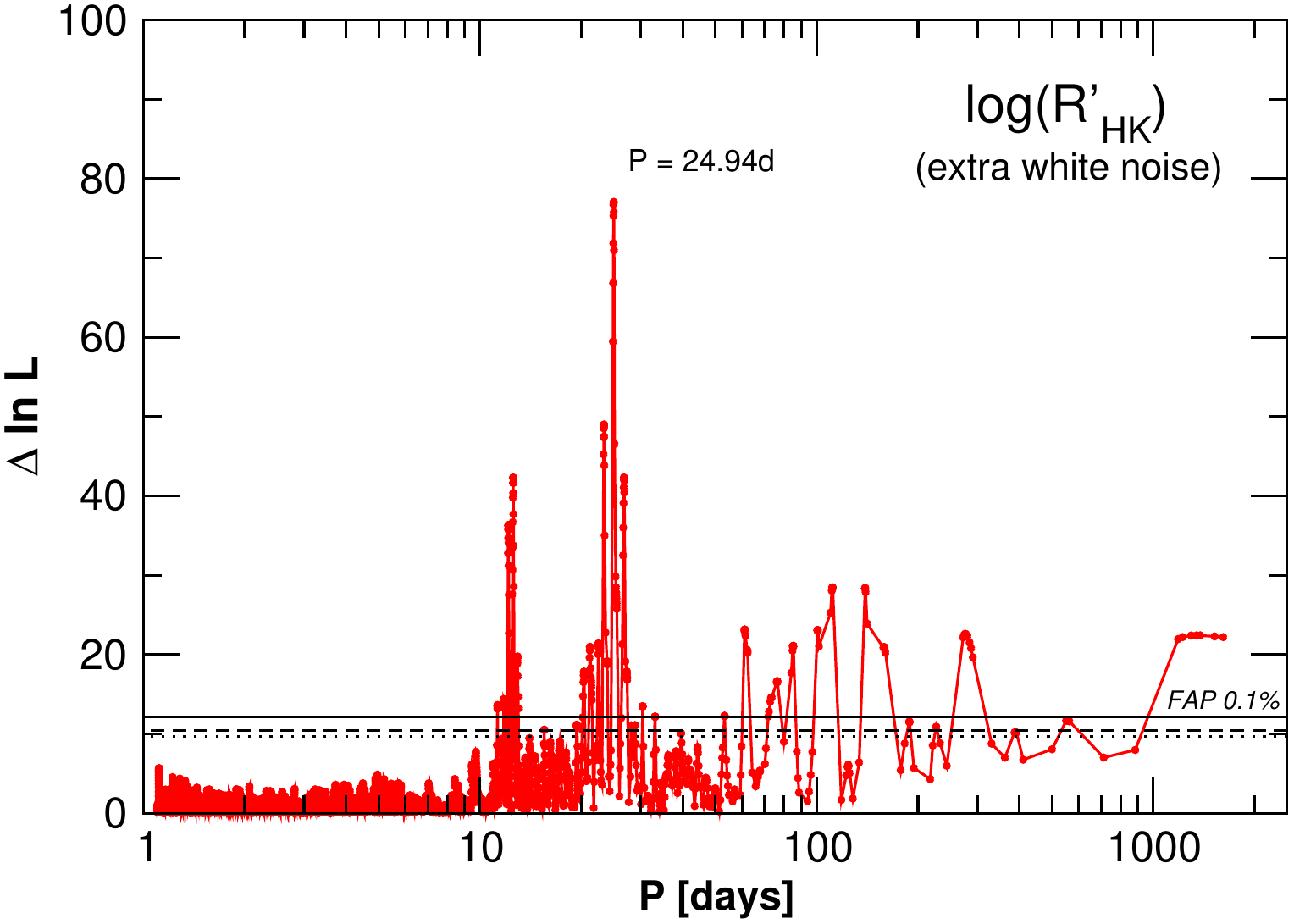}
\caption{Likelihood periodogram searches of the strongest periodic signal in the time series of the three activity indices provided (BIS SPAN, FWHM and \logrhk). Both the rotation period (most prominent in FWHM and \logrhk\,index) and its first harmonic (most prominent feature in BIS SPAN) are clearly seen in the activity time-series. In the three periodograms, the tested model contains sinusoid (ciruclar orbit), an offset, a linear trend and a extra white-noise jitter component.}
\label{fig:periodogram_tuomi}
\end{figure*}

\vspace{0.2cm}
{{\bf{Analysis of the RVs of system 2}}}
\vspace{0.2cm}

\noindent Team 3 analyzed the RVs with a statistical model including a linear trend
overtime, linear correlations with the three activity indicators given, and
correlated noise according to a first order moving average model. The
likelihood-ratio periodogram and signal searches without including correlation
terms would lead to a very different answer from the correct one. A model without including correlations 
would show a rather strong signal at the same period as the BIS SPAN and subsequent
inclusion of Keplerians requires fitting several sinusoids (all spuriously
generated by rotation and activity) before finally spotting the first real
planet candidate. The difference between the raw RVs and the RVs corrected from activity signal using linear correlations
is highlighted in Fig. \ref{fig:rv_tuomi}. A model including the linear correlation terms
directly spots three unambiguous Keplerian signals. Fig. \ref{fig:periodogram_tuomi2} shows
the likelihood periodograms with and without correlation terms for the first
signal search, and the subsequent likelihood periodograms obtained when
adjusting the signals under investigation together with all the model free
parameters. Likelihood periodograms are used to obtain a quick look at the
solution landscape when one new planet is added, but the actual search and
verification is then done using tempered delayed rejection and adaptive
Metropolis (DRAM) samplings \citep[][]{Haario-2001, Haario-2006} of the posterior density. This DRAM samplings allow to explore all the new
periods while allowing re-adjusting all the previously included Keplerian
signals. The posterior contours resemble the likelihood periodograms shown in 
Fig. \ref{fig:periodogram_tuomi2}. A more detailed description of the methodology used by team 3 is
given in the appendices of the paper.
\begin{figure*}
\center
\includegraphics[width=14cm]{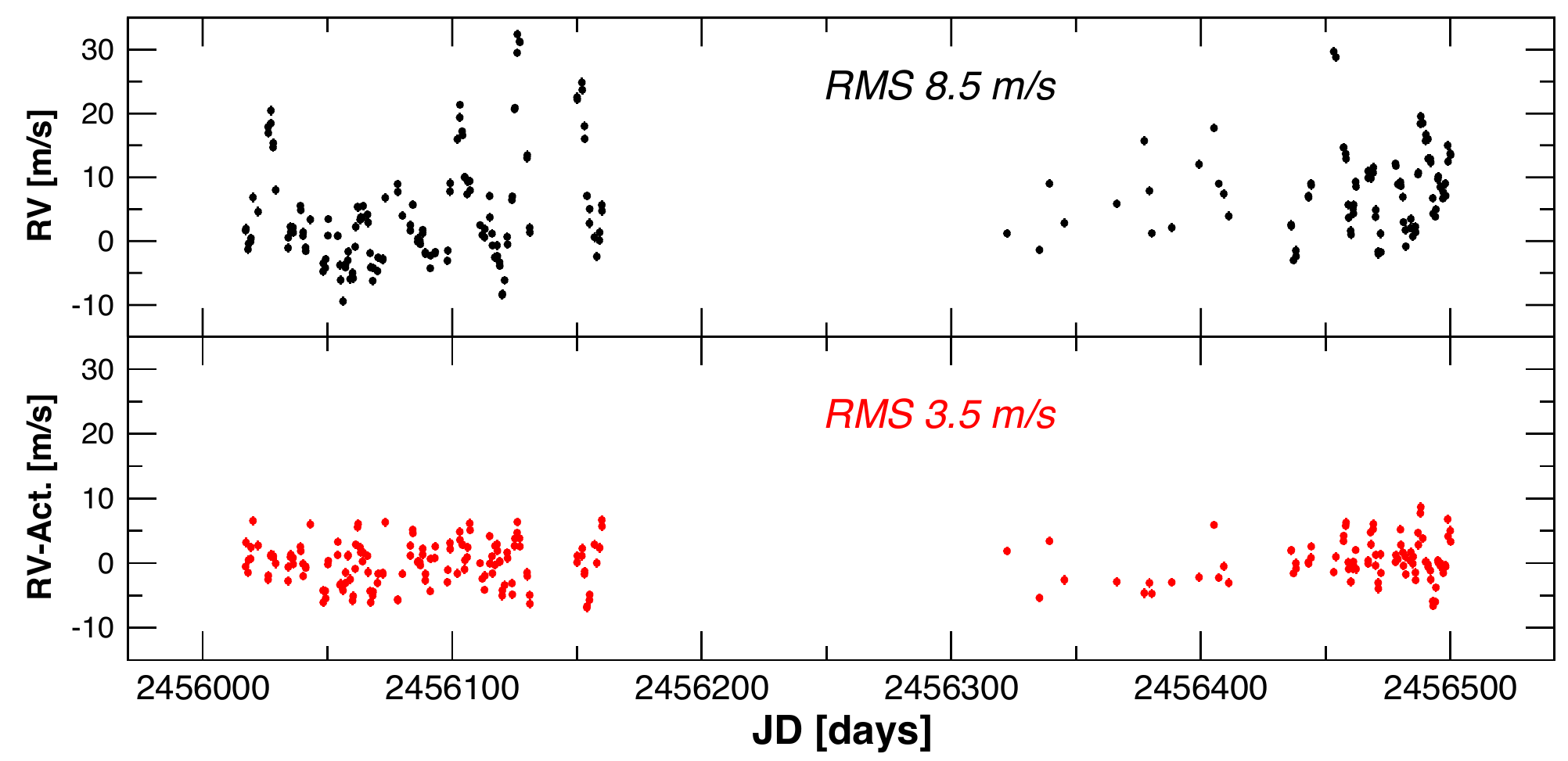}
\caption{Original Doppler measurements (top, black) and residual time-series (bottom, red) after adjusting a model with no periodic signal but with all activity correlation terms included. Even without subtracting any Keplerian signal, the reduction in the scatter in the RV time-series is apparent to the eye.}
\label{fig:rv_tuomi}
\end{figure*}
\begin{figure*}
\center
\includegraphics[width=6cm]{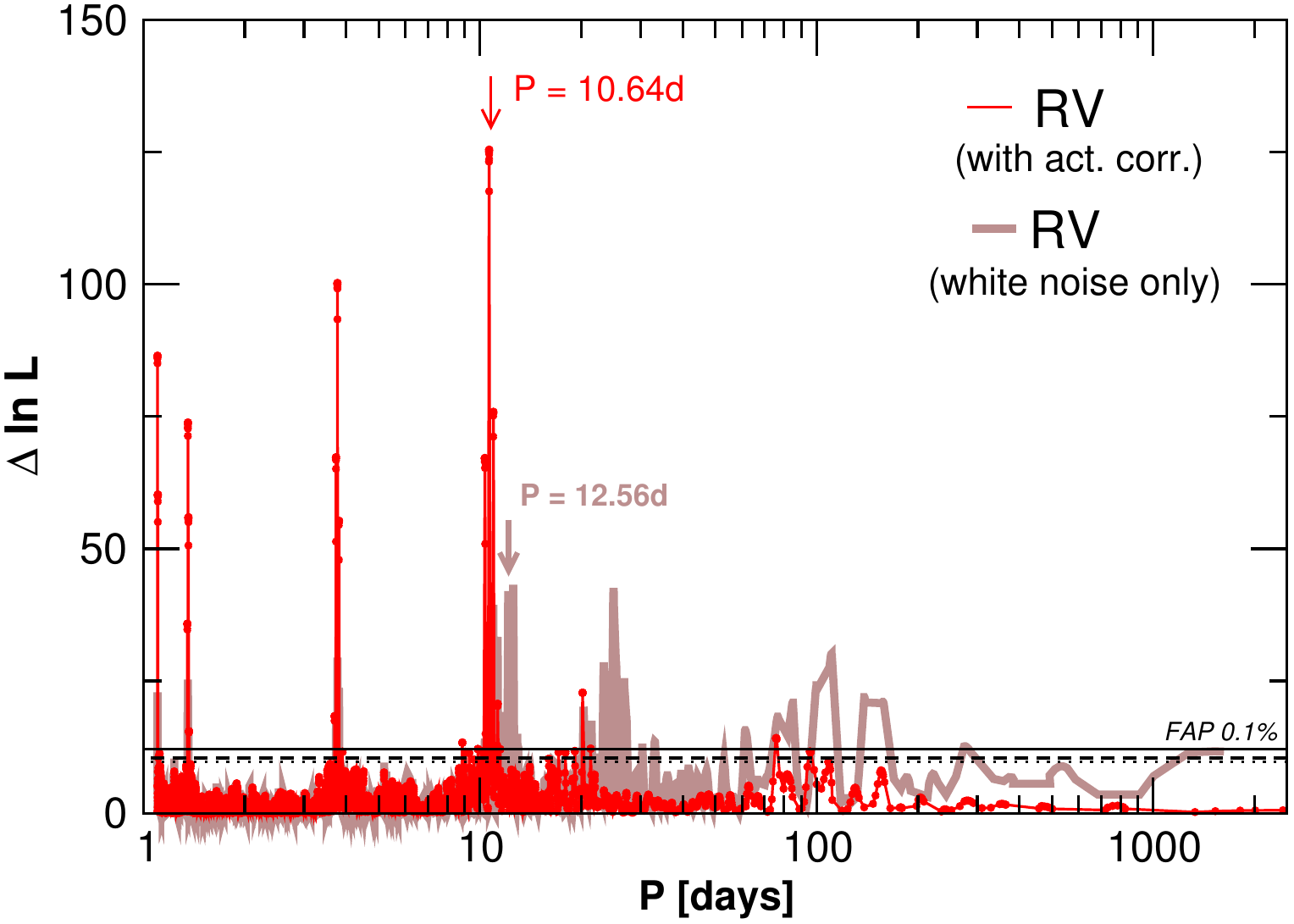}
\includegraphics[width=6cm]{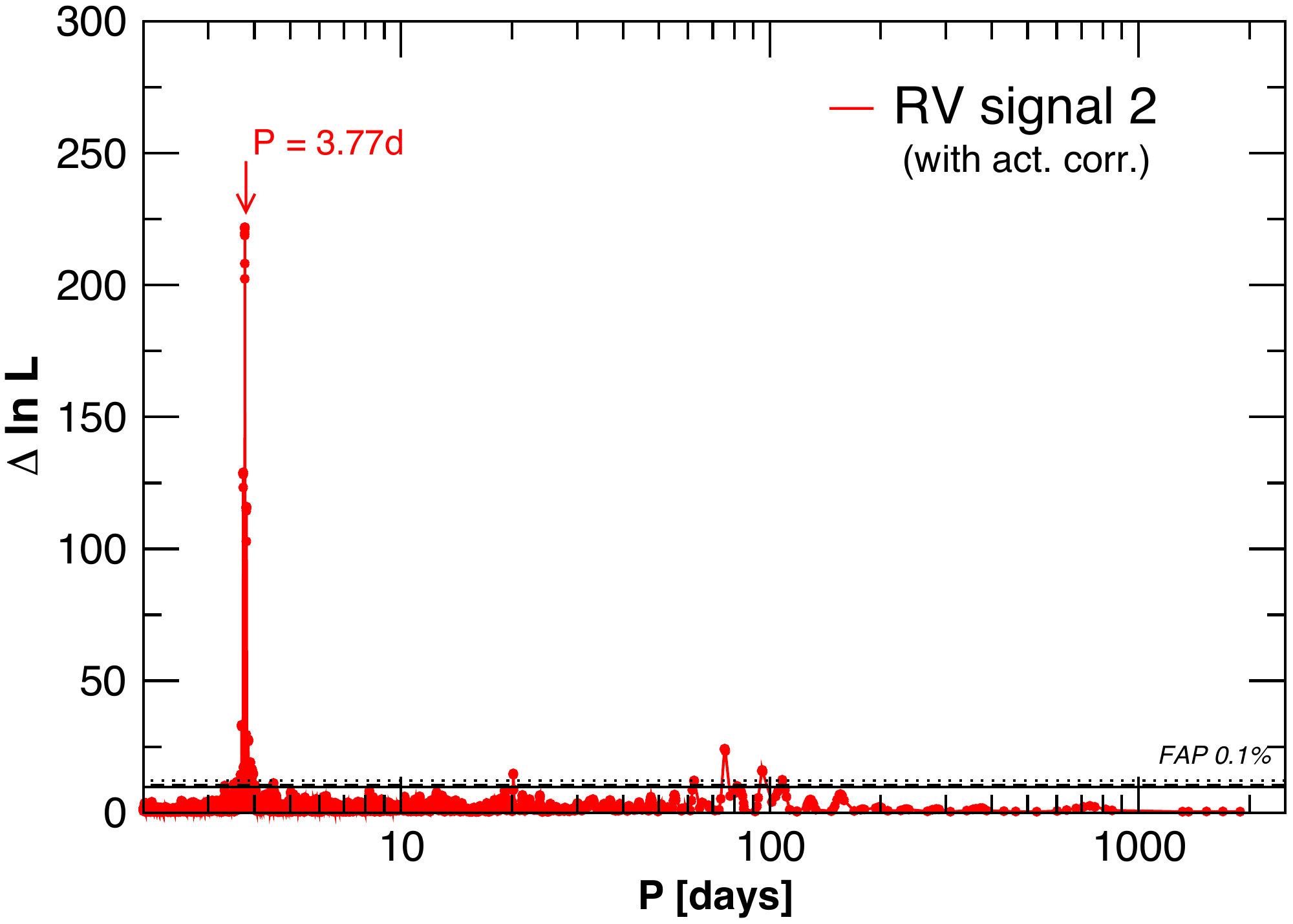}
\includegraphics[width=6cm]{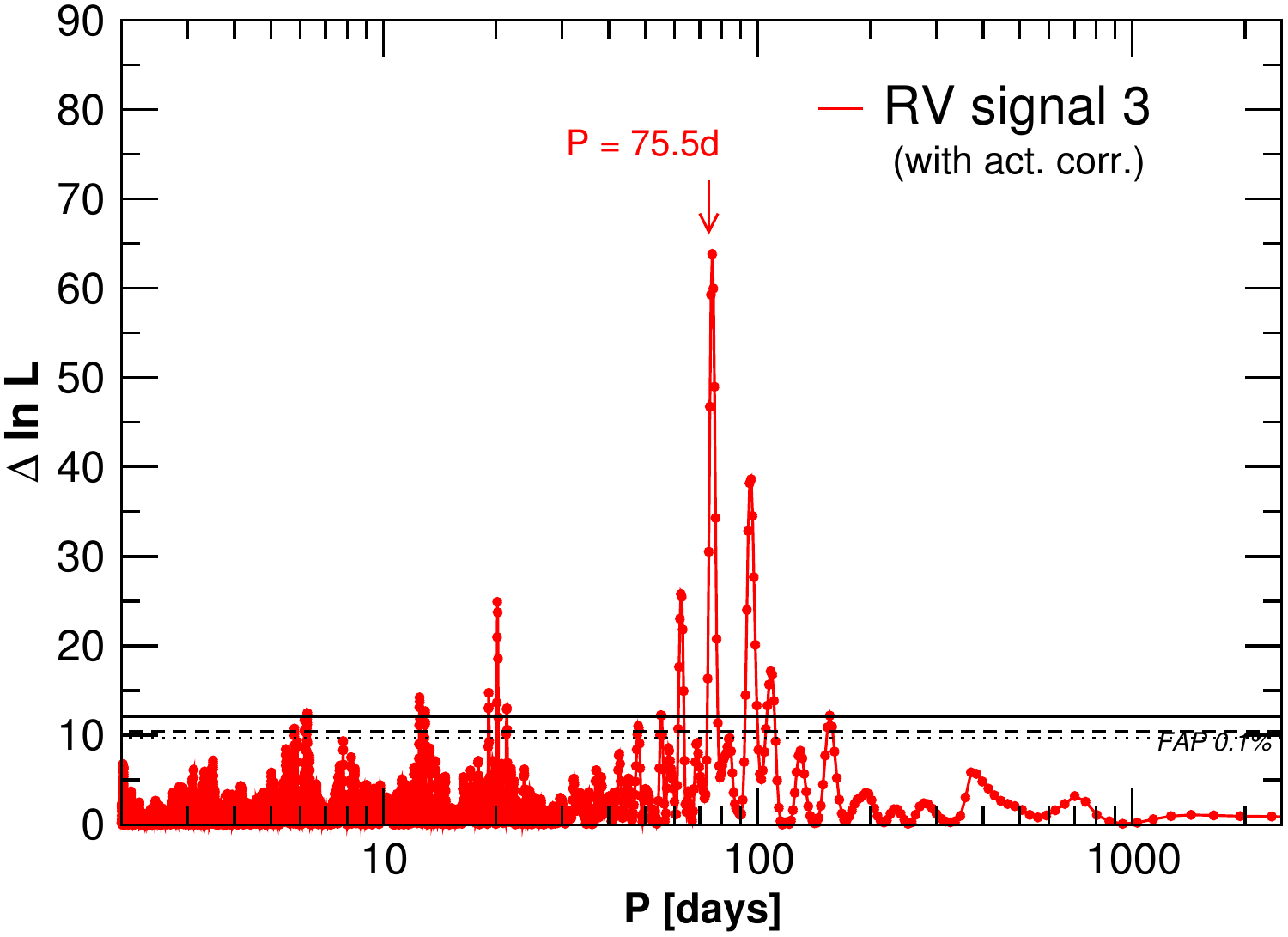}
\caption{Likelihood periodogram searches for the first, second and third signals in the time-series when adjusting the full noise model (first order moving average term plus linear correlations) at the same time as the Keplerian signals. In the first panel, we show the likelihood periodogram of the original RV time-series with and without accounting for correlations. In that case we would incorrectly conclude that the rotation period is a planet candidate. The correlations also reduce the impact of unstructured noise (eg. caused by high-frequency, sub-day jitter) which is made obvious by the much higher significance of the first planet candidate against the detection of the rotation period in the model without activity correlations.}
\label{fig:periodogram_tuomi2}
\end{figure*}

For a signal to be tagged as a planet candidate several conditions must be met:
\begin{itemize}
\item the period of the signal has to be well-constrained from above and below, 
\item the amplitude of the signal has to be statistically significantly different from
zero (the zero value must be excluded from the 99\% credibility interval), 
\item all other local maxima (peaks in posterior or likelihood periodogram) must be 100
times smaller than the preferred solution (uniqueness condition), and 
\item a model with this extra signal is statistically more significant than a model
without it when computing a Bayes factor using the mixture of posterior and
prior densities \citep[][]{Newton-1994}. 
\end{itemize}

As a threshold, it is often said that the
more complex model must have a Bayes factor 150 times higher than the 
simpler one \citep[][]{Kass-1995}, but team 3 uses a threshold of $10^4$ in Doppler time series
to acknowledge that the space of model is likely incomplete (eg. sinusoids 
fitting instrumental and activity features can still improve the model without
implying the presence of extra Keplerian signals). This threshold was considered
sufficient and adopted after examination and combination of data from the UVES
and HARPS spectrographs, that is, signals producing improvements on the model in
the UVES data below $10^4$ were often not confirmed when combining the
measurements with available HARPS data \citep{Tuomi-2014}. 
}

%#######################################################################
% GROUP 4
%#######################################################################

\subsection{Team 4: P. Gregory - Bayesian framework with apodized Keplerians to account for red noise} \label{sect:2-3}

Team 4 is composed of P. Gregory. He analyzed the first 5 systems of the RV fitting challenge. For each system, he reported the period, semi-amplitude, time of periastron passage, eccentricity and argument of periastron of the detected planets. He did not report stellar rotation periods.

\subsubsection{General framework}

P. Gregory analyzed the RV fitting challenge data set with a novel approach using apodized Keplerians. Stellar short-term activity creates semi-periodic signals due to stellar rotation and active region evolution, unlike the periodic signal induced by planets. He therefore decided to fit every significant signal in the RVs using Keplerians that could change their semi-amplitude as a function of time using a Gaussian apodization function. The two parameters that characterize each Gaussian apodization function, are fitted as free parameters. If the timescale appears to be much shorter than the time span of the RV measurements, it implies a non-stationary signal as a function of time, and therefore this signal is flagged as being induced by stellar activity. In addition, as team 3, P. Gregory also includes in its RV model a correlation with \logrhk\,to account for the RV effect of magnetic cycles in a statistically robust approach.

{A detailed step-by-step approach can be found in \citet{Gregory-2016}. We however give a small summary of the approach in the appendix of the paper (Section \ref{app:2-4}). In the following subsection, we give an example of how the method works for system 2.}

\subsubsection{Example for system 2}

\begin{figure*}
\begin{center}
\includegraphics[width=16cm]{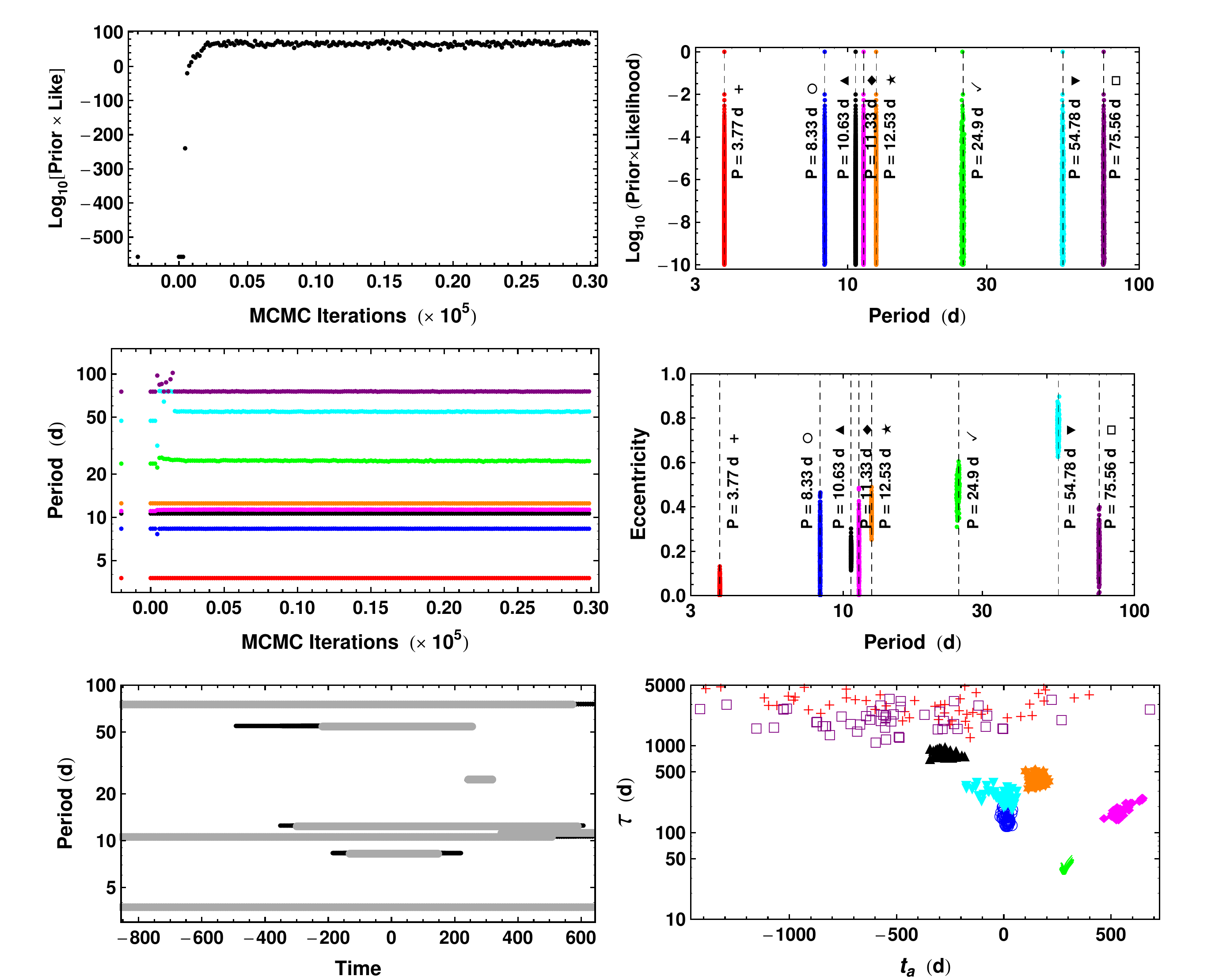}
\caption{\emph{Upper left:} Log$_{10}[\mathrm{Prior}\times \mathrm{Likelihood}]$ versus iterations for the 8 signal apodized Keplerian model used to fit system 2. \emph{Upper right:} Log$_{10}[\mathrm{Prior}\times \mathrm{Likelihood}]$ versus period showing the 8 periods detected. \emph{Middle left:} Values of the 8 unknown period parameters versus iteration number. \emph{Middle right:} Eccentricity parameters versus period parameters. \emph{Lower left:} Apodization window for each signal (gray trace for MAP values of the apodization time constant $\tau$ and the apodization window center time $t_a$, black for a representative set of samples which is mainly hidden below the gray). \emph{Lower right:} Apodization time constant versus apodization window center time for each signal (Credit: \citealt{Gregory-2016}).}
\label{fig:2-2-0}
\end{center}
\end{figure*}

Fig. \ref{fig:2-2-0} shows an example of the Bayesian Fusion MCMC results for system 2, for which apodized Keplerians were used to characterize both planetary and stellar activity signals.

\begin{figure*}
\begin{center}
\includegraphics[width=16cm]{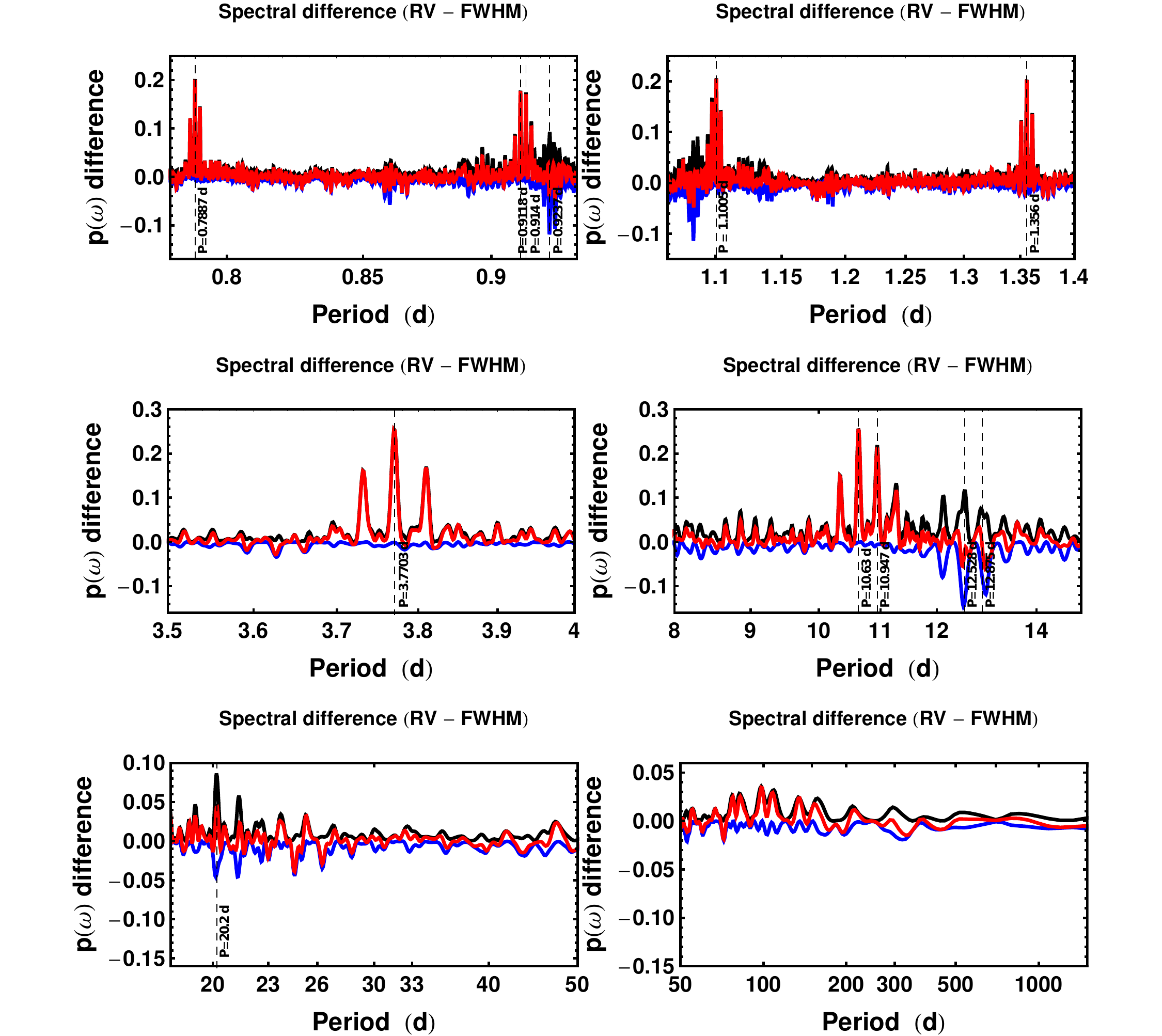}
\caption{Comparison between GLS periodograms of RV and FWHM data for system 2, each modified by the removal of the best-fit \logrhk\,correlation model {(hereafter called modified RV and FWHM)}. There are three traces in each of the 6 panels. The black trace is the modified RV periodogram, the blue trace is the negative of the modified FWHM periodogram, and the red trace is the difference, i.e., the black trace plus the blue traces. Each plot show an interesting portion of the periodogram. {As explained in the text, interesting signals that can be seen in both the modified RV (black) and the modified FWHM (blue) are not fitted as they probably are the result of stellar activity, while signals only present in the modified RVs are considered }(Credit: \citealt{Gregory-2016}).}
\label{fig:2-2-1}
\end{center}
\end{figure*}

Fig. \ref{fig:2-2-1} shows a comparison of the GLS periodograms of RV and FWHM data, each corrected by the removal of the best-fit \logrhk\,correlation model. There are three traces in each of the 6 panels. The black trace is the modified RV periodogram. The blue trace is the negative of the modified FWHM periodogram and the red trace shows the difference, i.e., the black trace plus the blue traces. One can clearly see blue trace counterparts to the black trace around periods of 12.5 and 20.2 days, indicating they are likely stellar activity signals. In contrast there is no blue trace counterpart to the peaks near 3.77 and 10.6 days.
%

%#######################################################################
% GROUP 5
%#######################################################################

\subsection{Team 5: Geneva team - Bayesian framework with white noise} \label{sect:2-4}

Team 5 is composed of, in order of contribution, R. D\'iaz, D. S\'egransan and S. Udry. Team 5 analyzed the two first systems of the RV fitting challenge. For each system, team 5 reported the period, semi-amplitude and eccentricity of the detected planets. Team 5 did not report stellar rotation periods.

\subsubsection{General framework}

Team 5 considered models including several Keplerians to represent planetary companions and velocity variations related to stellar rotation period. To account for the correlation between RV and \logrhk\,induced by magnetic activity cycles, team 5 fitted \logrhk\,using a third order polynomial. The posterior distributions of this fit are then used as priors for a third order polynomial added to the model used to fit the RVs. In other words, the low-frequency structure of the \logrhk\,is included in the model of the RVs. A source of additional white noise whose amplitude was set to scale linearly with activity (measured by the \logrhk\,index) was added to the model to account for the known correlation between stellar activity level and velocity jitter. The model is fully described in \citet{Diaz-2016}. 

After fitting several models with different numbers of Keplerians, team 5 used two estimation of the Bayesian evidence log\,$\mathcal{Z}$ to compare between models: the \citet{Chib-2001} and the \citet{Perrakis-2014} estimators. The best model is the one that exhibits a reasonable instrumental white noise component and not too-low evidence using the two estimators.

\section{Methods to deal with stellar signals without using a Bayesian Framework} \label{sect:3}

%#######################################################################
% GROUP 6
%#######################################################################

\subsection{Team 6: A. Hatzes - Pre-whitening} \label{sect:3-1}

Team 6 is composed of A. Hatzes. He analyzed the 14 systems of the RV fitting challenge. For each system, he reported the period and semi-amplitude of the detected planets, as well as the best estimate of the stellar rotation period.

\subsubsection{General framework}

A. Hatzes used the so-called \emph{pre-whitening} procedure.
One first computes the Discrete Fourier Transform (DFT) to find the dominant
peak in the Fourier amplitude spectrum. A least squares sine fit to the data
is made using this frequency and the resulting sine fit is subtracted
from the data. One then performs a DFT on the residual data to find
the next dominant peak. In finding a subsequent signal in the data, a simultaneous fit is made using all the previously found sine functions.
The process stops when the final peak amplitude is less than about four
times the mean amplitude of the surrounding noise peaks. \citet{Kuschnig-1997}
 established that this corresponds to a $p$-value of  about 1\%. Examples of this process performed
on RV data can be found for CoRoT-7 \citep[][]{Hatzes-2010} and GL 581 \citep[][]{Hatzes-2013b}. 
The DFT analysis was only performed out to the nominal Nyquist
frequency of 0.5 d$^{-1}$. This means that periods shorter than 2 days
were not actively searched for even if they were in the data.

Given the large number of time series, the program {\textsc{Period04}} \citep[][]{Lenz-2005}
was used to perform \emph{pre-whitening}. This program provides a convenient
environment for computing DFTs, selecting peaks in the amplitude spectrum,
fitting those, and searching for additional signals in the residual data.
The program also provides an option for computing the signal-to-noise
ratio (amplitude of a peak divided by the computed mean noise level).

The \emph{pre-whitening} procedure was performed on all time series. 
Significant peaks found in the RV data were compared to those
found in the activity indicators (\logrhk, BIS SPAN, and FWHM). If a significant
peak found in the RV did not have a corresponding peak in the activity
indicators it was identified as a planet. Signals found in the RVs and in the activity indicators
 were attributed to activity, with the dominant peak chosen as the rotation period. 

Note that in this case, only fitting sine waves is dangerous, because not removing the correct solution for a planet
or stellar signals can then perturb the residuals and lead to the detection of false positives.

%#######################################################################
% GROUP 7
%#######################################################################

\subsection{Team 7: Brera team - Filtering in frequency space} \label{sect:3-2}

Team 7 is composed of, in order of contribution, F. Borsa, G. Frustagli, E. Poretti and M. Rainer, from INAF - Brera Astronomical Observatory. Their activities are in the framework of the {\it Global Architecture of Planetary Systems} (GAPS) project \citep[e.g.][]{Poretti-2015}. Team 7 analyzed the 14 systems of the RV fitting challenge. For each system, team 7 reported the period, semi-amplitude, time of periastron passage, eccentricity and argument of periastron of the detected planets. Team 7 did not report stellar rotation periods.

\subsubsection{General framework}

Team 7 decided to try an approach that is not model dependant for stellar signals. This approach is based on filtering signals in the frequency domain of the RVs, using the frequency information found in the different activity indicators. {The details about the method used by team 7 to analyze the data of the RV fitting challenge can be found in the appendix of the paper (Section \ref{app:2-7}). In the next subsection we illustrate the method using as example system 2.}

\subsubsection{Example for system 2}

\begin{figure*}
\begin{center}
\includegraphics[width=18cm]{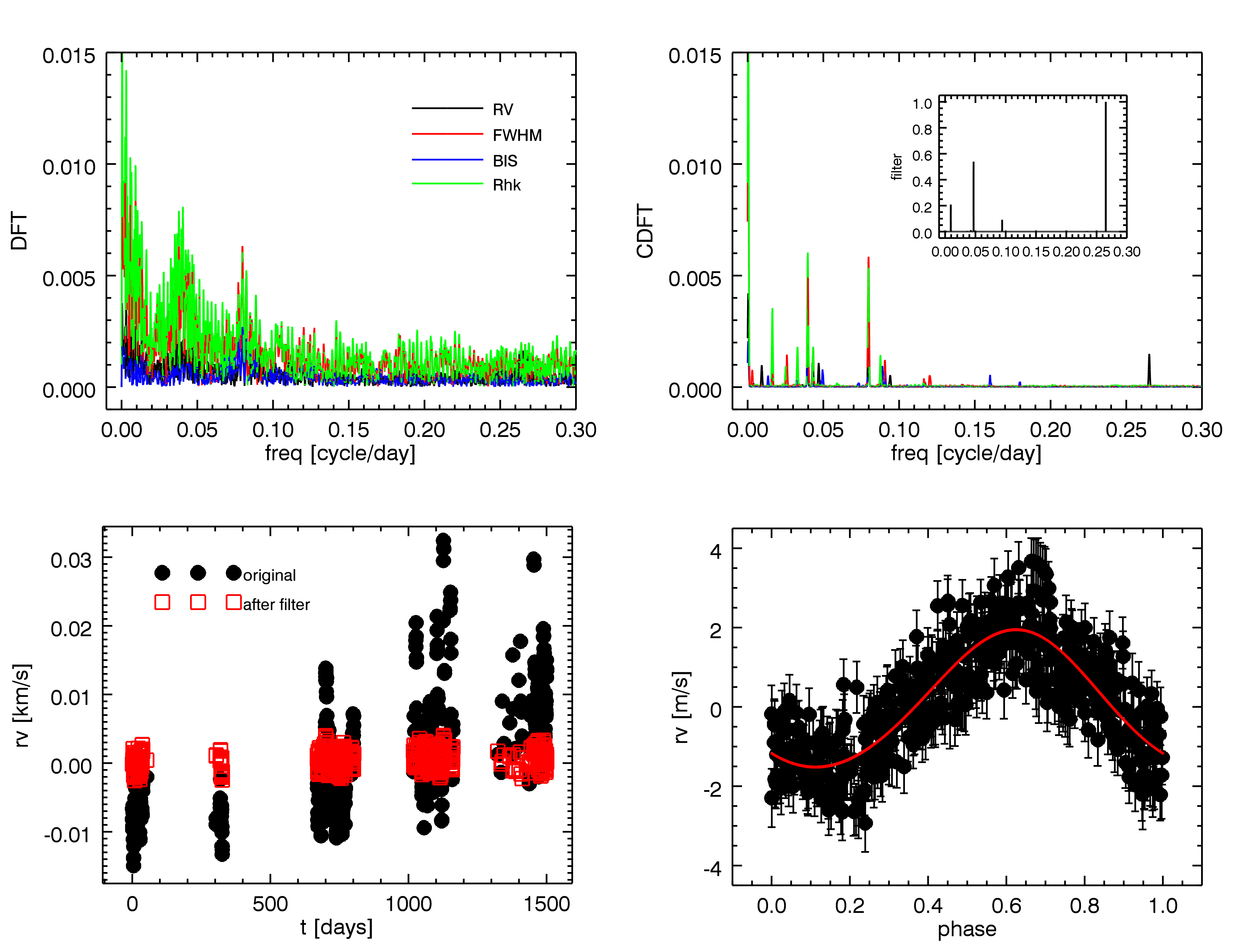}
\caption{Different steps used by team 7 to detect the 3.77-day planetary
signal present in System 2. \emph{Upper left panel:} DFT of the RVs, FWHM,
BIS SPAN and \logrhk . \emph{Upper right panel:} Cleaned DFT (CDFT)
obtained
after applying the CLEAN algorithm to the DFTs of all observables. In the
inset, the resulting pass-planet filter in the frequency domain, that is
later applied to the DFT of the RVs to mitigate the effect of stellar
signals. \emph{Lower left panel:} The RVs before (black circles) and after
(red squares) applying the pass-planet filter. \emph{Lower right panel:}
The phase-folded result of the Keplerian fit to the 3.77-day planetary
signal found in the RVs.}
\label{fig:brera}
\end{center}
\end{figure*}

As an example, Fig. \ref{fig:brera} shows the different steps used by team 7 to
detect the 3.77-day planetary signal present in System 2.

%#######################################################################
% GROUP 8
%#######################################################################

\subsection{Team 8: IMCCE team - Compressed sensing and frequency filtering} \label{sect:3-3}

Team 8 is composed of, in order of contribution, N. Hara, F. Dauvergne and G. Bou\'e, from the Institut de M\'ecanique C\'eleste et de Calcul des \'Eph\'em\'erides in Paris (IMCCE). Team 8 analyzed the 14 data sets of the RV fitting challenge. For each system, team 8 reported the period and semi-amplitude of the detected planets, as well as the best estimate of the stellar rotation period.

\subsubsection{General Framework}

The IMCCE team used an approach based on Compressed Sensing~\citep[or Compressive Sampling, see][]{donoho2006_2,candes2006} and frequency filtering. This method was devised to avoid fitting the planets one by one. Indeed, after removing a certain number of signals, the tallest peak of the periodogram of the residuals might not correspond to a real planet. One might even face the case where the maximum of the periodogram of the raw data is significant but spurious. The usual way to circumvent this issue is to fit a complete model accounting for several planets and sometimes noise parameters. In that case, one uses MCMC methods or genetic algorithms to explore the whole parameter space and avoid being trapped in a suboptimal local minimum. The compressed sensing framework allows in a certain sense to search all the planets at once while considering an objective function which has only one minimum. As the minimization problem is convex, one can design fast algorithms. 

The key is to use an \textit{a priori} information: the signal is supposed to be ``simple'' in a certain sense. Here, it means that there exists a set of vectors, termed the dictionary, such that a linear combination of a few of its entries reproduces the signal. For instance, dictionaries made of wavelets are appropriate to represent most images. This feature is exploited for the JPEG2000 format~\citep{taubman2002}. The image is stored via its significant wavelet coefficients, which are a few compared to the total number of pixels. The MP3 and AAC audio formats rely on the same principles.

In our case, the movement of a star due to its planets is quasi-periodic. In other words, it is a linear combination of a few sine functions $\exp^{-\mathrm{i}\omega t}$ and $\exp^{\mathrm{i}\omega t}$. However, we do not measure the motion of the star only. The signal is also contaminated - and in the present challenge, dominated - by the stellar activity. To account for this effect, in addition to the sine functions, frequency-filtered FWHM, bisector span and $\log R_{HK}'$ were incorporated into the dictionary. 

Once the dictionary is defined, one searches for a combination of a few of its element which is close to the observations. The output of that procedure are the coefficients of the linear combination of dictionary elements reproducing the data within a certain tolerance. This vector is plotted versus the frequency (or the period), just like a GLS periodogram. To avoid the confusion with this one, the figure obtained is termed $\ell_1$-periodogram~(see Fig. \ref{imcce_rvchallenge2}).

The procedure applied to the RV fitting Challenge data was at an intermediate stage of development and is outlined in subsection~\ref{example_imcce} on an example. This preliminary method was not very robust, and was greatly improved afterwards. For a precise description of the most recent version, see~\citep[][]{Hara-2016}, where system 2 of the RV Fitting Challenge is treated in detail along with real radial velocity signals.

\subsubsection{Example for system 2}
\label{example_imcce}

\begin{figure*}[!th]
\noindent
%\centering
%\hspace{-0.4cm}
%\begin{tikzpicture}
\includegraphics[width=18.7cm]{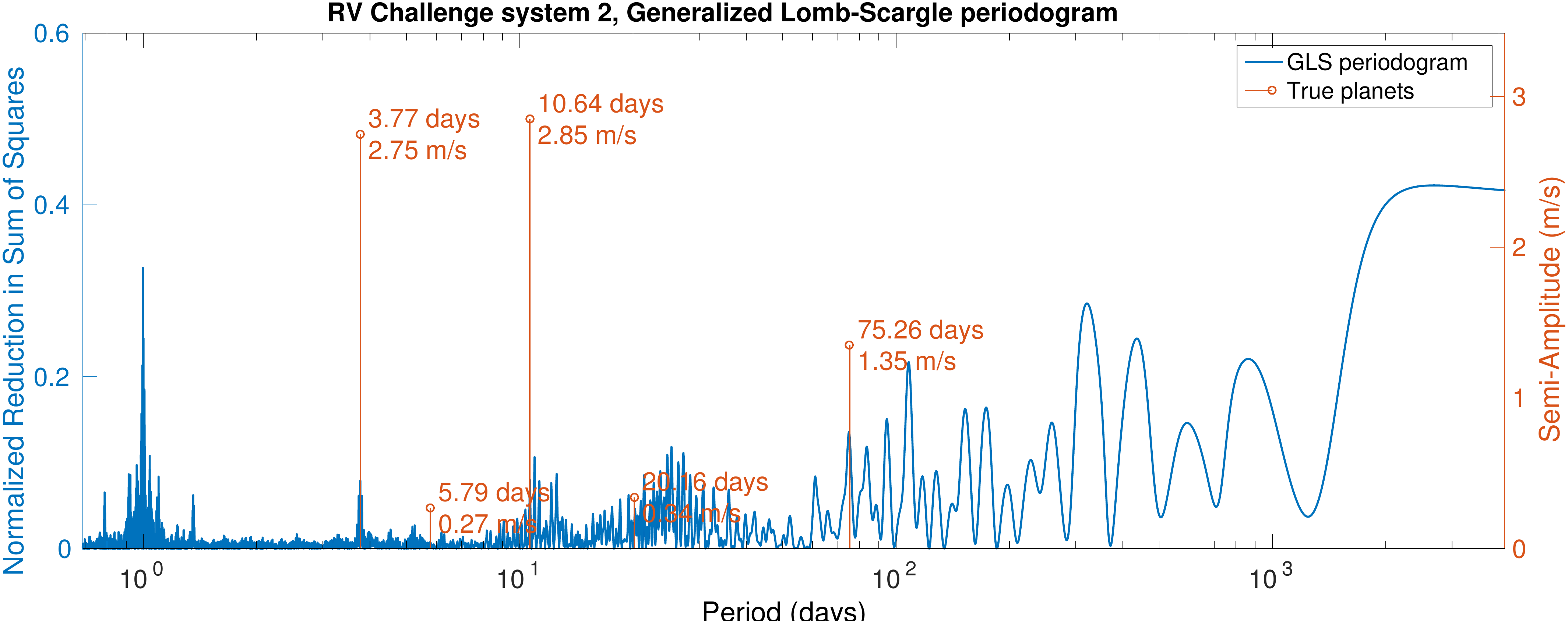}
\includegraphics[width=18cm]{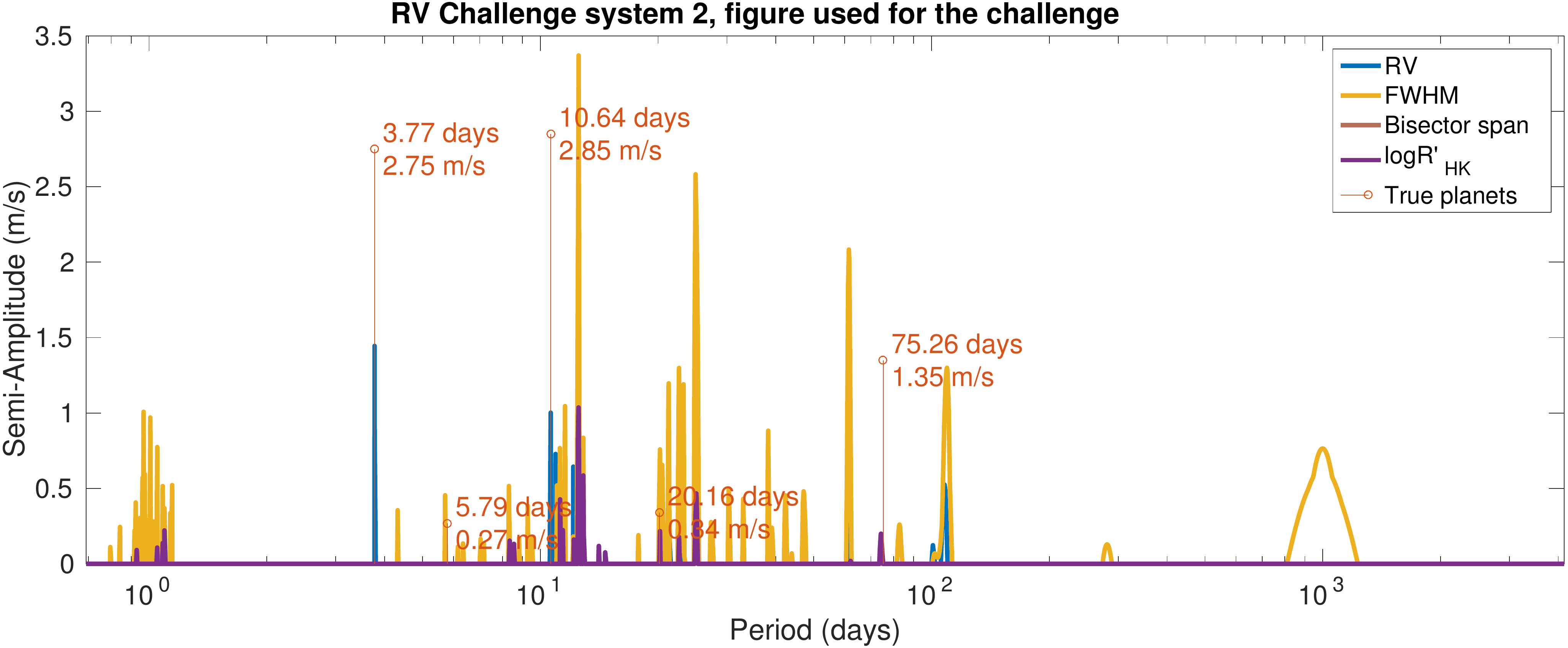}
\includegraphics[width=18cm]{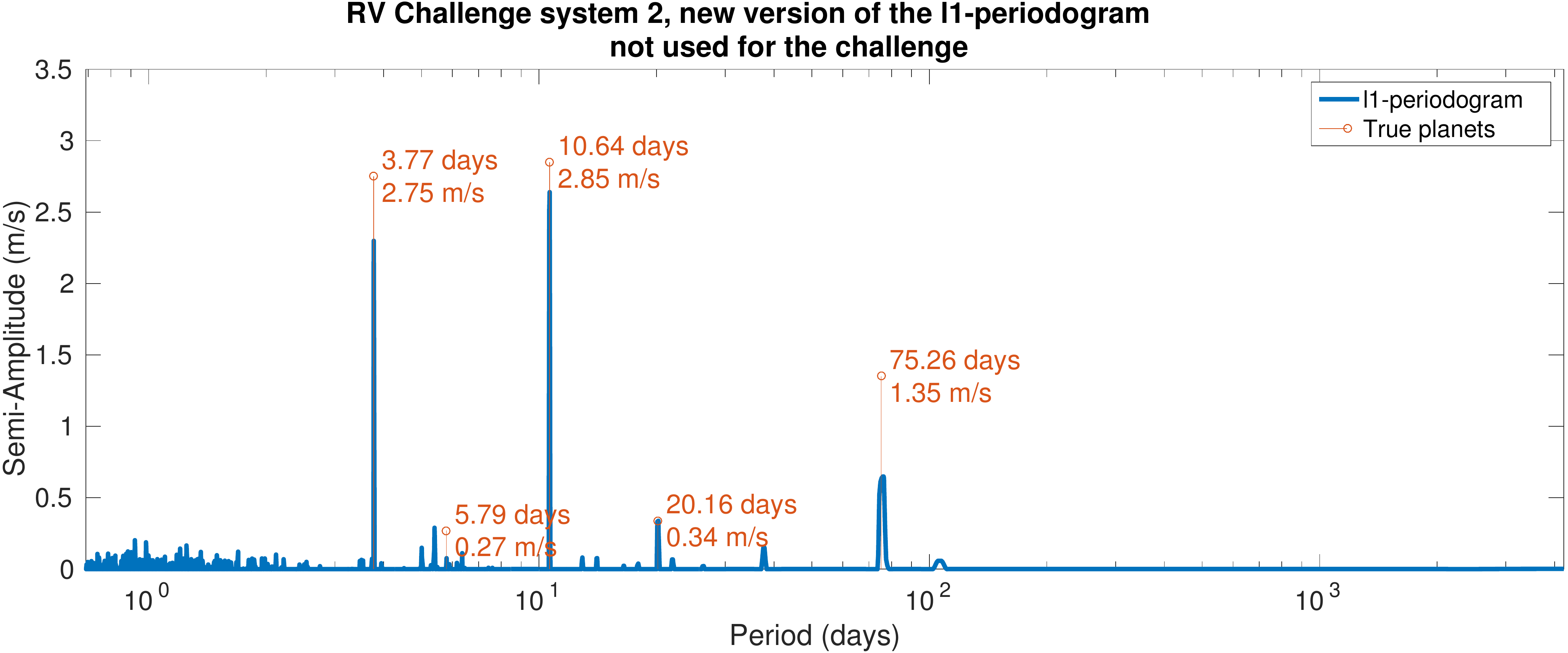}

%\path (0.,0) node[above right]{\includegraphics[scale=0.43]{imcce_rvchal1_glsraw.pdf}};
%\path (1.2,7.15) node[above right]{a)};
%\begin{scope}[yshift=-7.6cm]
%\path (0,0) node[above right]{\includegraphics[scale=0.43]{imcce_rvchal1_oldl1.pdf}};
%\path (1.2,7.12) node[above right]{b)};
%\end{scope}
%\begin{scope}[yshift=-15.4cm]
%\path (0.,0) node[above right]{\includegraphics[scale=0.43]{imcce_rvchal1_newl1.pdf}};
%\path (1.2,7.2) node[above right]{c)};
%\end{scope}
%\end{tikzpicture}
\caption{{\emph{Top: }GLS of the RV fitting challenge system 2 (raw time series). The red horizontal lines correspond to the true planetary signals injected into the data. \emph{Middle: }Figure used for the challenge, blue: $\ell_1$-periodogram of the RVs, yellow: $\ell_1$-periodogram of FWHM, purple: $\ell_1$-periodogram of \logrhk. Only the 3.77-day signal was detected by team 8. \emph{Bottom:} New version of the $\ell_1$-periodogram \citep[][]{Hara-2016}.}}
\label{imcce_rvchallenge2}
\end{figure*}

{In this section the results of the method applied to the second system of the RV Fitting Challenge are presented. The system contains five planets, whose periods and true amplitudes are represented in red in Fig. \ref{imcce_rvchallenge2}. The blue curve on the top plot of Fig. \ref{imcce_rvchallenge2} is the GLS periodogram of the raw RV data~\citep[][]{Zechmeister-2009}, displayed for comparison.

The dictionary is made of sine functions $\exp^{-\mathrm{i} \omega_k t}$ and $\exp^{\mathrm{i} \omega_k t}$  for $n=3.10^5$ frequencies, $\omega_k =  3 k\pi/n$ radian per day, $k=0..n-1$. To obtain a representation of the activity, each of the FWHM, bisector span and \logrhk\,signals are bandpass filtered by projection onto five families of orthonormal polynomials. The family $j$, $j=1..5$ is made of $D_j-d_j+1$ polynomials of degrees $d_j$ to $D_j$. Here $D_{j}=d_{j+1}-1$, and $d_1=0$, $d_2=15$, $d_3=60$, $d_4=160$, $d_5=300$.

The $\ell_1$-periodograms - in a version used for the RV Fitting Challenge - of the RV, FWHM and bisector span are represented in the middle plot of Fig.\ref{imcce_rvchallenge2}. We then selected planetary signals following this principle: If a "high" peak of the $\ell_1$-periodogram of the radial velocity data is sufficiently far from peaks of the  $\ell_1$-periodogram of the FWHM and bisector span, it is retained. If this peak is too close to at least one of the peaks of the FWHM or bisector span $\ell_1$-periodogram, it is discarded.  Here, the three smallest peaks do not appear. The 10.64 days periodicity does show up, but was discarded due to its proximity to features of the other signals. Finally, we see clearly the 3.77 days periodicity, which was indeed selected. 

As said above, the IMCCE team kept on working on the method. If one subtracts the estimated activity of the star before performing the $\ell_1$ minimization and with further improvements, one obtains the bottom plot in Fig. \ref{imcce_rvchallenge2}. In this case, the four strongest signals appear without ambiguity. There are also two signals close to 5.4 and 37 days, which are signatures of the first harmonics of the eccentric orbits at 10.64 and 75.26 days. We however could not see clearly the 5.79 days periodicity, which seems to be buried in the noise. A further study shows that the five planets plus the harmonic of the 10.64 days orbit are statistically significant in some sense. This system is treated in detail in~\citep[][]{Hara-2016}.}

%#######################################################################
% Results
%#######################################################################

\section{Results} \label{sect:4}

In this section, we analyze the results of the different teams, in term of stellar rotation periods found, planetary signals detected and false positives announced. We also discuss the accuracy of orbital parameters recovered, as well as the realism of the simulated systems generated for the purpose of the RV fitting challenge. Because a wrong estimate of the stellar rotation period can lead to the detection of false positives, this is the first point we discuss.

\subsection{Detection of stellar rotation periods} \label{sect:4-0}

As described in detail in \citet{Dumusque-2016a}, the data of the RV fitting challenge include planetary signals, but also stellar signals, i.e., oscillations, granulation, short-term activity and long-term activity signals. Among all these stellar signals, the most difficult to deal with is short-term activity, induced by active regions, i.e., spots and plages, rotating with the stellar surface. Because several active regions rotating with the star are present simultaneously on the stellar surface, the observed RV signal induced by short-term activity is characterized by signals at the stellar rotation period $P_{\mathrm{rot}}$, and its harmonics \citep[$P_{\mathrm{rot}}/2$, $P_{\mathrm{rot}}/3$, \dots,][]{Boisse-2011}. Therefore, one of the very important aspects to differentiate between planetary signal and short-term activity signal is the detection of the stellar rotation period. If a signal is found in the RVs with a periodicity similar to $P_{\mathrm{rot}}$ or its harmonics, it is very likely that this signal is induced by active regions. For teams that used a GP regression to model short-term activity, it is essential for them to have a good guess of the stellar rotation period, otherwise the flexibility of a GP applied to the RVs could model planetary signals.
\begin{figure*}
\begin{center}
\includegraphics[width=14cm]{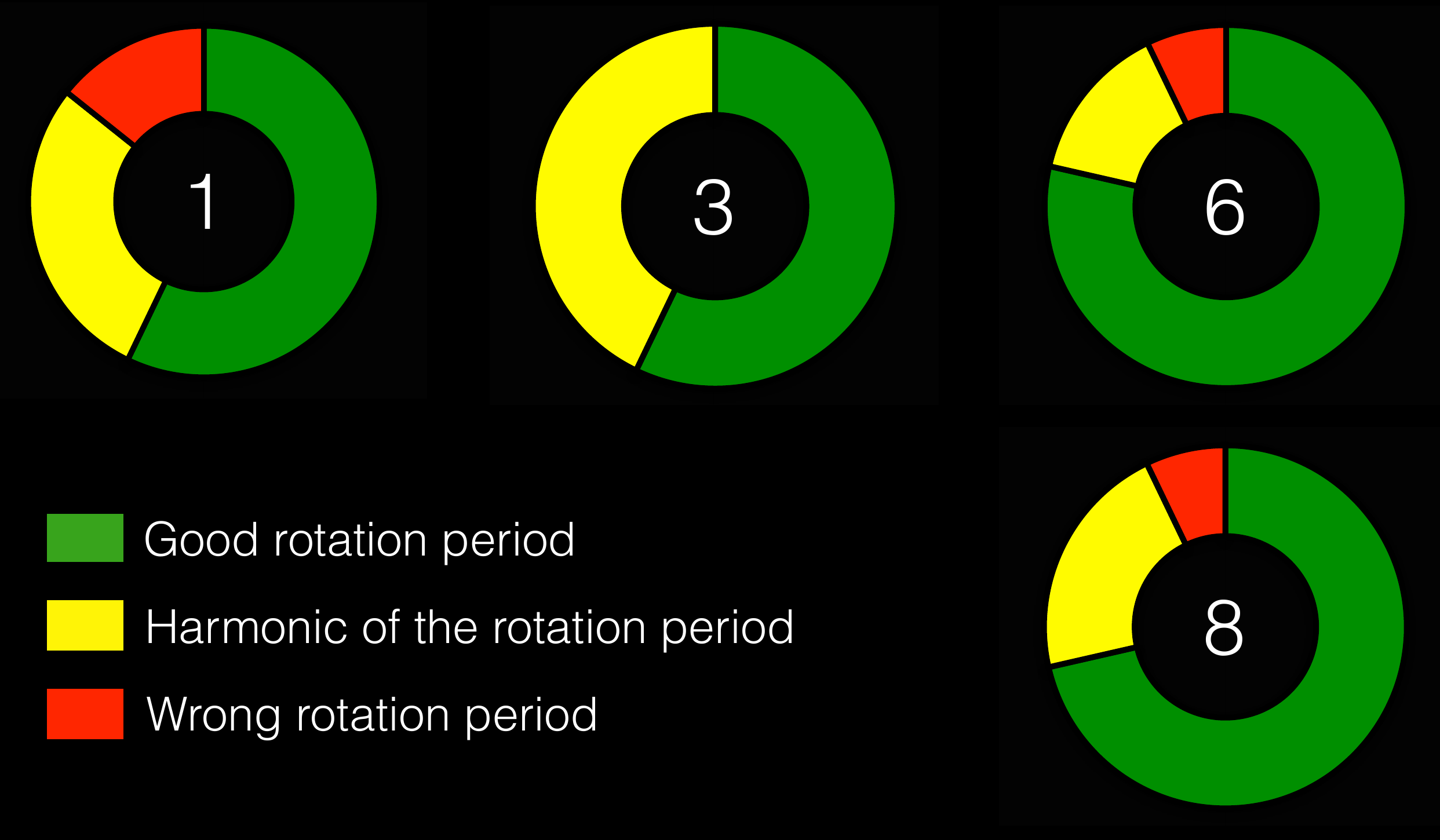}
\caption{Stellar rotation periods detected by teams 1, 3, 6 and 8 for the 14 systems of the RV fitting challenge.}
\label{fig:4-0-1}
\end{center}
\end{figure*}

Teams 1, 3, 6 and 8 reported stellar rotation periods for all the RV fitting challenge systems, while the other teams did not explicitly derive such an estimate. All the teams that performed this analysis compared the signals found in the RVs with the ones found in the other observables sensitive to activity, i.e., the calcium activity index \logrhk, the BIS SPAN and the FWHM. A clear detection at the same period in the RVs and any other observables was assigned to a non-planetary component, because certainly due to short-term activity. Team 1 used a GLS periodogram to have a first estimate of the stellar rotation period, and then fitted a GP using a MCMC starting at this first estimate (see Equation \ref{eq:2-0}). Team 3 smooths the time series of the different activity observables using a moving average, and then looked for the stellar rotation period using a GLS periodogram. Team 6 looked at significant peaks in the DFT of the RVs and different activity observables, and finally team 8 at significant peaks in the GLS and $\ell_1$-periodograms of the RVs, the BIS SPAN and the FWHM.

In Figs. \ref{fig:summary_detection1} to \ref{fig:summary_detection6}, we show for each team and system, the stellar rotation period found. In Fig. \ref{fig:4-0-1}, we summarize those results by classifying them as:
\begin{itemize} 
\item correct rotation period detected in green,
\item detection of an harmonic of the true rotation period in yellow, 
\item and wrong rotation period announced in red). 
\end{itemize}
There are only a small number of wrong periods detected, which is positive. We will see their impact in Section \ref{sect:4-1} when analyzing the detection of planetary signals. However, we can see that in 20 to 45\% of the cases, the different teams detected a harmonic of the stellar rotation period, and not the true period used to model the data. This can be a problem as we expect short-term activity to induce signals at $P_{\mathrm{rot}}$, $P_{\mathrm{rot}}/2$, $P_{\mathrm{rot}}/3$ and so on, but not at 2$P_{\mathrm{rot}}$ and 3$P_{\mathrm{rot}}$. Therefore, if the detected stellar rotation period is in fact $P_{\mathrm{rot}}/2$, it is possible to confuse an activity signal found at $P_{\mathrm{rot}}$ with a planet. We will see further that this case happened when team 1 analyzed systems {5, 9, 10, 11 and 12, team 3 analyzed system 13 and team 7 analyzed system 5.}

We know that depending on the active region configuration on the stellar surface, and depending on the sampling of the data, the first harmonic of the rotation period ($P_{\mathrm{rot}}/2$) can have more power that the fundamental \citep[$P_{\mathrm{rot}}$,][]{Boisse-2011}. To prevent confounding a signal due to short-term activity with a planetary signal when the detected stellar rotation period is a harmonic of the real period, an easy solution is simply to reject signals at 2$P_{\mathrm{rot}}$ and 3$P_{\mathrm{rot}}$. In addition, we can use the average activity level of a star {and its spectral type} to guess its rotation period. First demonstrated by \citet{Noyes-1984}, and then updated by \citet {Mamajek-2008}, a relation exists between the average \logrhk\,level, {the spectral type} and the stellar rotation period, with a few day error. Therefore, when analyzing the RVs of an old star, for which the rotation period is longer than 20 days, using such a relation can tell us if the period detected is the true stellar rotation period, or a harmonic of it. {The spectral type of the stars were not given for the RV fitting challenge, therefore the different teams could not use this relation to estimate rotation periods. This was done on purpose because, as explain in detail in \citet{Dumusque-2016a}, only the variation of the \logrhk\,was properly simulated for the RV fitting challenge, and not the absolute value of it. Therefore using the average \logrhk\,level given in the RV fitting challenge dataset to calculate rotation periods would give wrong rotational period estimates.} If this would have been possible, several yellow detections in Fig. \ref{fig:4-0-1} would turn green, therefore only leaving a few mistakes. This is something that should be taken into account for any further RV fitting challenges.

%Regarding mistakes for the determination of stellar rotation periods:
%%
%\begin{itemize}
%\item team 1 made 2 mistakes, on system 9 and 12, 46 days instead of 36-40 days, and 26 days instead of 40 days, respectively, 
%\item team 3 made no mistakes,
%\item team 6 made 1 mistake, on system 7, 55.2 days instead of 40 days,
%\item team 8 made 1 mistake on system 1, 29.6 days instead of 25 days.
%\end{itemize}
%%
Because the different teams did not make mistakes on the same systems, it is difficult to conclude on the origin of these mistakes. However, among all the teams, team 3 performed the best as it reported no mistakes. Therefore, the technique used by this team to estimate stellar rotation periods from the activity observables (\logrhk, FWHM and BIS SPAN), consisting on first modeling correlated noise using a moving average and then analyzing the residuals to find the stellar rotation period, seems to be the most robust (see Section \ref{sect:2-2}).

%HOW TUOMI FOUND TEH Prot
%I modeled correlated noise in the activity indices (Rhk, FWHM and BIS, to enable comparing them), as well as any trends and the white (Gaussian) noise, and simply had a look at the residual periodograms. Three attached for the first test set, indicating a clear signal. This is different from looking at the periodograms of the activity data as such, as I have removed intrinsic correlations from the data that occur in a time-scale of few days.
%4 day exponential decay

\subsection{Detection of planetary signals} \label{sect:4-1}

In this section, we analyze the results of the different teams in terms of planetary detection. {In total, 14 planetary systems were given, including a total of 45 simulated planetary signals and 6 published planetary signals probably present in real datasets 9,10,11 and 14 ($\alpha$ Centauri Bb and Corot-7b, c and d). As we can see in \citet{Dumusque-2016a}, and in Figs.  \ref{fig:summary_detection1} to \ref{fig:summary_detection6}, the semi-amplitude of the planetary signals was ranging between 0.16 and 5.85 \ms, with rather low eccentricities}. The RV fitting challenge time series for systems 9, 10, 11, and 14 are real observations obtained with HARPS, while the time series of all the other systems were simulated using the modelization of stellar signals described in \citet{Dumusque-2016a}. Note that no planetary signal were present in simulated systems 4, 8 and 13. In addition, no planetary signal was injected in real system 9, however the time series used for this system are the published HARPS measurements of $\alpha$ Centauri B that led to the discovery of an Earth-mass planet \citep[][]{Dumusque-2012}, therefore a 0.5 \ms planetary signal {might} be recovered when analyzing this system.

\vspace{0.2cm}
\subsubsection{Comparing the results obtained by the different teams} \label{sect:4-1-0}
\vspace{0.2cm}

{To compare the results between the different team, we define the $K/N$ ratio as:
\begin{equation} \label{eq:4-1-0}
K/N = \frac{K_{\mathrm{pl}}}{\mathrm{RV}_{\mathrm{rms}}}  \sqrt{N_{\mathrm{obs}}},
\end{equation}
where $K_{\mathrm{pl}}$ is the semi-amplitude of each planetary signal, $N_{obs}$ is the number of observation in each system, and $\mathrm{RV}_{\mathrm{rms}}$ is the RV rms} of each system once the best-fit of a model consisting of a linear correlation with \logrhk\,plus a second order polynomial as a function of time was removed. This model allows removal of the effect of magnetic cycles \citep[][]{Meunier-2013, Dumusque-2011} and any long-term drift in the RVs due to binary companions.
\begin{figure*}
\begin{center}
\includegraphics[width=9cm]{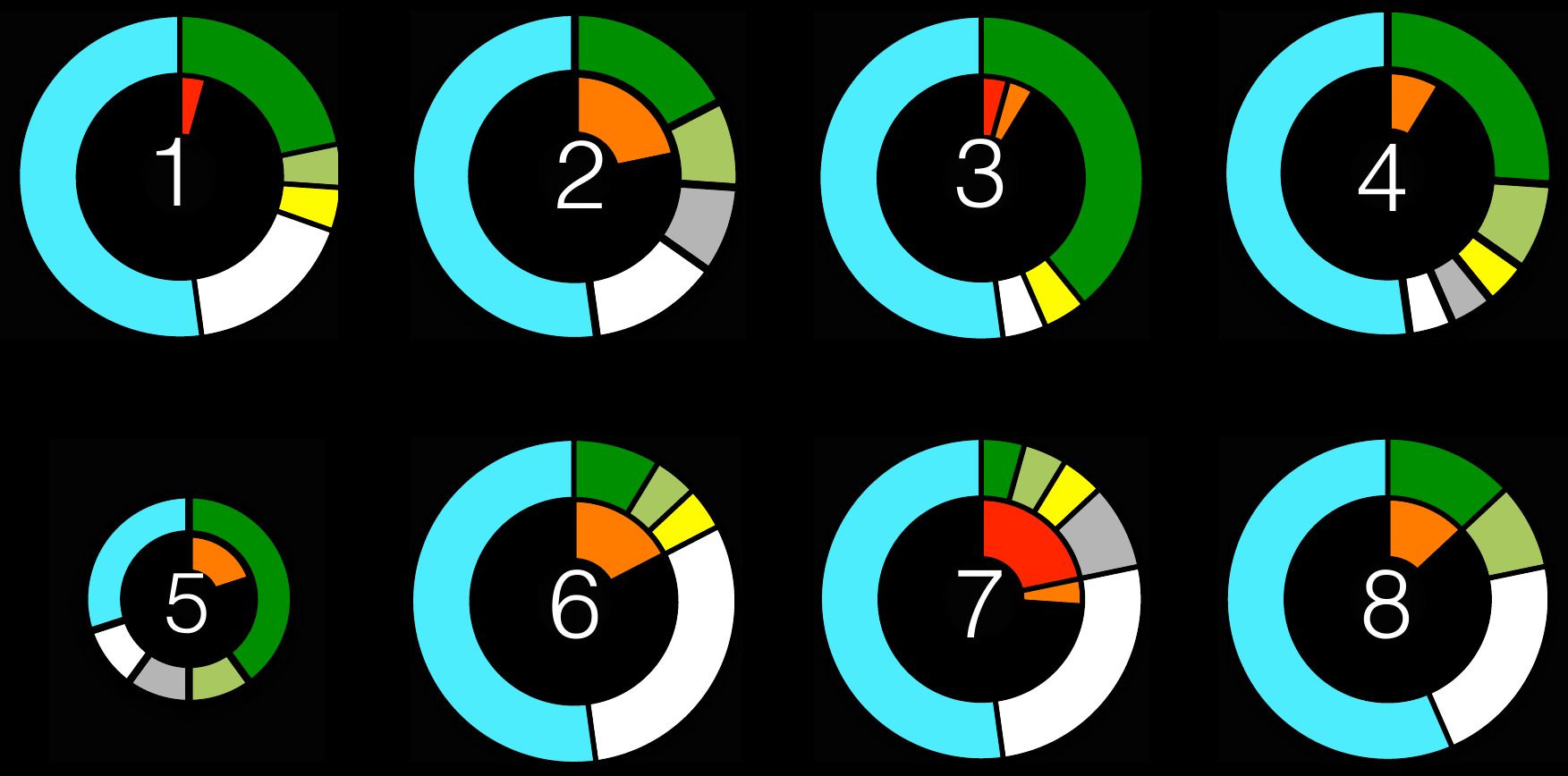}
\includegraphics[width=14cm]{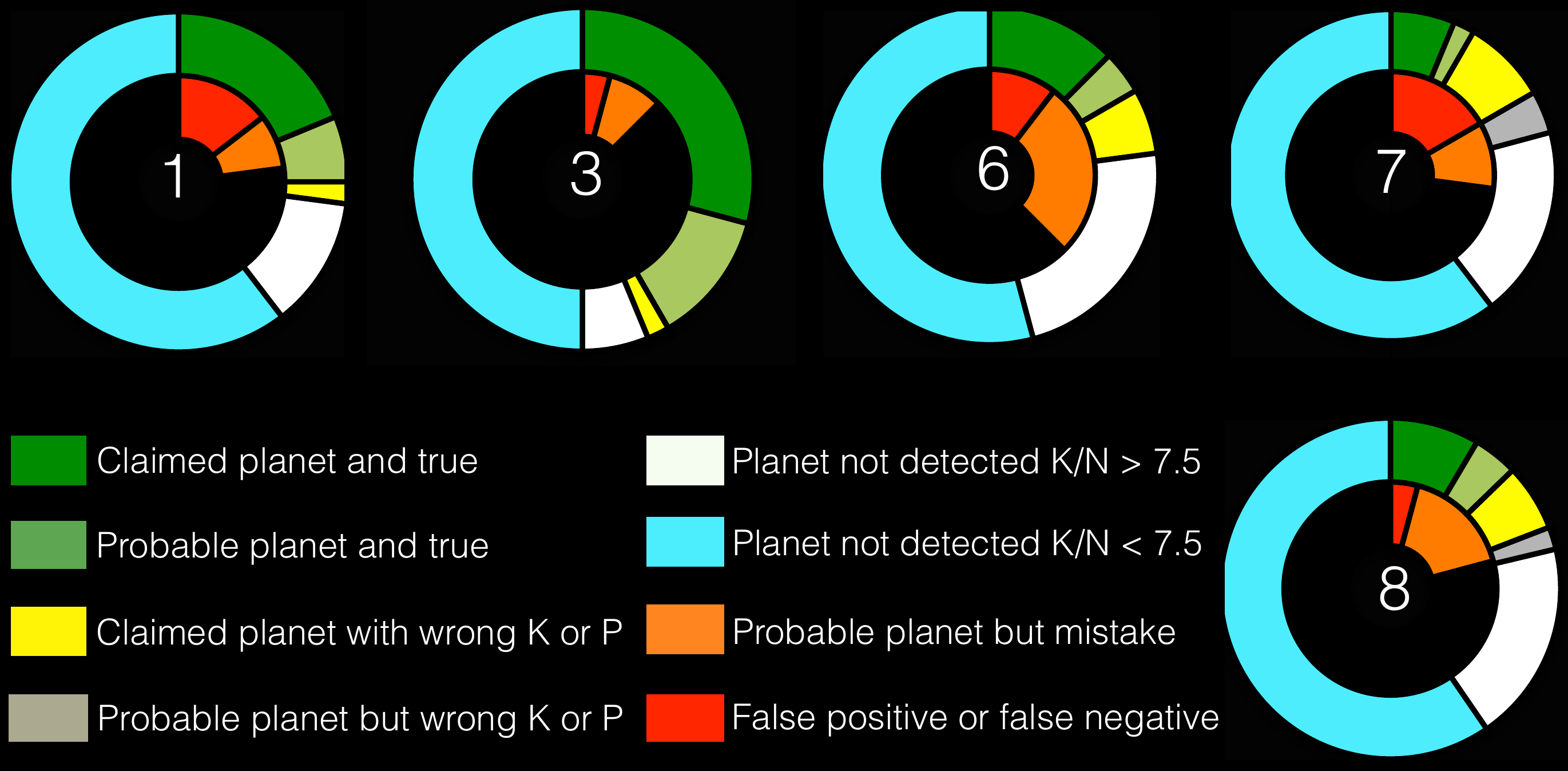}
\caption{\emph{Top:} Summary of the signals detected in the first 5 systems of the RV fitting challenge data set. All the teams have analyzed those system, expect team 5 that only looked at system 1 and 2. The different color flags are defined in the legend and in more details in the second paragraph of Section \ref{sect:4-1}. For each team, the outer circle diagram represents true planetary signals that were present in the data, and show how well the different teams could recover those. The inner circle diagram represents signals announced by the teams, but that were not present in the provided RV measurements. \emph{Bottom:} Same but considering all the systems in the RV fitting challenge. Only teams 1, 3, 6, 7 and 8 performed this analysis. Size of the circle diagrams represents the number of systems analyzed: large size for all 14 systems, medium size for the first 5 systems and small size for the first 2 systems.}
\label{fig:4-1-0}
\end{center}
\end{figure*}

In Fig. \ref{fig:4-1-0}, we summarize the results of the different teams when analyzing only the first five systems and all the systems. For each signal detected, we assign a different color flag depending on the true signals present in the data. The different possibilities are:
\begin{itemize}
\item dark green: the team recovered a planetary signal that exists in the data and would have published the result.
\item light green: the team recovered a planetary signal that exists in the data but is not confident enough in its detection for publication.
\item yellow: the team recovered a planetary signal that exists in the data and would have published the result, however the semi-amplitude or period is wrong compared to the truth or an alias of the true signal was detected.
\item grey: the team recovered a planetary signal that exists in the data but is not confident enough in its detection for publication. The semi-amplitude or period is wrong compared to the truth or an alias of the true signal was detected.
\item white: non-detected planetary signal for which $K/N>7.5$.
\item cyan: non-detected planetary signal for which $K/N\le7.5$.
\item orange: the team recovered a planetary signal that does not exist in the data but is not confident enough in its detection for publication.
\item red: false positive or false negative, i.e., the team recovered a planetary signal that does not exist in the data and would have published the result, or the team rejected with confidence the detection of a true signal, respectively.
\end{itemize}

To study planetary population, we believe that the most important criteria are publishable planets with correct parameters (dark green flag), false positives or false negatives (red flag) and non-detection of planetary signals (white flag for $K/N>7.5$, cyan flag for $K/N\le7.5$). The selection of the threshold $K/N=7.5$ will be discussed in the next paragraph. Publishable signals that are slightly wrong (yellow flag) represent only a small fraction of the detections and therefore should not strongly bias planetary population statistics. All the other signals flagged as light green, grey, and orange would not have been published. Therefore, in Fig. \ref{fig:4-1-0}, the most successful teams in terms of planet detection should have a large dark green region, while having small red, white and cyan regions. Given those criteria, we can separate the teams in two different groups: teams 1 to 5, and teams 6 to 8. {This delimitation separates teams that used a Bayesian framework with red-noise models, which allows to compare between different solutions and model stellar signals, from teams that used other frameworks (see Table \ref{tab:0} and Sections \ref{sect:2} and \ref{sect:3} for more details about the different techniques used).}

\begin{figure*}
\begin{center}
\includegraphics[width=18cm]{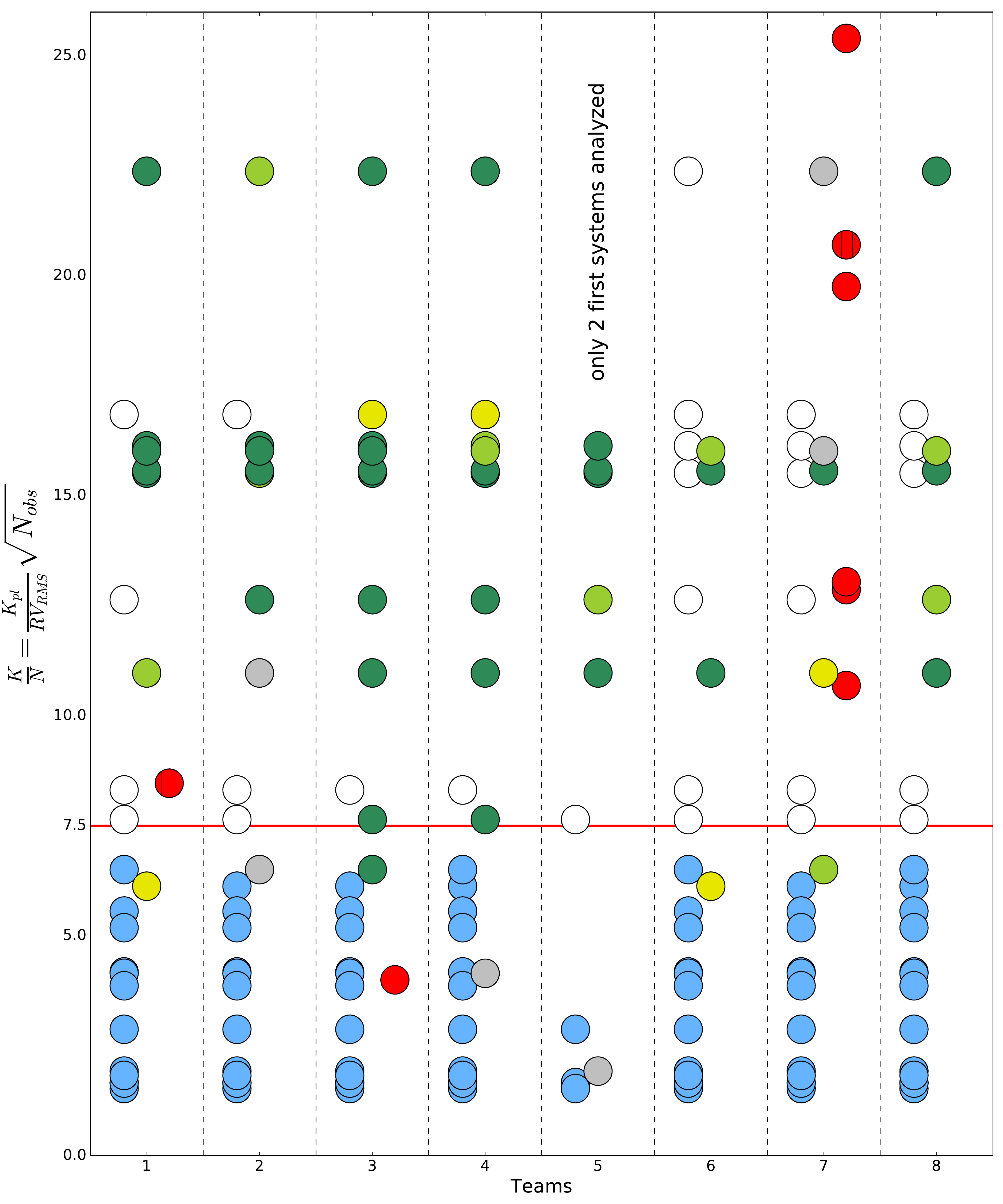}
\caption{{K/N ratio for all the planets present in the 5 first systems of the RV fitting challenge, in addition to the false positives and the false negatives announced for those 5 first systems. All the teams analyzed those 5 first systems, except team 5 that only looked at the first two systems, which explain why there is less dots corresponding to planetary signals. The different color flags are defined in the legend of Fig. \ref{fig:4-1-0} and in more details in the second paragraph of Section \ref{sect:4-1}. We separate the false positives or false negatives appearing in red in two categories. Either they cannot be explained easily (plain red dots), or the activity signal at the stellar rotation period has been confused with a planetary signal despite the fact that the correct stellar rotation period was found a priori (hatched red dots). The red horizontal line corresponds to a K/N ratio of 7.5. Note that the RV rms used to calculate K/N is the rms of the raw RVs once the best-fit of a model consisting of a linear correlation with \logrhk\,plus a second order polynomial as a function of time was removed. This model allows removing the effect of magnetic cycles and any long-term drift in the RVs. We removed from this plot the 2 planets in system 11 and 12 that have an orbital period longer than 3000 days, much longer than the timespan of the data, which explain why they were not detected by any team despite their large K/N values (see Section \ref{sect:4-1-3}).}}
\label{fig:4-1-1}
\end{center}
\end{figure*}

\begin{figure*}
\begin{center}
\includegraphics[width=18cm]{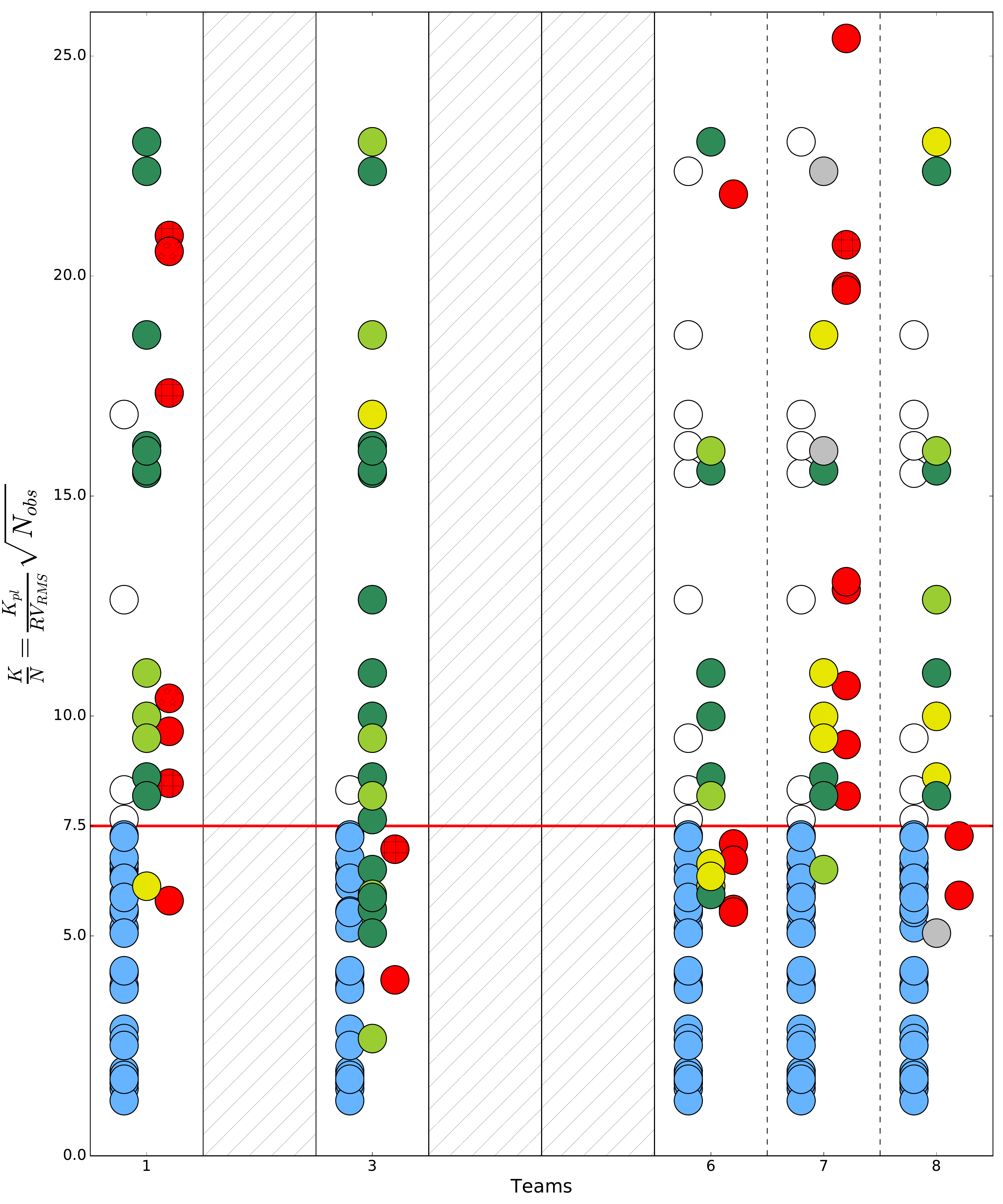}
\caption{{Same as Fig. \ref{fig:4-1-1} but for all the signals announced by the different teams in the entire dataset of the RV fitting challenge. Only team 1, 3, 6, 7 and 8 performed this full analysis. We separate the false positives or false negatives appearing in red in three categories. Either they cannot be explained easily (plain red dots), or the activity signal at the stellar rotation period has been confused with a planetary signal despite the fact that the correct stellar rotation period was found a priori (hatched red dots) or the activity signal at the stellar rotation period has been confused with a planetary signal when a wrong stellar rotation period was found a priori (red dots with stars).}}
\label{fig:4-1-2}
\end{center}
\end{figure*}

{In Figs. \ref{fig:4-1-1} and  \ref{fig:4-1-2}, we plot the different planets that have been detected by the different teams as a function of the $K/N$ ratio. We also highlight the false positives and false negatives. Note that only one false negative was announced: the true planetary signal in system 11 with a $K/N$ ratio of 6 rejected by team 8 (see Fig. \ref{fig:summary_detection5} and the lowest red dot for team 8 in Fig. \ref{fig:4-1-2}). Except for this signal, all the other mistakes correspond to false positives therefore we will only discuss false positive in the rest of the section.

Looking at the results of the different teams for the first five systems (Figs. \ref{fig:4-1-1}), we see that teams 1, 3, 6 and team 4 were able to detect confidently (i.e. color flags dark green and yellow) planetary signals with a $K/N$ ratio as low as 6 and 7.5, respectively. Excluding false positives found at the stellar rotation period when the correct rotation period was detected a priori, because those mistakes could have been avoided (hatched red dots), those teams did not detect any false positives above a $K/N$ level of 5. For the other teams, i.e. 2, 5, 7 and 8, it was more difficult to detect confidently planetary signals with a small $K/N$ ratio, and the threshold between detecting and not detecting a planetary signal is closer to $K/N=10$. We note also that team 7 detected a lot of false positives, therefore the filtering technique in frequency space they used does not seem optimal to prevent false positives.

On the first 5 systems of the RV fitting challenge, there was not many planetary signals with a $K/N$ ratio between 5 and 10. It is therefore worth analyzing the results of teams 1,3, 6, 7 and 8 with the entire data set to get a better idea at which threshold in $K/N$ planets start to be detected. In Fig. \ref{fig:4-1-2}, all the teams that analyzed the entire data set were able to confidently detect planetary signals with a $K/N$ level above 7.5. However teams 1 and 3 detected most of the planetary signals above this threshold, which is not the case for teams 6, 7 and 8. We therefore see here a significant difference between techniques using a Bayesian framework with model comparison in addition to red noise models and technique that do not. 

In Figs. \ref{fig:4-1-2}, we distinguish between three types of false positives: those that cannot be explained easily (plain red dots), those that correspond to the stellar rotation period when the correct rotation period was detected a priori (hatched red dots), and those that correspond to the stellar rotation period when a wrong rotation period was detected a priori (red dots with stars). The second type of false positive could have been avoided assuming that all signals close to the rotation period should be excluded, and the third type could have been avoided using a better algorithm to estimate the stellar rotation period\footnote{For the two cases of third-type false positive, only detected by team 1, all the other teams were able to find the correct stellar rotation period.}. We therefore decided to exclude the second and third-type false positives discussed jsut above from the following discussion. When doing so, we see that team 1 detected 2 false positives at $K/N$ ratios of 6 and 9.7, therefore the confidence level to detect a planetary signal without risking a false positive is close to $K/N=10$. For team 3 this level is closer to $K/N=5$, and for team 6 and 7, closer to $K/N=7.5$. Finally, team 7 have detected too many false positives, up to $K/N=25$, making it impossible to estimate a threshold between confidently detecting a planetary signal without risking a false positive. As a general conclusion, the limit between confident and non-confident detections is somewhere close to $K/N=7.5$. Only the method used by team 3 allows to detect a few candidates with a $K/N$ ratio between 5 and 7.5, without risking of announcing a false positive.
}
%we see that the threshold between confidently detecting a planetary signal and non detecting it is {somewhere close to $K/N=7.5$ for the exercise of the RV fitting challenge. Above this threshold of $K/N=7.5$, 18 planetary signals were present in the RV fitting challenge data set, and 15 where confidently discovered, including only 2 detections for which the orbital parameters are not fully correct compared to the truth (yellow flag). Two out of the three non-detected planetary signals have periods longer than the time span of the data (period $>3000$\,days), which can explain why the different teams were not able to find them (see Section \ref{sect:4-1-2}). Below this threshold of $K/N=7.5$, 30 planetary signals were present in the data, and only 9 were detected, including 4 detections for which the orbital parameters are not fully correct compared to the truth.} Note that those values are when including the results of all the teams together. 
%When analyzing the detection of each individual team in Section \ref{sect:4-1-0}, we will see that the recovery rate of planetary signal with $K/N\le7.5$ is actually much smaller.
%

{In Table \ref{tab:4-1-0}, we report for each team the recovery rate of planetary signals detected and publishable with $K/N$ ratios above and below 7.5}. Planetary signals detected correspond to color flags dark green, light green, yellow and grey, and publishable planets to color flag dark green and yellow only (see definition of color flags in the second paragraph of Section \ref{sect:4-1}). We show the results when studying only the first 5 systems, analyzed by all the teams except team 5 that only worked on system 1 and 2, and when studying all the RV fitting challenge system, analyzed by teams 1, 3, 6, 7, and 8. 
\begin{table*}
\begin{center}
\caption{Recovery rate of planetary signals detected (dark green, light green, yellow and grey color flags), of publishable planets with correct orbital parameters (dark green and yellow color flags) and of false positives and false negatives (red color flag) for each team. {Recovery rates between 0 and 33, 33 and 66, and 66 and 100\% are highligthed in red, yellow and green, respectively.}} 
\label{tab:4-1-0}
\begin{tabular}{ccccccccc}
\hline\hline
 & \multicolumn{5}{c}{Bayesian framework + red-noise models} & \multicolumn{3}{c}{Other techniques}\\
 & 1: Torino & 2: Oxford & 3: Tuomi & 4: Gregory & 5: Geneva & 6: Hatzes & 7: Brera & 8: IMCCE\\
\hline
\multicolumn{2}{c}{{\bf Detected planetary signals $K/N>7.5$}} & & & & & & &\\
5 first systems (total 10)        & {\color{mygreen}80\%} (8)   & {\color{mygreen}70\%} (7) & {\color{mygreen}90\%} (9)   & {\color{mygreen}90\%} (9) & {\color{mygreen}83\%} (5/6)  & {\color{myred}30\%} (3)   & {\color{myorange}40\%} (4)  & {\color{myorange}50\%} (5)\\
all systems (total 18) & {\color{mygreen}68\%} (12) & -              & {\color{mygreen}83\%} (15) & -             & -                              & {\color{myorange}39\%} (7)  & {\color{myorange}50\%} (9)  & {\color{myorange}50\%} (9)\\
\hline
\multicolumn{2}{c}{{\bf Publishable planetary signals $K/N>7.5$}} & & & & & & &\\
5 first systems (total 10)       & {\color{myorange}50\%} (5)   & {\color{myorange}40\%} (4) &  {\color{mygreen}90\%} (9)   &  {\color{mygreen}70\%} (7)  &  {\color{mygreen}67\%} (4/6)  &  {\color{myred}20\%} (2)   &  {\color{myred}20\%} (2) & {\color{myred}30\%} (3)\\
all systems (total 18) &             {\color{myorange}50\%} (9)   & -              &  {\color{myorange}61\%} (11) & -             & -                  & {\color{myred}28\%} (5)  & {\color{myorange}39\%} (7)  & {\color{myorange}39\%} (7)\\
\hline
\multicolumn{2}{c}{{\bf Detected planetary signals $K/N\le7.5$}} & & & & & & &\\
5 first systems (total 13)        & {\color{myred}8\%} (1)  & {\color{myred}8\%} (1)       & {\color{myred}8\%} (1)   & {\color{myred}8\%} (1)    & {\color{myred}25\%} (1/4)  & {\color{myred}8\%} (1)   & {\color{myred}15\%} (2) &  {\color{myred}0\%} \\
all systems (total 30)             & {\color{myred}3\%} (1)    & -               & {\color{myred}20\%} (6) & -               & -                 & {\color{myred}13\%} (4) & {\color{myred}7\%} (2)   &  {\color{myred}3\%} (1)\\
\hline
\multicolumn{2}{c}{{\bf Publishable planetary signals $K/N\le7.5$}} & & & & & & &\\
5 first systems (total 13)       & {\color{myred}0\%}       & {\color{myred}0\%}  & {\color{myred}8\%} (1)      & {\color{myred}0\%}        & {\color{myred}0\%}            & {\color{myred}8\%} (1)     & {\color{myred}8\%} (1)   & {\color{myred}0\%}\\
all systems (total 30)            & {\color{myred}3\%} (1)  & -       & {\color{myred}13\%} (4)   & -              & -                 & {\color{myred}13\%} (4)   & {\color{myred}3\%} (1)   & {\color{myred}0\%}\\
\hline
%\multicolumn{2}{c}{{\bf false positive or negative $K/N>7.5$}} & & & & & & &\\
%5 first systems           & 1 & 0 & 0 & 0 & 0 & 0 & 5 & 0\\
%all systems                & 6 & -  & 0 & -  & - & 1 & 8 & 0\\
%\hline
%\multicolumn{2}{c}{{\bf false positive or negative $K/N\le7.5$}} & & & & & & &\\
%5 first systems          & 0 & 0 & 1 & 0 & 0 & 0 & 0 & 0\\
%all systems               & 1 & -  & 2 & -  & -  & 4 & 0 & 2\\
%\hline
\end{tabular}
\end{center}
\end{table*}

When looking at detected planetary signals with $K/N>7.5$, it is clear that teams 1 to 5, which used a Bayesian framework with model comparison in addition to red noise models, were more successful at finding those type of planetary signals.
There is however a difference between detecting planetary signals, and being confident in those to publish them. Only publishable results will be used for planetary statistics, therefore those are the most important. 
When looking at publishable planetary signals with $K/N>7.5$, we arrive to a similar ranking in performance. However, except in the special case of team 3 when analyzing the 5 first models, 20 to 30\% of planetary detections with $K/N>7.5$ will not be good enough to lead to publications. Those signals are however detected, which is a valuable argument to get more data for a system and thus publish it at a later stage.

As we can see in Fig. \ref{fig:4-1-2}, compared to team 1 and 3, team 6 to 8 have a significantly larger proportion of detections flagged as yellow, i.e. planetary signals for which the teams are confident in the detection, however the period or semi-amplitude differs from the true solution, or the detection corresponds to an alias of the true signal. Including those solutions would bias any statistical analysis on planetary semi-amplitude and period distributions. Therefore, when searching for planetary signal for which $K/N>7.5$, techniques using Bayesian model selection with red-noise models allow to recover more signals and give better estimates of the orbital period and semi-amplitude (see Section \ref{sect:4-1-1} for more details).

When looking at detected planetary signals with $K/N\le7.5$, we have very few number of detections and even less of publishable planetary signals. It is therefore difficult to draw out strong conclusions. Only team 3 and 6 were able to find a significant number of candidates, 6 and 4 respectively. However, 3 planetary signals out of the 4 found by team 6 have incorrect period or semi-amplitude (yellow color flag), {in addition to 3 false positives announced. The results found by team 3 are therefore more robust.}

{The $K/N$ ratio is used here as a measure of the detectability of planetary signals that were present in the RV fitting challenge dataset. In the case of the RV fitting challenge, systems 1 to 13 were very similar, with:
\begin{itemize}
\item a large number of measurements, between 433 and 527,
\item planetary signals much shorter than the timespan of the data,
\item planetary signals with a good phase coverage,
\item and stellar signals similar to what is observed for the Sun, with a RV rms ranging from 1.8 to 5\,\ms once the best-fit of a model consisting of a linear correlation with \logrhk\,plus a second order polynomial as a function of time was removed to the raw RVs.
\end{itemize}
For those systems, we show that for most of the teams it was possible to confidently detect planetary signals above a threshold in $K/N$ of 7.5, without announcing false positives. However teams that used a Bayesian framework with model comparison in addition to red noise models where able to detect nearly all the planetary signals above this threshold, which was not the case for the other teams.

Systems 14 and 15 exhibit a higher level of stellar signals, with a RV rms of 8.9 and 7.6\,\ms once the best-fit of a model consisting of a linear correlation with \logrhk\,plus a second order polynomial as a function of time was removed to the raw RVs. In addition, those two systems presented only 170 measurements. However, in these very different cases compared to system 1 to 13, most of the team were able to detect the signals of Corot-7c and Corot-7d with a $K/N$ ranging from 8 to 10, while only team 1 was able to detect the $K/N=5$ signal of Corot-7b (see Fig. \ref{fig:summary_detection6} in the appendix). Therefore, it seems that this $K/N$ threshold of 7.5, and probably 5 for team 3, can be applied to quite different set of data. We however only have a few systems in the RV fitting challenge to test this hypothesis and a detailed study of the behavior of this threshold as a function of number of measurements, ratio of the planet period to the timespan of the measurements, phase coverage of the signal and level of stellar signals would be something extremely useful to explore. 

This threshold $K/N=7.5$, or 5 for team 3, is the best that can currently be done by the different teams when analyzing the RV fitting challenge data set. We expect that this level goes down with ongoing progress in the different methods used to detect planetary signals in the presence of stellar signals.
}

\subsubsection{Accuracy of estimated planetary period and semi-amplitude} \label{sect:4-1-1}

Detecting a true planetary signal and being confident in its veracity is a difficult task, even more when $K/N\le$7.5. We discussed in the previous sections that some techniques to deal with stellar signals are performing better. In this section, we look at the parameters found for each planet detected by the different teams, and compare them to the true parameters that were used to generate those planetary signals. When dealing with real data including planetary signals like in system 14, we compared with the latest published parameters. 

\begin{figure*}
\begin{center}
\includegraphics[width=8cm]{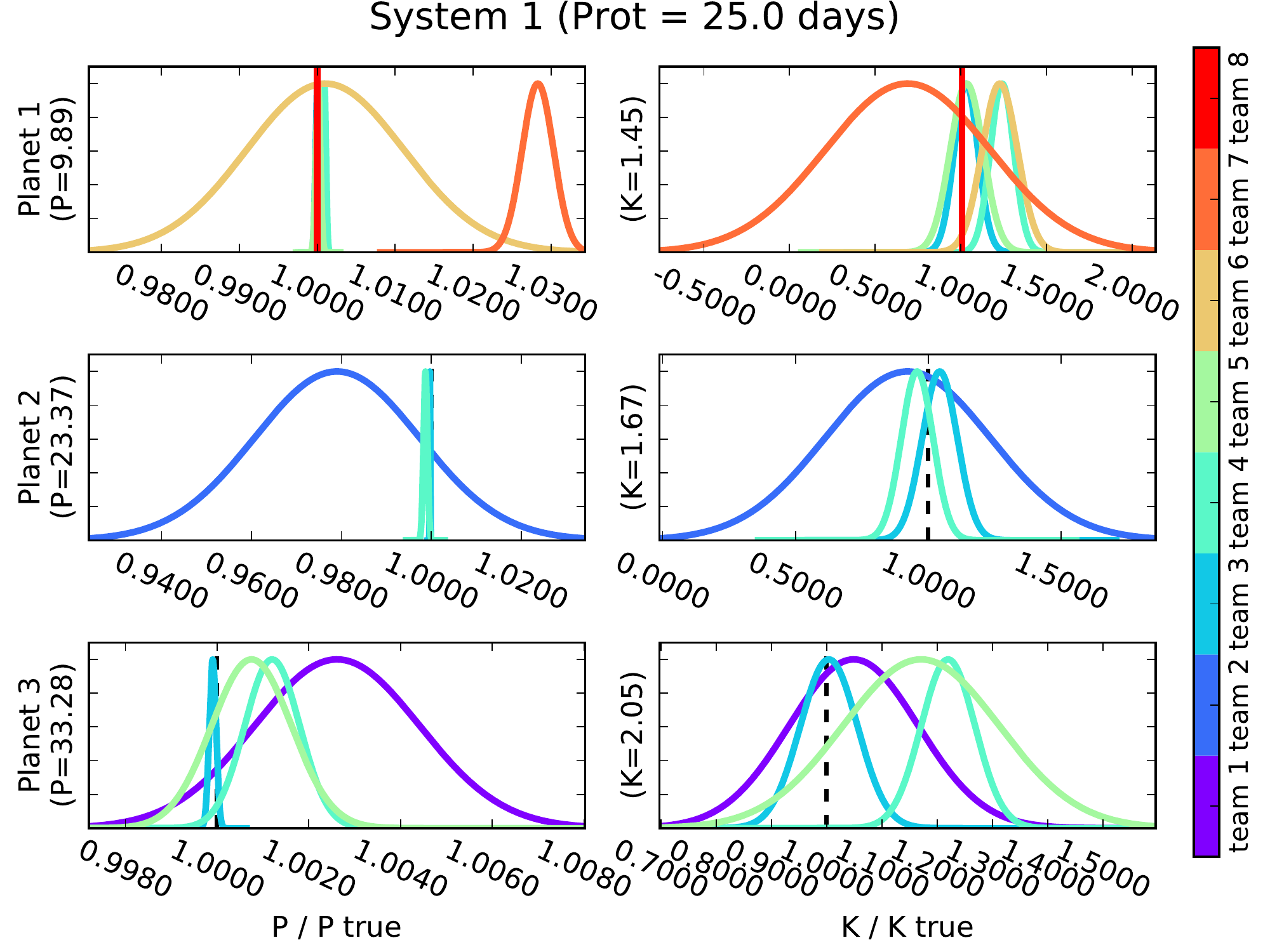}
\includegraphics[width=8cm]{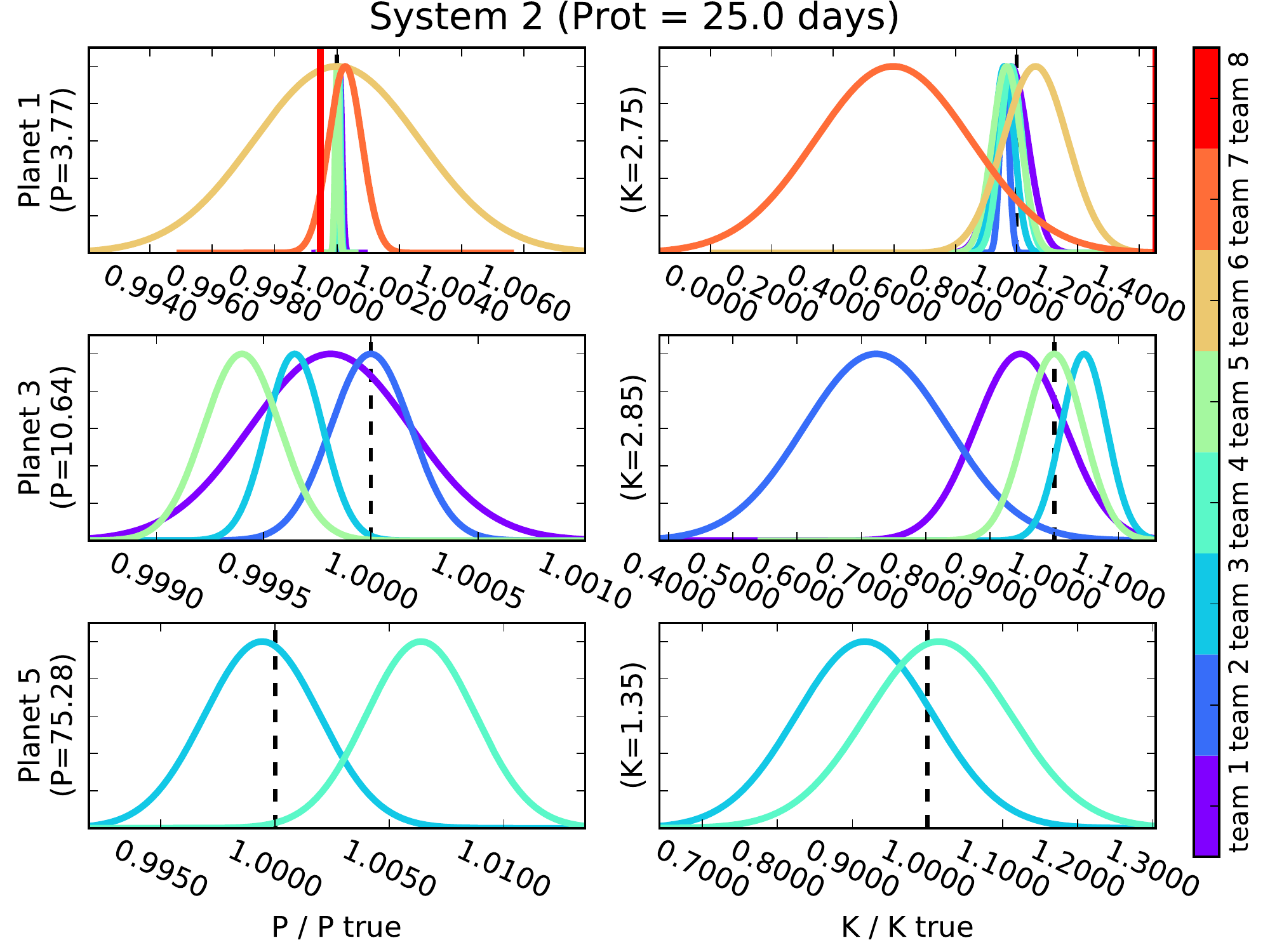}
\includegraphics[width=8cm]{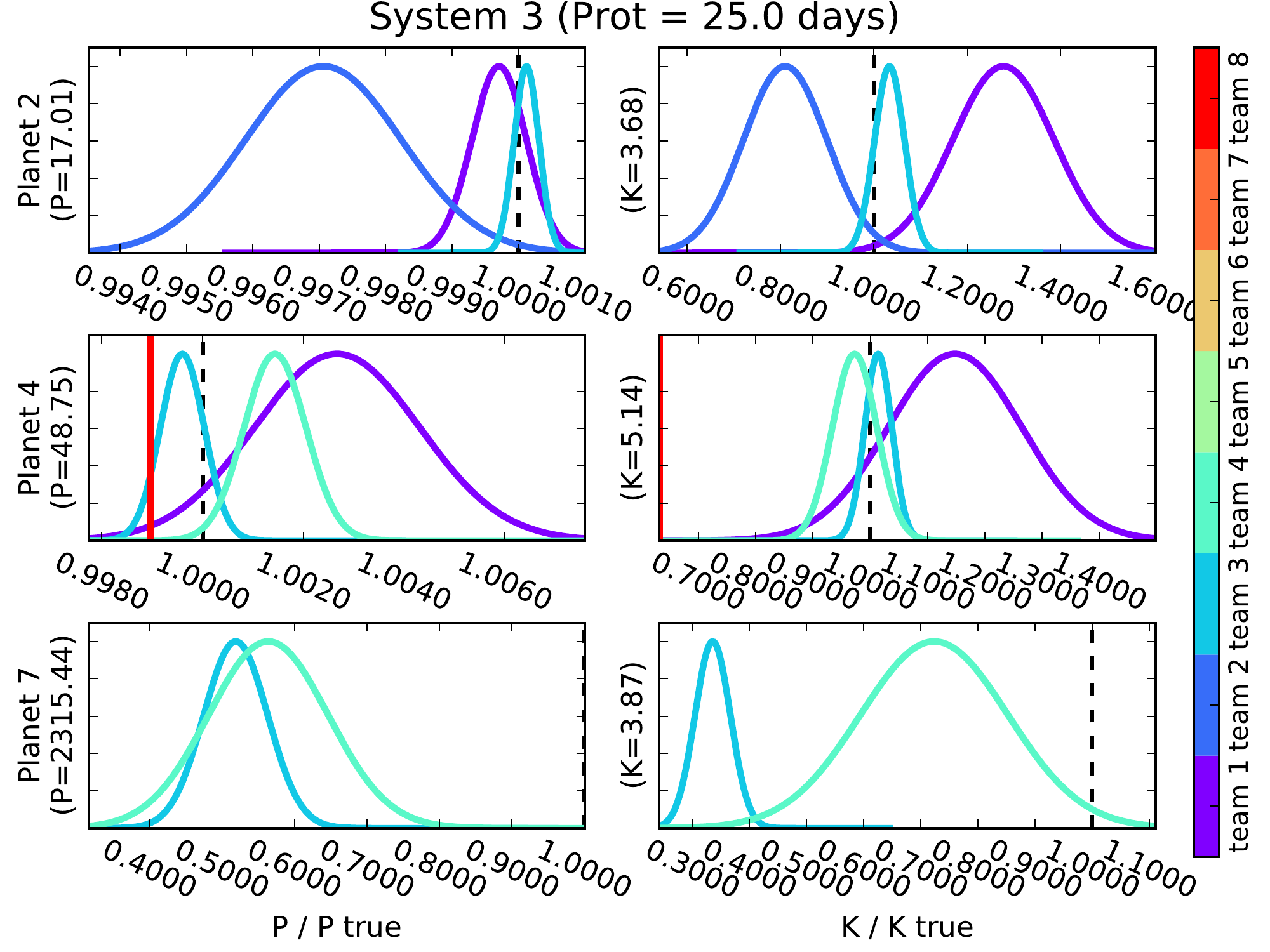}
\includegraphics[width=8cm]{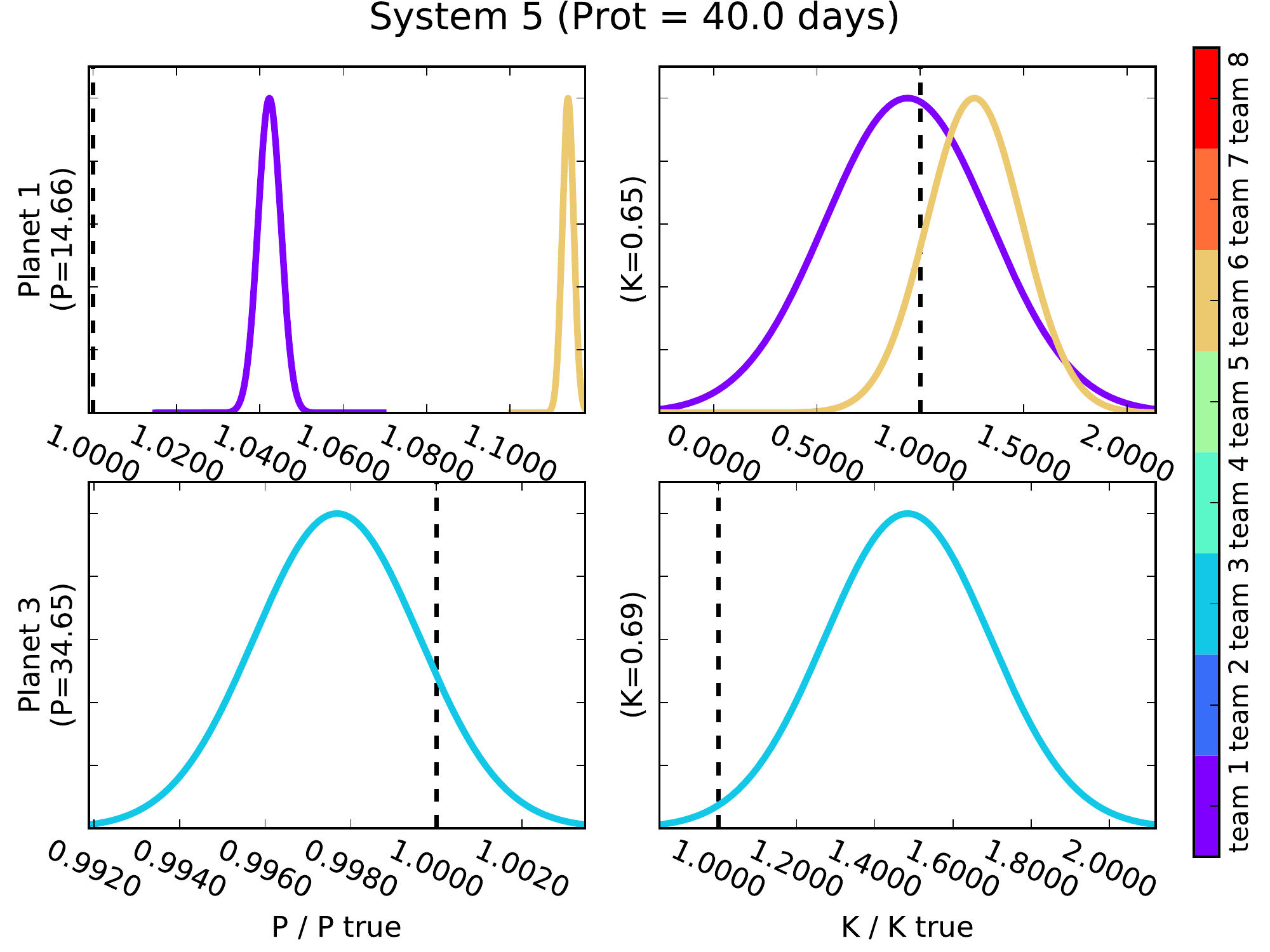}
\includegraphics[width=8cm]{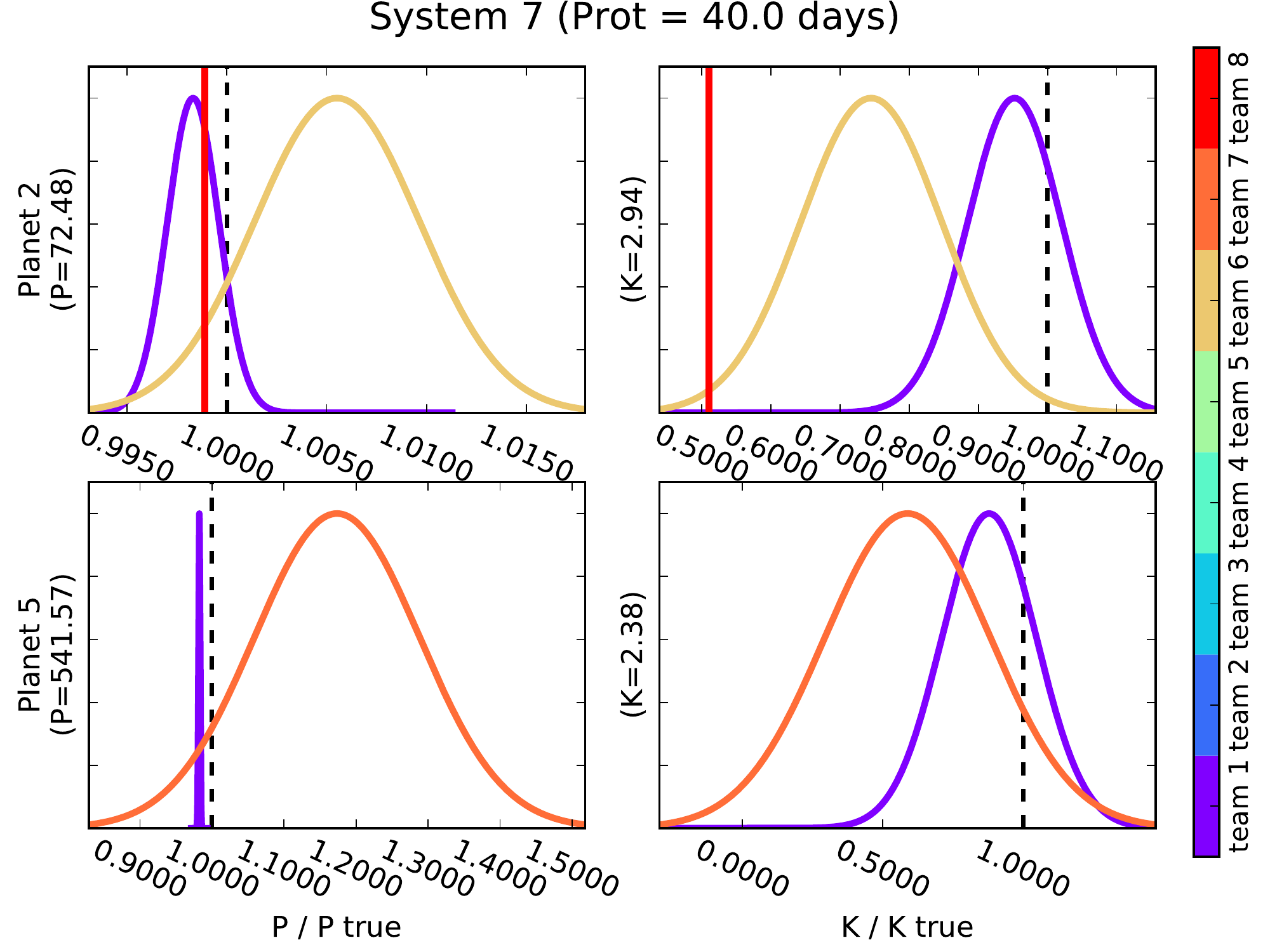}
\includegraphics[width=8cm]{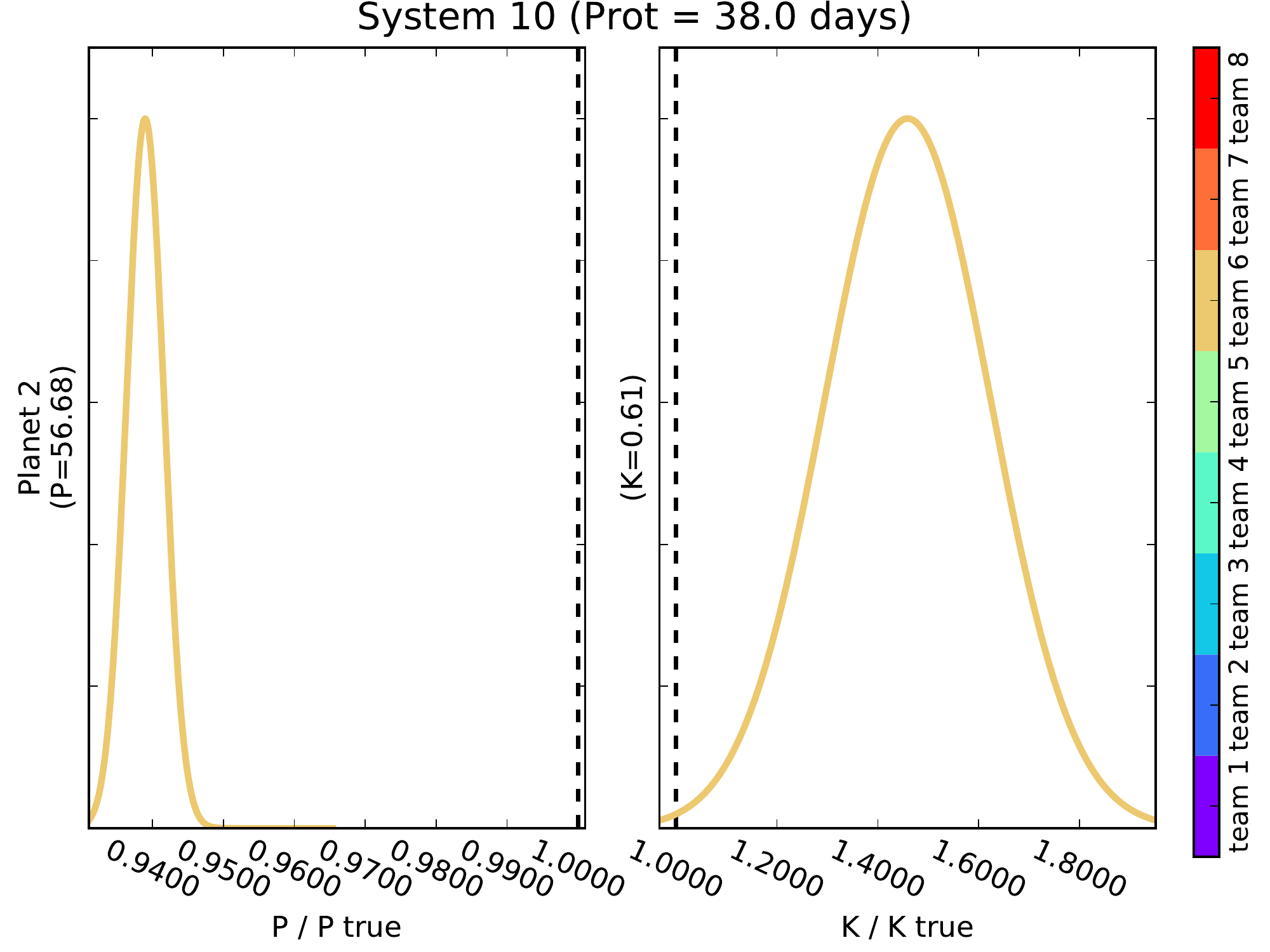}
\includegraphics[width=8cm]{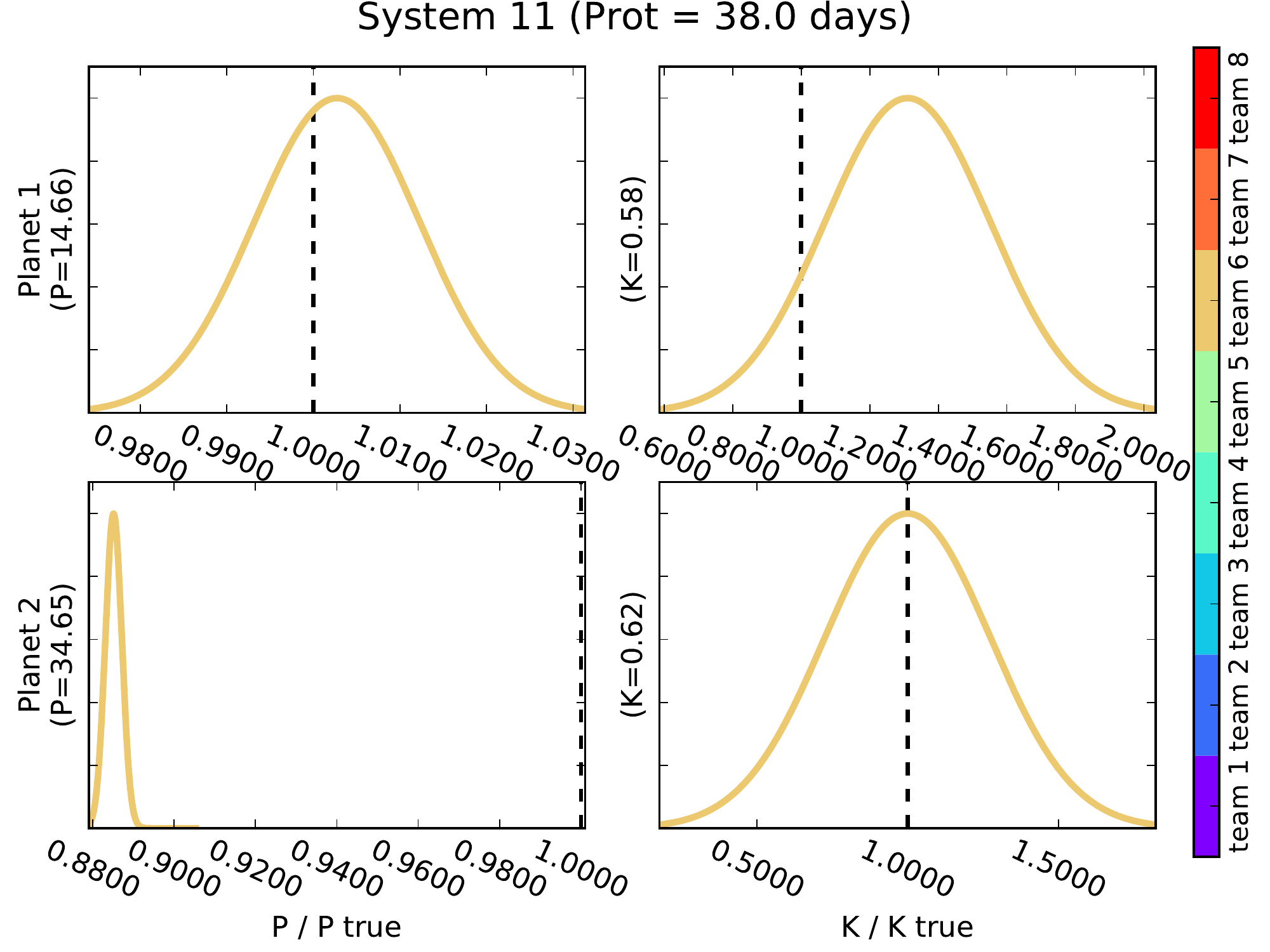}
\includegraphics[width=8cm]{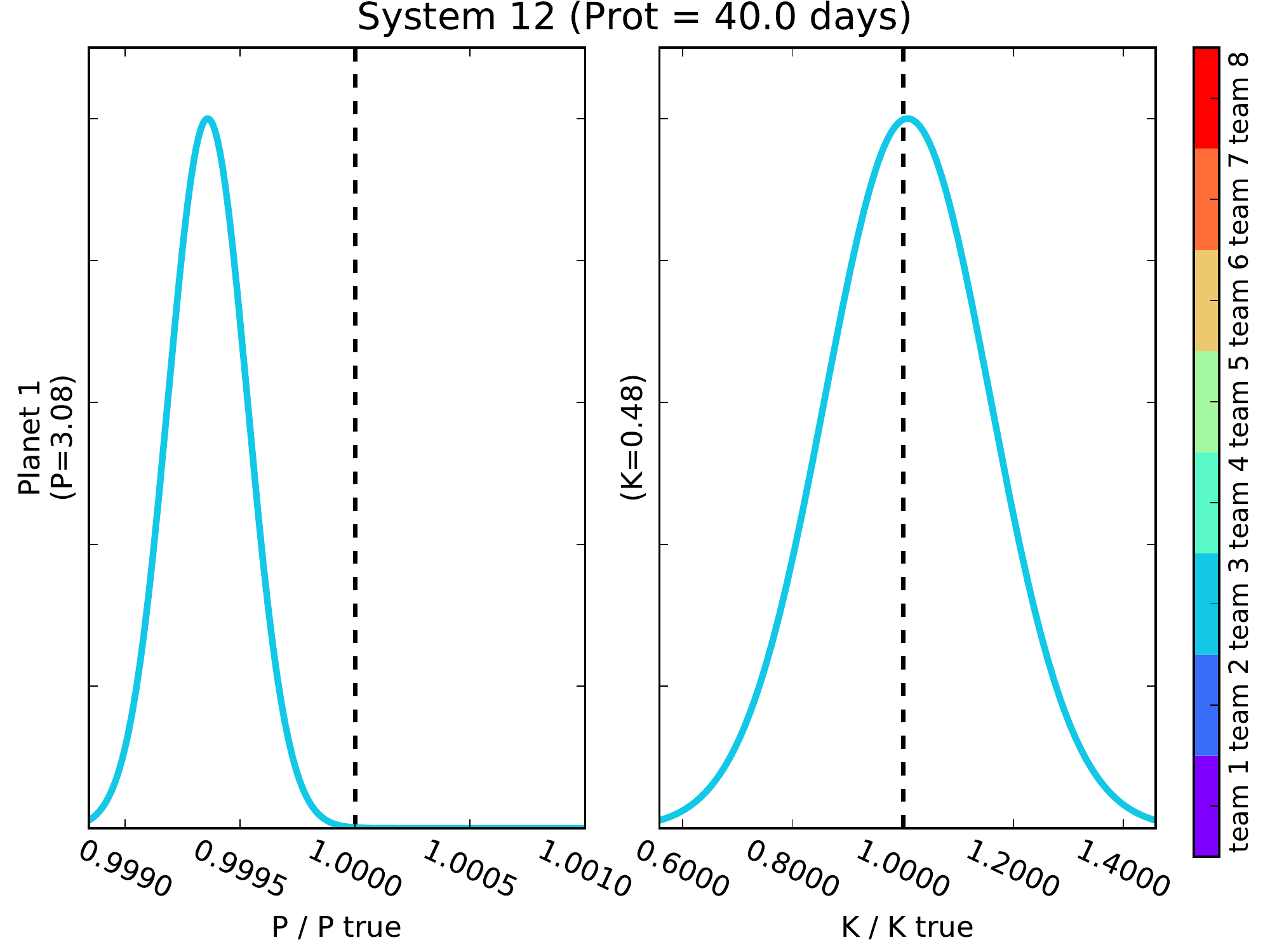}
\caption{Periods and semi-amplitudes reported by each team for the planets detected in systems 1, 2, 3, 5, 7, 10, 11 and 12. We divided those parameters by the values of the true signals, so that a perfect estimate would fall on one. Team 8 was the only one not reporting error bars on their parameters, therefore we just show their best estimates as red vertical lines.}
\label{fig:4-1-4-0}
\end{center}
\end{figure*}

\begin{figure*}
\begin{center}
\includegraphics[width=8cm]{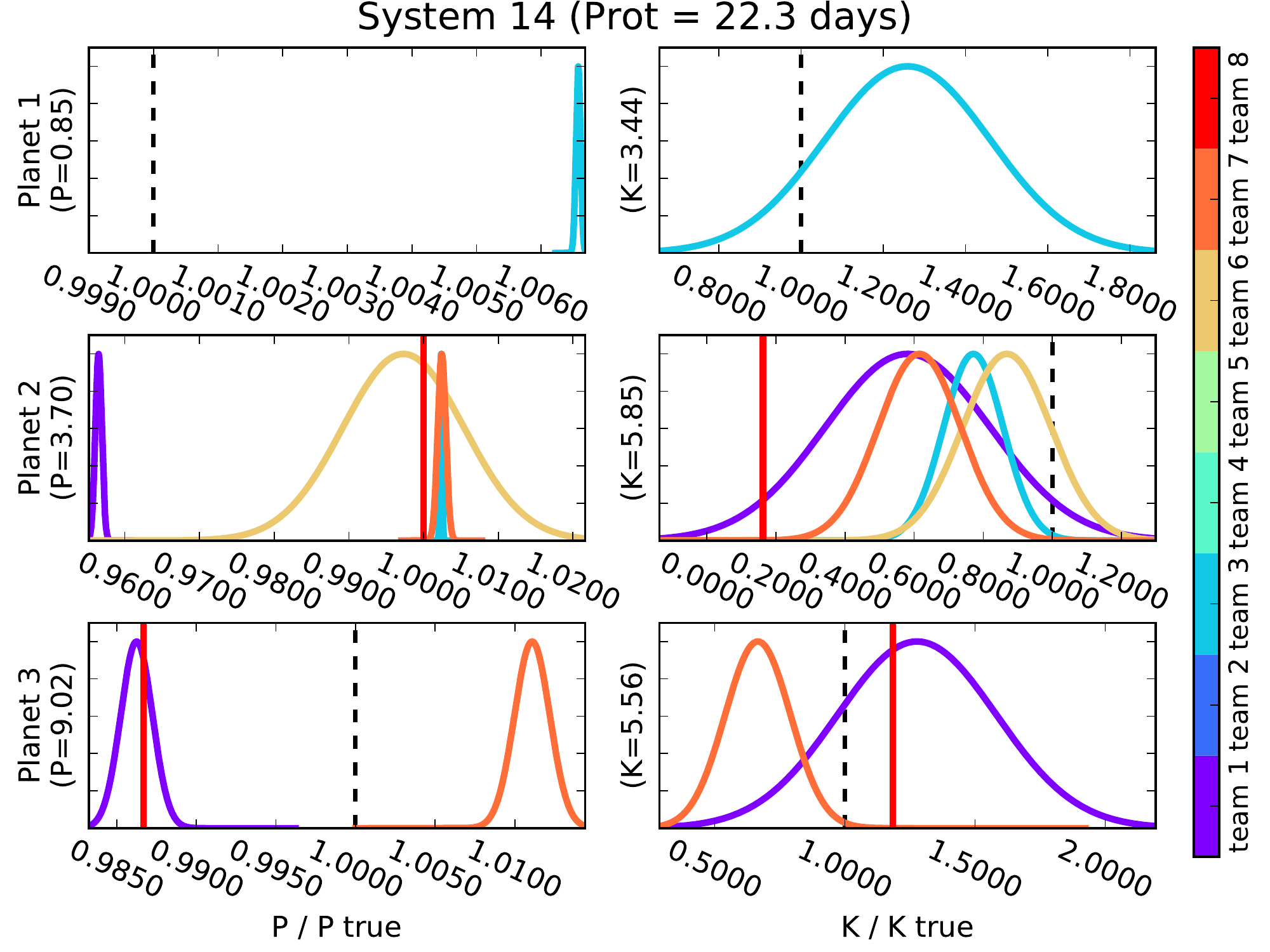}
\includegraphics[width=8cm]{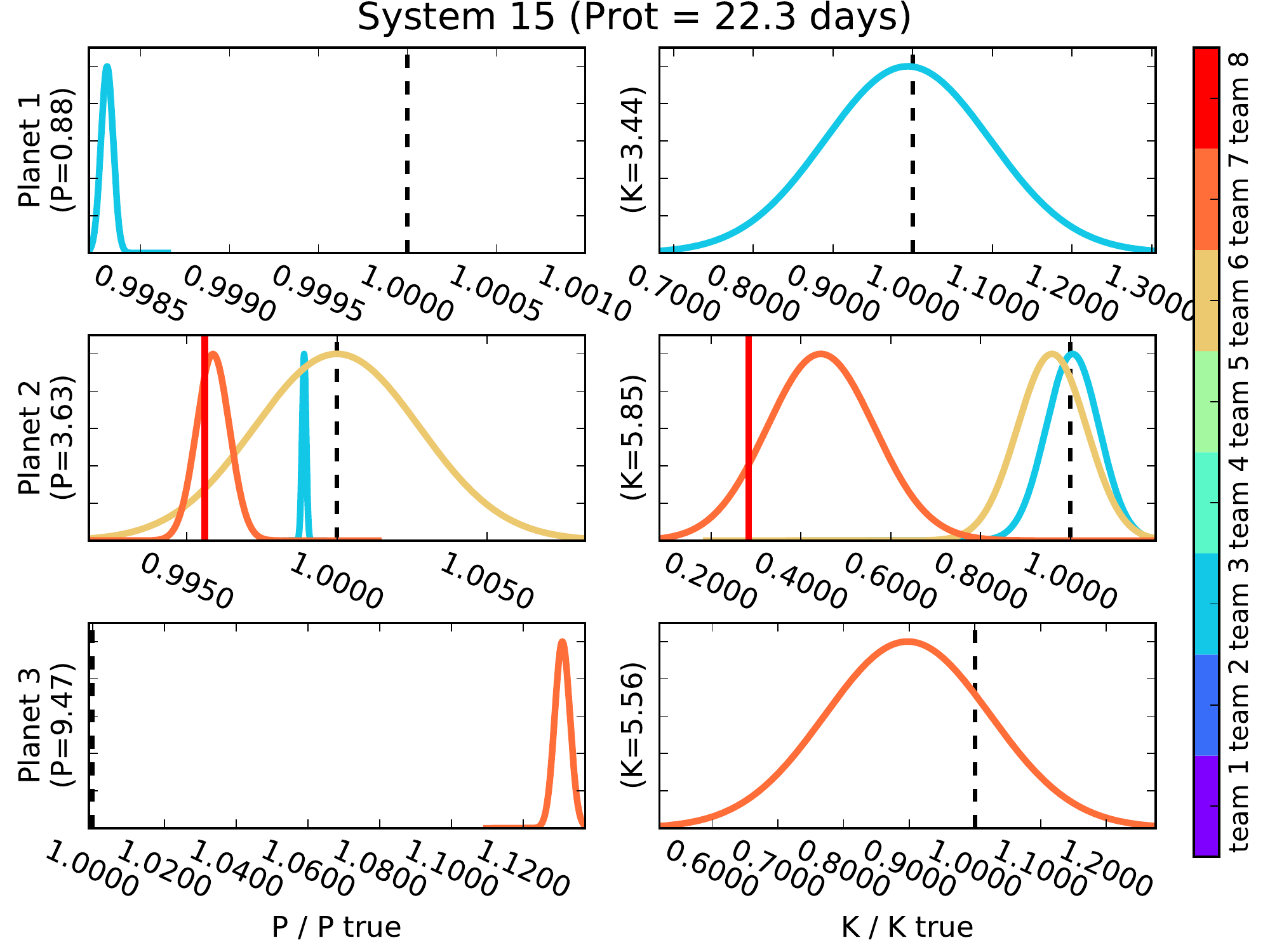}
\caption{Same as Fig. \ref{fig:4-1-4-0} for systems 14 and 15.}
\label{fig:4-1-4-1}
\end{center}
\end{figure*}

In Figs.  \ref{fig:4-1-4-0} and \ref{fig:4-1-4-1}, we show for each system and each planet detected, the period and semi-amplitude parameters found by each team. We divided those parameters by the values of the true signals, so that a perfect estimate would fall on one. Team 8 was the only team that did not report error bars on their measurements, therefore we just show their best estimates as red vertical lines in Figs.  \ref{fig:4-1-4-0} and \ref{fig:4-1-4-1}. Note that we included in those figures all the signals for which the teams were confident in (dark green and yellow color flags).

It is difficult to conclude which team have recovered the best period and semi-amplitude parameters for the planetary signals present in the RV fitting challenge data set, as some teams discovered more signals than others. However, if we look at system 1, 2, 3, 14 and 15, for which many teams detected a lot of signals, team 3 found the best estimate for the different parameters. Therefore using a moving average model to account for stellar signals seems the best approach to deal with the correlated noise induced by RV stellar signals. This does not mean that team 3 always found the best parameters, and as a general conclusion, the errors on the period and semi-amplitude parameters are often underestimated, certainly due to the fact that the models used to account for stellar signals are not perfect.

\vspace{0.2cm}
\subsubsection{Comparing the results obtained by teams using a GP regression and teams using other red-noise modelings} \label{sect:4-1-2}
\vspace{0.2cm}

\noindent When looking in Fig. \ref{fig:4-1-1} at the first 5 systems, we notice that GP regression techniques (team 1 and 2) could not confidently recover 3 planetary signals with $K/N>7.5$ compared to teams that used other red-noise models (team 3 and 4). This is probably due to the fact that stellar signals do not have the same covariance in RVs than in the activity observables. Just as an example, an equatorial spot on a star seen equator-on will induce a sinusoidal variation with a period of $P_{\mathrm{rot}}/2$ when the spot will pass on the visible hemisphere \citep[e.g.,][]{Dumusque-2014b}. Then no signal is observed when the spot is behind the star. For the signal in \logrhk, projection effect comes into play; \logrhk\,increases when a spot moves from the limb to the stellar disc center, and symmetrical decreases when a spot moves from the disc center to the opposite side of the limb. Therefore, the observed signal is half of a sine wave with the stellar rotation period. Then, like for the RVs, no signal is seen when the spot is behind the star. Although the signals in RVs and \logrhk\,seems to have different periods, they do not as after a full stellar rotation period, the same signal will appear again in RVs and \logrhk\,if there is no spot evolution. However, because many spots are present on the stellar surface at the same time, in addition to their evolution, the RVs of stars spot-dominated will tend to have a significant signal at period $P_{\mathrm{rot}}/2$, while the \logrhk\,will present a significant signal at $P_{\mathrm{rot}}$. This example shows that the RVs and the activity observables can have a different covariance and therefore could explain why teams 1 and 2 were confident in fewer true planetary signals than team 3 and 4 when analyzing the first five systems. We believe that further investigations should be done to confirm or reject this argument.

{When looking at the 14 systems of the RV fitting challenge, team 1 announced 6 false positives with $K/N>7.5$ (see Fig. \ref{fig:4-1-2}). Out of those 6 false positives, one cannot be explained easily (plain red dots), three are due to a confusion with the stellar rotation period, thus stellar activity, despite the fact that the correct stellar rotation period was found a priori (hatched red dots), and the two last one due to a confusion with the stellar rotation period knowing that a wrong stellar rotation period was found a priori (red dots with stars). Although it is difficult to draw any conclusion on the first and the two last false positives, for the three other ones, it seems that the GP regression used by team 1 was not able to fully model stellar activity and that an extra Keplerian with a period close to stellar rotaion was needed to better explain the observed RV variations. This is therefore something that should be explored in detail as we do not want GP regression to create some false-positives. Team 2 used GP regression with a different formalism, unfortunately it is not possible to compare the results of team 1 and 2 because team 2 only analyzed the first five systems, for which only one false positive was announced by team 1.}

\subsubsection{Detection of long period planets} \label{sect:4-1-3}

In the different systems of the RV fitting challenge, we injected long-period signals to see if they could be recovered despite the RV stellar signal induced by magnetic cycles.
In total 6 planetary signals with periods longer than 500 days were present in the data: 
\begin{itemize}
\item 596 and 2315 days for system 3, with $K/N=8.3$ and 16.9, respectively ($K$=1.91 and 3.87 \ms),
\item 616 days for system 5, with $K/N=5.2$ ($K$=0.55 \ms),
\item 542 days for system 7, with $K/N=18.7$ ($K$=2.38 \ms),
\item 3245 days for system 11, with $K/N=15.8$ ($K$=1.54 \ms),
\item 3407 days for system 12, with $K/N=19.2$ ($K$=1.64 \ms).
\end{itemize}
Regarding our previous discussion about signal smaller than $K/N=7.5$ (see Section \ref{sect:4-1-0}), it is clear that the 596-day period signal in system 3, and the long-period signal injected in system 5 are difficult to find. However, all the other signals have $K/N>15$ and could have been discovered by the different teams, mainly those using a Bayesian framework with red-noise models to mitigate the impact of stellar signals.

The long-period signals in systems 3, 11 and 12 are close to 6, 9 and 9 years, much longer than the 4-year time span of the RVs for these systems. Therefore, the RVs do not cover an entire phase of those signals, which makes it very difficult to characterize them, {as in general orbital periods need closure for correct parameter estimation \citep[e.g.][]{Black-1982}}. Team 3 and 4 reported a signal at 1202 and 1306 days for system 3, which is half of the period of the real signal. Because the RVs do not cover an entire phase, the power of the signal is transferred to its first harmonic, i.e., half of its period. The two other planetary signals were not detected, and this can be explained by the fact that the different teams added in their RV model a polynomial up to the second order to account for any drift in the data, which absorbs any long-period planetary signals if the time span of the data is much shorter than the orbital period of the planets. 
%Note that because the orbits of these extremely long-period planetary signals are not complete, modeling them using Keplerians is useless, as it leads to non-reliable orbital parameter estimation.

The 542-day signal in system 7 has a shorter period than the time span of the data and was thus confidently announced by team 1 and 7, and detected but not confidently by team 3.

\subsubsection{Detection of short period planets} \label{sect:4-1-4}

In opposition to long-period planetary signals, 6 signals in the RV fitting challenge have periods shorter than 5 days:
\begin{itemize}
\item 3.77 days for system 2 with $K/N=15.6$ ($K$=2.75 \ms),
\item 1.12 days for system 3 with $K/N=4.2$ ($K$=0.96 \ms),
\item 0.82 day for system 10 with $K/N=7.3$ ($K$=0.67 \ms),
\item 3.08 days for system 12 with $K/N=5.6$ ($K$=0.48 \ms),
\item 0.85 day for system 14 with $K/N=5.1$ ($K$=3.44 \ms),
\item and 0.88 days for system 15 $K/N=5.9$ ($K$=3.44 \ms).
\end{itemize}
Out of these 6 signals, all the teams recovered the one in system 2, which can be explained by the high $K/N$ ratio. The 5 other signals all have $K/N\le7.5$ and therefore, following our discussion in Sections \ref{sect:4-1-0}, they were difficult to find. Three of them, in systems 12, 14 and 15, were confidently recovered only by team 3. {This shows that the technique used by team 3 seems more efficient at finding short-period planets.} A probable explanation is that the moving average model used by team 3 includes correlation between points on short-period timescales, which therefore reduces the effect of granulation (timescale up to 2 days) and of stellar short-term activity over a few day timescale that can be important for active stars (systems 14 and 15). {Short-period planets are therefore easier to find.}

The short-period planets in systems 14 and 15 are the true and simulated version of Corot-7b, the first Earth-radius planet ever detected. This planet was first found by photometry \citep[][]{Leger-2009}, and then confirmed with the RV measurements of system 14 \citep[][]{Queloz-2009}. It is however interesting to see that only team 3 was able to recover this planet without imposing as priors the period and time of transit derived from photometry.

The planets recovered by team 3 in system 14 and 15 have a very similar periods and a smaller $K/N$ ratio than the planet not detected in system 10. Therefore the $K/N$ ratio is not the only criterion to separate detections from non-detection. The detection of the planets in system 14 and 15 and not in system 10 can be explained by two effects here: i) for system 14 and 15, the signals from short-term activity is dominating the other sources of stellar signal; short-term activity is better characterized and therefore easier to model with the moving average, and ii) the 0.82-day planetary signal has an amplitude much larger than the expected perturbations induced by the other sources of stellar signal, i.e., granulation and stellar oscillations.

When comparing the $K/N$ ratio of the planets in systems 10 and 12, we would guess that the one in system 10 is easier to recover. However, team 3 could not recover it but could confidently detect the other. This can be explained by the longer period of the planet in system 12, 3 days, compared to close to 1 day for the planet in system 10. Indeed, this difference in period implies a better phase coverage of the planet with a longer orbital period. Nine measurements per orbit can be obtained for a 3-day period planet when observing with a strategy of 3 measurements per night ({similar to the sampling of the different systems in the RV fitting challenge}) compared to only 3 for a 1-day period planet. In addition, planetary signals close to 1 day are more affected by granulation signal that affects RV measurement on a timescale smaller than 2 days, which makes them harder to find. Team 3 was the only team to recover the 3-day signal, and this is probably because it is the only team that considered, with its moving average model, correlation between points on short-period timescales. The moving average reduces the impact of granulation on RV measurements, and therefore increases the significance of planetary signals with similar or shorter periods.

Speaking about planetary signals with periods shorter than 5 days, we need to discuss system 9,10 and 11, for which the RVs have been extracted from the HARPS measurements that led to the detection of $\alpha$ Centauri Bb. The RVs and the activity observables for system 9 are the raw data published in \citet{Dumusque-2012}. We only reversed time, added a Gaussian noise of 0.05 \ms, and changed the gamma velocity of the star, so that the time series for this system could not be recognized \citep[][]{Dumusque-2016a}. These modifications should not perturb the 0.5 \ms\,planetary signal of $\alpha$ Centauri Bb present in the data. This planetary signal should also be present in system 10 and 11, however for those systems we added extra planets, which can perturb the detection of this small semi-amplitude signal. {The $K/N$ ratio for $\alpha$ Centauri Bb in system 9, 10 and 11 would be 5.7, 5.4 and 5.1 respectively, implying a very challenging detection according to the discussion in Section \ref{sect:4-1-0}. None of the teams were able to recover the signal of $\alpha$ Centauri Bb. However, team 3 was able to confidently recover the simulated planetary signal of $\alpha$ Centauri Bb in simulated system 12. Based on this result, team 3 should have been able to detect the signal of $\alpha$ Centauri Bb in systems 9,10 and 11. It is therefore possible that the signal of 
$\alpha$ Centauri Bb announced in \citet{Dumusque-2012} is in fact a spurious one induced by a combination of the sampling of the data and of the model used to fit stellar activity, as questioned by \citet{Rajpaul-2016}. The signal of $\alpha$ Centauri Bb is however at the limit of what can be done with current methods to deal with stellar signals and more RV measurements are needed to really conclude on the existence or not of $\alpha$ Centauri Bb.}

\subsection{Results for real RV data compared to simulated ones} \label{sect:4-3}

Testing the efficiency of different techniques on simulated data can be useless, if those simulated data are not realistic.

To be able to test the realism of simulated data, \citet{Dumusque-2016a} included in the data set of the RV fitting challenge some real observations done with HARPS, and then simulated RVs as close as possible to those real data. Thus systems 6 and 7, 9 and 13, 11 and 12, and 14 and 15, including real and simulated data, can be compared. Unfortunately, as discussed in \citet{Dumusque-2016a}, system 6 cannot be used. The comparison of the other systems is described below {and each time we refer to the RV rms, this one is calculated on the raw RVs once the best-fit of a model consisting of a linear correlation with \logrhk\,plus a second order polynomial as a function of time was removed:}

\vspace{0.2cm}
{\textsc{{{\textbf{System 9 and 13}}}}}
{The two systems have a similar RV rms, 1.82 and 2.06 \ms, respectively, therefore the level of stellar signal present in the simulated data seems realistic. For real system 9, only team 1 could not find the correct stellar rotation period, and team 1 and 6 announced 3 false positives. For system 13, only team 3 could find the correct stellar rotation period while the other teams found the first harmonic. Team 3 and 8 announced 2 false positives. It is difficult to conclude as no similar mistake was done on the two systems, however it seems that it was as difficult to analyze the real and the simulated data.}

\vspace{0.2cm}
{\textsc{{{\textbf{System 11 and 12}}}}}
The simulated data seems to have a realistic level of stellar signal as the two systems have a similar RV rms, 2.04 and 1.78 \ms, respectively.
Regarding stellar rotation period, all the teams could recover the correct value, except team 1 for simulated system 12. By making this mistake, team 1 announced a false positive at $P_{\mathrm{rot}}/2$. Four mistakes were done on system 12, while only two were done on system 11. In addition, team 3, 6 and 8 could recover the $K/N=5.95$ signal at 15 days orbiting system 11 (note however that team 8 announced it as false-negative), while no one could recover the same signal in system 12. From the comparison of these two systems, it seems that is was more difficult for the different teams to find planetary signals in the simulated data, and easier to make some mistakes.

\vspace{0.2cm}
{\textsc{{{\textbf{System 14 and 15}}}}}
These two systems exhibit similar RV rms, 8.86 and 7.64 \ms, respectively, therefore implying at first order a correct modelization of stellar signals. From the comparison of those two systems presenting the real and simulated data of Corot-7, we find that it was easier to detect the correct rotation period in the real time series. Regarding false positives, none was announced for system 14, while 2 were detected in system 15. {Except team 3 that found exactly the same solution}, the different teams were more confident in the signals found in system 14 than in system 15. 

{In general, it was slightly more difficult for the different teams to analyze simulated data. However, we note that team 3, that performed the best at the exercise of the RV fitting challenge, found very similar solutions when analyzing real and simulated data. We therefore believe that even if not perfect, the simulated data are realistic enough to be used to test the efficiency of techniques to recover planetary signature despite stellar signals.}

\section{Conclusion} \label{sect:5}

In total, 8 different teams participated in the analysis of the RV fitting challenge data set. They all used different techniques to find the {low eccentricity planets that were hidden inside stellar signals}. Except system 14 and 15, that present the real and simulated RVs of the active star Corot-7, all the other systems present a typical level of stellar signal for inactive G-K dwarfs. Those stars are the typical targets of most high-precision RV surveys searching for low-mass planets, and therefore the conclusions made here can be applied to most of the RV measurements gathered up to now.

With 14 different systems, 48 planets with semi-amplitude ranging between 0.16 and 5.85 \ms, and different modelizations of stellar signals, the number of parameter is huge, and it is difficult to draw some strong conclusions with the analysis of only 8 different teams. In addition, the data set of the RV fitting challenge was given to the different teams 8 months before the deadline. {Techniques used by team 1 to 5, based on a Bayesian framework with red-noise models, required significantly more computational time than the other techniques used by teams 6 to 8. As a result, team 2 and 4 could only analyze the first five systems out of 14, team 5 only the first two, and teams 1 to 5 used statistical shortcuts to find planetary signals in the data, or could not test all possible models, taking the risk of biasing their final results. Readers should therefore be aware that the results presented in this paper are preliminary, and depends on (1) how much time each team was able to invest in the challenge, (2) how mature their analytical methods were, and (3) how experienced the team members were with such analyses. Looking at the results presented in this paper, it seems that some techniques work better at recovering planets despite stellar signals, however further investigation need to be performed to be confident in the conclusions presented here. Note that the best techniques all require intensive computational efforts}.

A first important step before finding planets is the detection of the stellar rotation period. For team 1 and 2, this period is used in their model that accounts for short-term activity, for the other teams, this period and its harmonics defines regions in period space were planetary signal should be excluded because likely due to short-term activity. Finding the correct stellar rotation period is therefore crucial to reduce the number of false positives in the end, {and team 3, using its moving average model to account for stellar signals, performed the best at this exercise.} Among all the teams that reported explicitly a stellar rotation period, we notice that only a small number of mistakes were done. However in many cases, a harmonic of the stellar rotation period was found, which can be dangerous because then a signal at the true rotation period can be confounded with a planet. To distinguish between the true stellar rotation period and a harmonic of it, an activity level-rotation calibration as the one developed by \citet{Mamajek-2008} can be used. This was however not possible here due to lack of information in the RV fitting challenge data set, {but this is something that people analyzing RV data should strongly consider to prevent false positives (see Section \ref{sect:4-0}).}

When looking at the recovery rate of planetary signals for each team, teams can be separated in two groups. Teams 1, 2, 3, 4 and 5 that used a Bayesian framework with red-noise models and teams 6, 7 and 8 that used \emph{pre-whitening}, {compressed sensing} and/or filtering techniques in the frequency domain to deal with stellar signals. The first group discovered more true planetary signals than the second one, and also made fewer mistakes. In addition, when asked if those detections are significant enough to lead to publications, the first group of teams was also more confident in announcing a planetary signal. {The planets for which the $K/N$ ratio (see Eq. \ref{eq:4-1-0}) was above 7.5 were nearly all recovered by the best teams. Below this threshold, the detection rate drops to 20\% at best. Note that team 3 was able to find the smallest $K/N$ ratio true planetary signals, with $K/N$ ratios between 5 and 7.5, without announcing false positives. Below $K/N=5$, no planetary signals were confidently recovered, it is therefore a lower limit for planetary detections using data with similar properties as those of the RV fitting challenge (see Section \ref{sect:4-1-0}).}

%Team 3, that performed slightly better in the detection of planetary signals with a S/N$>$0.5 compared to teams 1, 2, 4 and 5, could recover all the planetary signals, except two long-period planets for which the orbital periods were longer than the time span of the data. Out of these 13 planetary signals with S/N$>$0.5, team 3 was confident in 9 of those to publish them, like team 1. Therefore, even if the best techniques known today to deal with stellar signals are used to analyze some RVs, about 10$-$30\% of planetary detections with S/N$>$0.5 will not be good enough to lead to publications. Those signals are however detected, which is a valuable argument to get more data for a system and thus publish the detection at a later stage. Analyzing the results of teams 6 to 8, it seems that filtering stellar signals in the frequency domain allows to recover slightly more planetary signals with S/N$>$0.5 than \emph{pre-whitening}. It is the contrary for planetary signals with S/N$\le$0.5, however note that in this case, most of the detection done by team 6 that used \emph{pre-whitening} have either a wrong estimate of the period or of the semi-amplitude (yellow flag). Therefore difficult to conclude if \emph{pre-whitening} allows to detect smaller S/N planetary signals.

Regarding accuracy when estimating the best orbital parameters for planetary signals qualified as publishable most of the teams recovered the correct orbital parameters within 3$-\sigma$ from the truth. A few signals were however out of the 3$-\sigma$ limit, which is probably due to the fact that the models used in this paper to account for stellar signals are not perfect. {This is not surprising as models to account for stellar activity are not perfect, however those are the best we have so far (see Section \ref{sect:4-1-1}).}

{\bf Besides recovering real planetary signal in the data and giving correct orbital parameters, it is very important that the false positive rate stays low. Above a threshold of 7.5 in $K/N$ ratio, team 7 announced nine false positives, team 1 six, team 6 one, and the other teams none. The technique use by team 7 is therefore prone to false positive and cannot be used to reliably detect planets. Team 1 also announced several false positives, however a few of them} correspond to the stellar
rotation period, despite the fact that the correct rotation period was
found a priori. Therefore, although their GP regression has the correct
stellar rotation period, it seems that the GP regression cannot fully
model stellar signals and that an extra sinusoidal signal is needed.
Further investigation on GP modeling needs therefore to be performed to be
sure that GP regression does not create false-positives. For the time
being, signals close to the stellar rotation period or its harmonics
should always be associated to stellar activity to prevent false
positives (see Section \ref{sect:4-1-2}).

For planetary signals with periods longer than 500 days, several effects make their detection difficult. It is common that drifts in the data are observed due to magnetic cycle effects and long-period binaries. To remove such long-period signals, the different teams corrected the RVs from magnetic cycle effects by using the observed long-term correlation between the RVs and the different activity observables (\logrhk, BIS SPAN, FWHM), and removed the effect of binaries by fitting polynomials as a function of time. {People analyzing RVs data should be aware that such a model can absorb the signal of planets that have orbital periods similar or longer than the time span of the data, and that orbits need closure before inferring planet parameters (see Section \ref{sect:4-1-3}).}

When analyzing the recovery of planets with periods shorter than 5 days, teams 3 found 4 out of 6 planets, including 3 with $K/N\le7.5$, while all the other teams found only the planet for which $K/N>7.5$. It seems therefore that the moving average model used by team 3 is more sensitive to short-period planets because such a model consider measurement correlation on short-period timescales, which therefore mitigate the effect of granulation on quiet stars, and the strong short-timescale effect of short-term activity on active stars like Corot-7. We would therefore encourage people using GP modeling, or apodized Keplerians, to add on top of their model a correlation between measurements on short-period timescales, as this seems critical to detect short-period planetary signals with small $K/N$ ratios (see Section \ref{sect:4-1-4}).

{The RV rms of real and simulated systems was similar, going in the direction that the different sources of stellar signals were realistically taken into account. Team 3, that performed the best at the exercise of the RV fitting challenge found very similar solutions between real and simulated data. However, it was slightly more difficult for the other teams to analyze simulated data. We therefore believe that even if not perfect, the simulated data are realistic enough to be used to test the efficient of techniques to recover planetary signature despite stellar signals (see Section \ref{sect:4-3}} 

With more time, each technique can be improved, and the different teams are making progress \citep[see][]{Gregory-2016,Hara-2016}. {The Oxford team also made some important progresses (priv. comm.). Now they are able to perform a full Bayesian marginalisation over all parameters (planets + GP), which give them much more reliable Bayesian model evidences.} Following a private communication with N.C Hara from team 8, It seems that their method is now delivering similar performances in terms of planetary detection as Bayesian framework techniques using red-noise models {and with a much shorter computational time} \citep[see bottom plot in Fig. \ref{imcce_rvchallenge2} and][]{Hara-2016}. However, following the first results of the RV fitting challenge presented here, techniques using a Bayesian framework and red-noise models seem the most efficient at modeling the effect of stellar signals, and therefore detecting true planetary signals while limiting the number of false positives. Moving average, GP regression and apodized Keplerian modelizations should be investigated further, to see the sensitivity of these models to planets at short and long-periods, to planets with a similar period than stellar rotation, to planet with high and low $K/N$ ratios, to multi-planet systems. 

The goal of the RV fitting challenge was to test the efficiency of the different techniques to recover planets in RV data given the presence of stellar signals, while limiting the number of false positives. As we can see in the different discussions above, the Bayesian framework and moving average model used by team 3 performed the best. {Then, in second position comes the Bayesian framework and apodized Keplerian model used by team 4, followed by the Bayesian framework and GP model used by team 1 in third position. Although team 1 performed well in analyzing system 6 to 15 in terms of true planetary signals detected, they announced a lot of false-positives at the stellar rotation period. Further investigation need to be performed to test if those false positives originate from the GP regression they used, or from another part of their method.}

{Team 3 was able to confidently discover a few planetary signals with $K/N$ ratios between 5 and 7.5 without announcing false positives, and nearly all the planetary signal with $K/N>7.5$.
Team 4 and 1 detected confidently most of the signals for which $K/N>7.5$, and none below this threshold. In conclusion, for RV measurement similar to those of the RV fitting challenge, a ratio $K/N=7.5$ seems to be a threshold separating confident detection from non-detection of planetary signals. Note however that the method used by team 3 could confidently detect $\sim$20\% of the planetary signals with $K/N$ ratios as low as 5, without announcing false positives.} 

\begin{acknowledgements}
We thanks the two referees of this paper for their valuable comments that greatly improved the paper and made the results more impactful and significant. 

We are extremely grateful to A. Correia, D. A. Fischer, E. B. Ford, J. S. Jenkins, A. Santerne, S. H. Saar, and G. Scandariato for interesting discussions about the design of the RV fitting challenge data set and the best way to organize it.

X. Dumusque is grateful to the Society in Science$-$The Branco Weiss Fellowship for its financial support. 

S. Aigrain acknowledges support from the Leverhulme Trust and the UK Science and Technology Facilities Council.

The Brera team acknowledges financial support from INAF through the
"Progetti Premiali" funding scheme of the Italian Ministry of Education,
 University, and Research. 
 
M. Damasso acknowledges funding from Progetto Premiale INAF "Way to Other Worlds" (WoW), decreto 35/2014.
A. S. Bonomo is grateful to the European Union Seventh Framework Programme (FP7/2007-2013) for funding under Grant agreement number 313014 (ETAEARTH).

R. D. Haywood gratefully acknowledges a grant from the John Templeton Foundation. The opinions expressed in this publication are those of the authors and do not necessarily reflect the views of the John Templeton Foundation.
 
M. Tuomi, G. Anglada-Escud\'e and H. R. A. Jones acknowledge the Leverhulme Trust funding scheme for grants RPG-2014-281 and STFC ST/M001008/1.

\end{acknowledgements}

\bibliographystyle{aa}
\bibliography{dumusque_bibliography}

\begin{appendix}

\section{Summary of the planet detection results obtained by the different teams}

\begin{figure*}
\begin{center}
\includegraphics[width=24cm,angle=90]{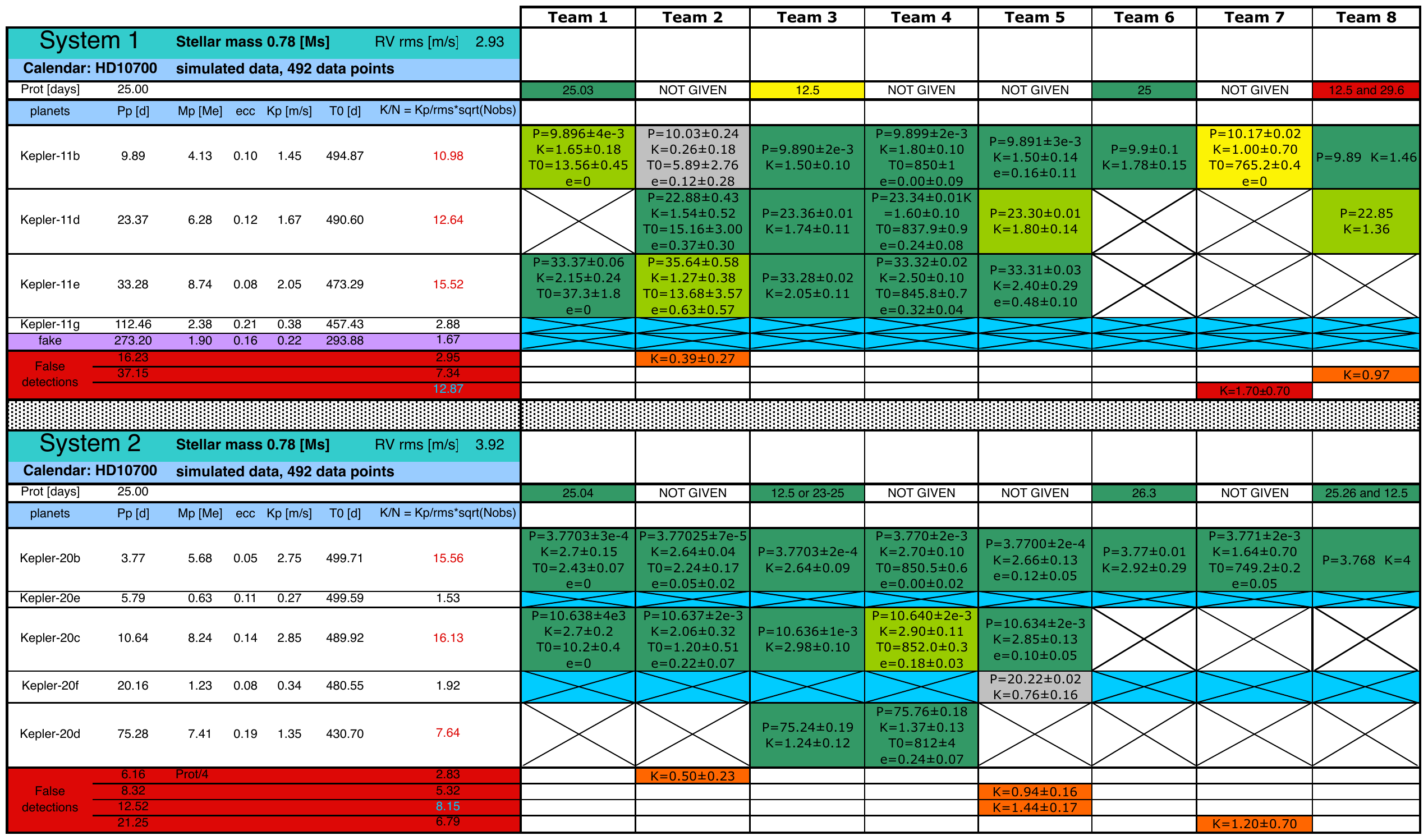}
\caption{Summary of signal detection for RV fitting challenge systems 1 and 2 reported by the different teams. Color flags are defined in the legend of Fig. \ref{fig:4-1-0} and in more details in the second paragraph of Section \ref{sect:4-1}. {Note that the RV rms shown here is the one obtained from the raw RVs once the best-fit of a model consisting of a linear correlation with \logrhk\,plus a second order polynomial as a function of time was removed.}}
\label{fig:summary_detection1}
\end{center}
\end{figure*}

\begin{figure*}
\begin{center}
\includegraphics[width=24cm,angle=90]{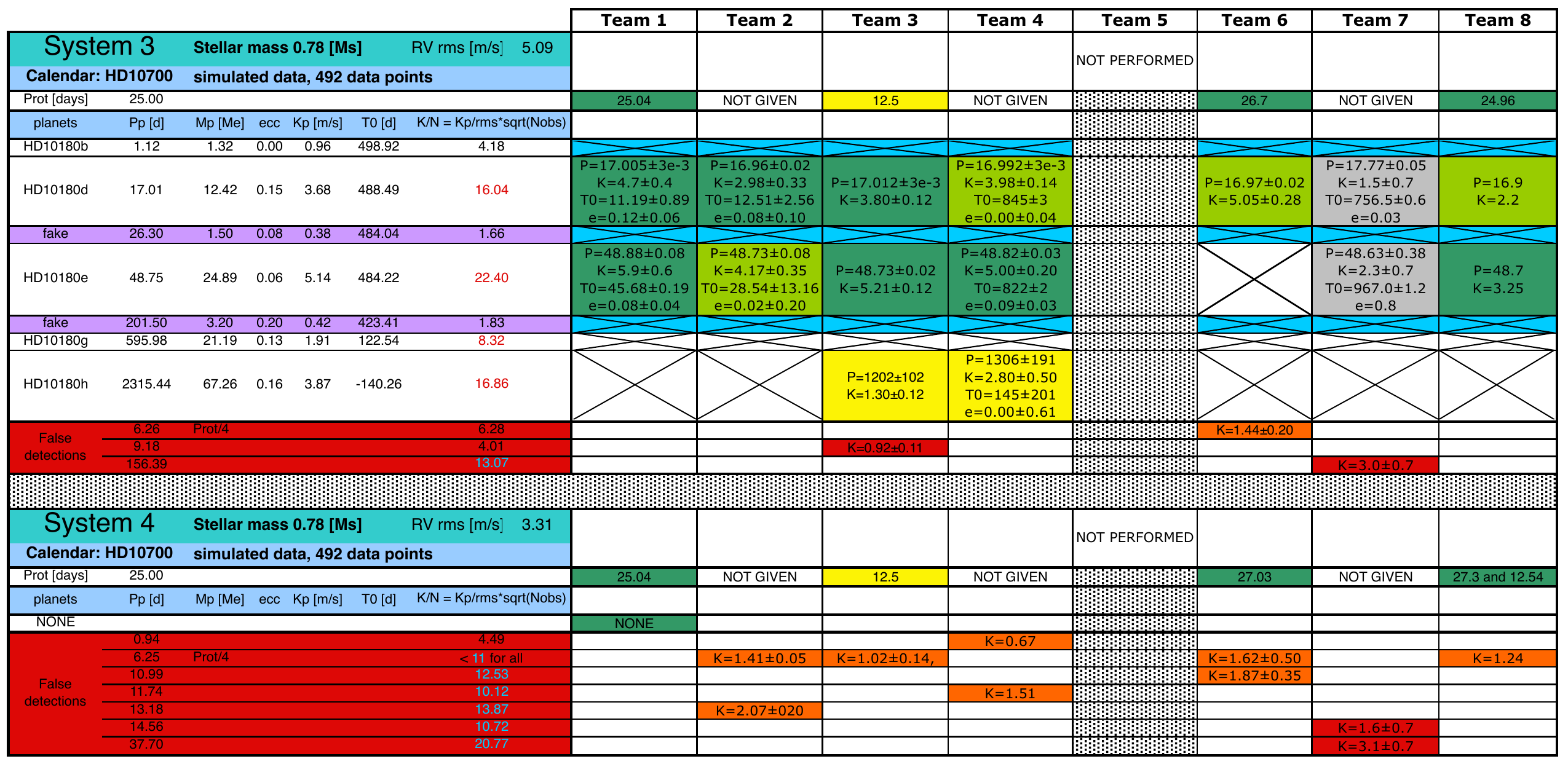}
\caption{Same as Fig. \ref{fig:summary_detection1} but for RV fitting challenge systems 3 and 4.}
\label{fig:summary_detection2}
\end{center}
\end{figure*}

\begin{figure*}
\begin{center}
\includegraphics[width=24cm,angle=90]{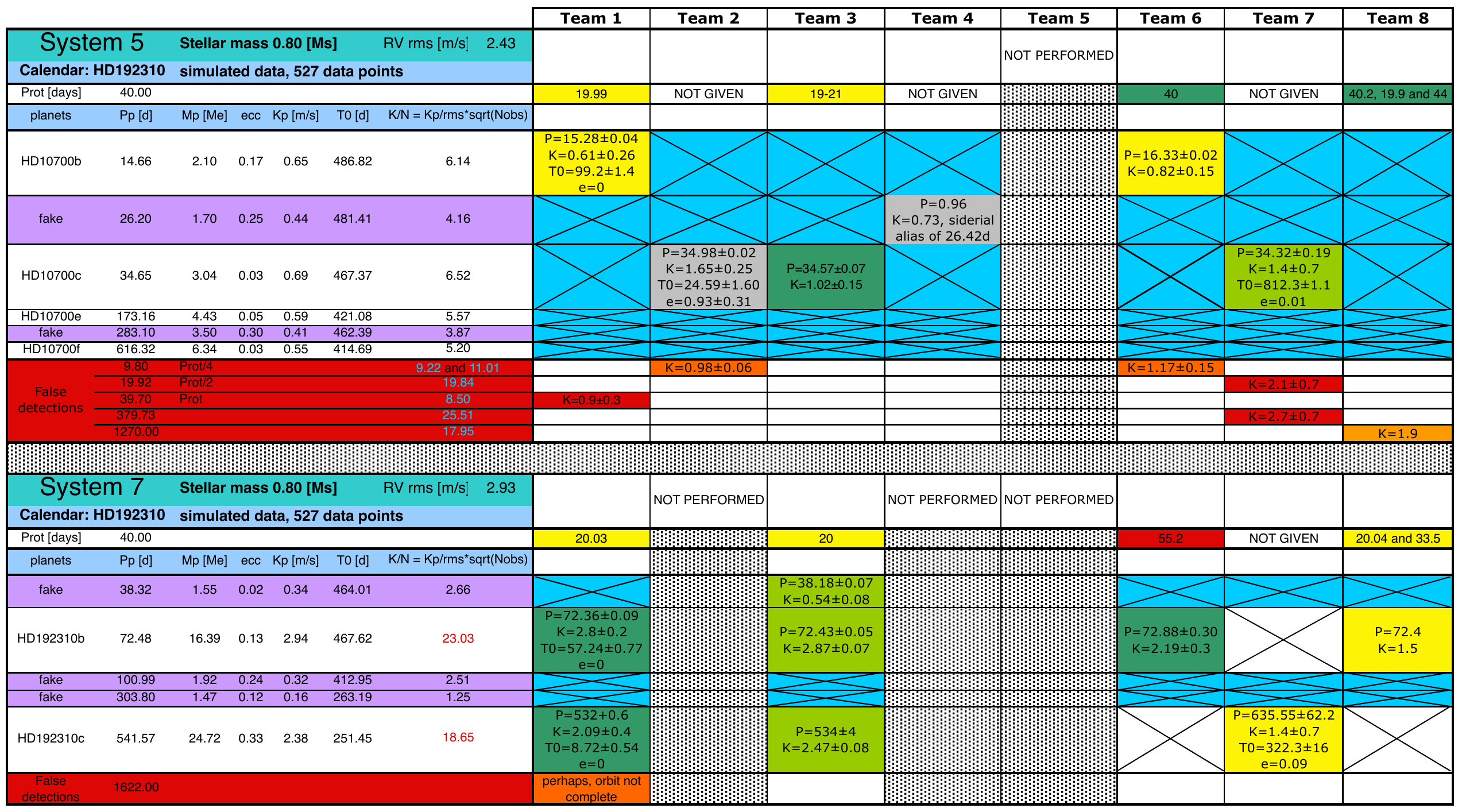}
\caption{Same as Fig. \ref{fig:summary_detection1} but for RV fitting challenge systems 5 and 7.}
\label{fig:summary_detection3}
\end{center}
\end{figure*}

\begin{figure*}
\begin{center}
\includegraphics[width=24cm,angle=90]{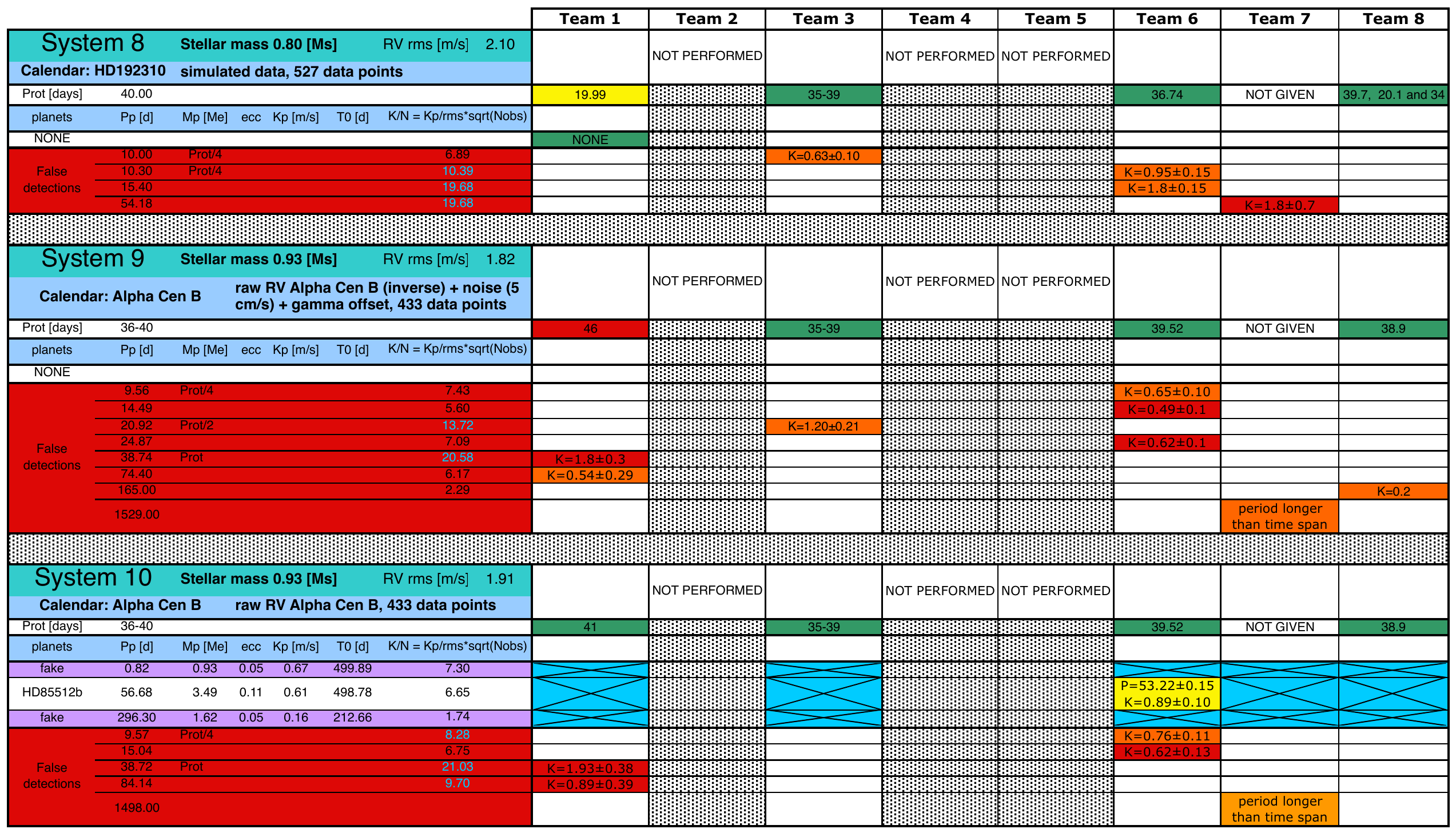}
\caption{Same as Fig. \ref{fig:summary_detection1} but for RV fitting challenge systems 8, 9 and 10.}
\label{fig:summary_detection4}
\end{center}
\end{figure*}

\begin{figure*}
\begin{center}
\includegraphics[width=24cm,angle=90]{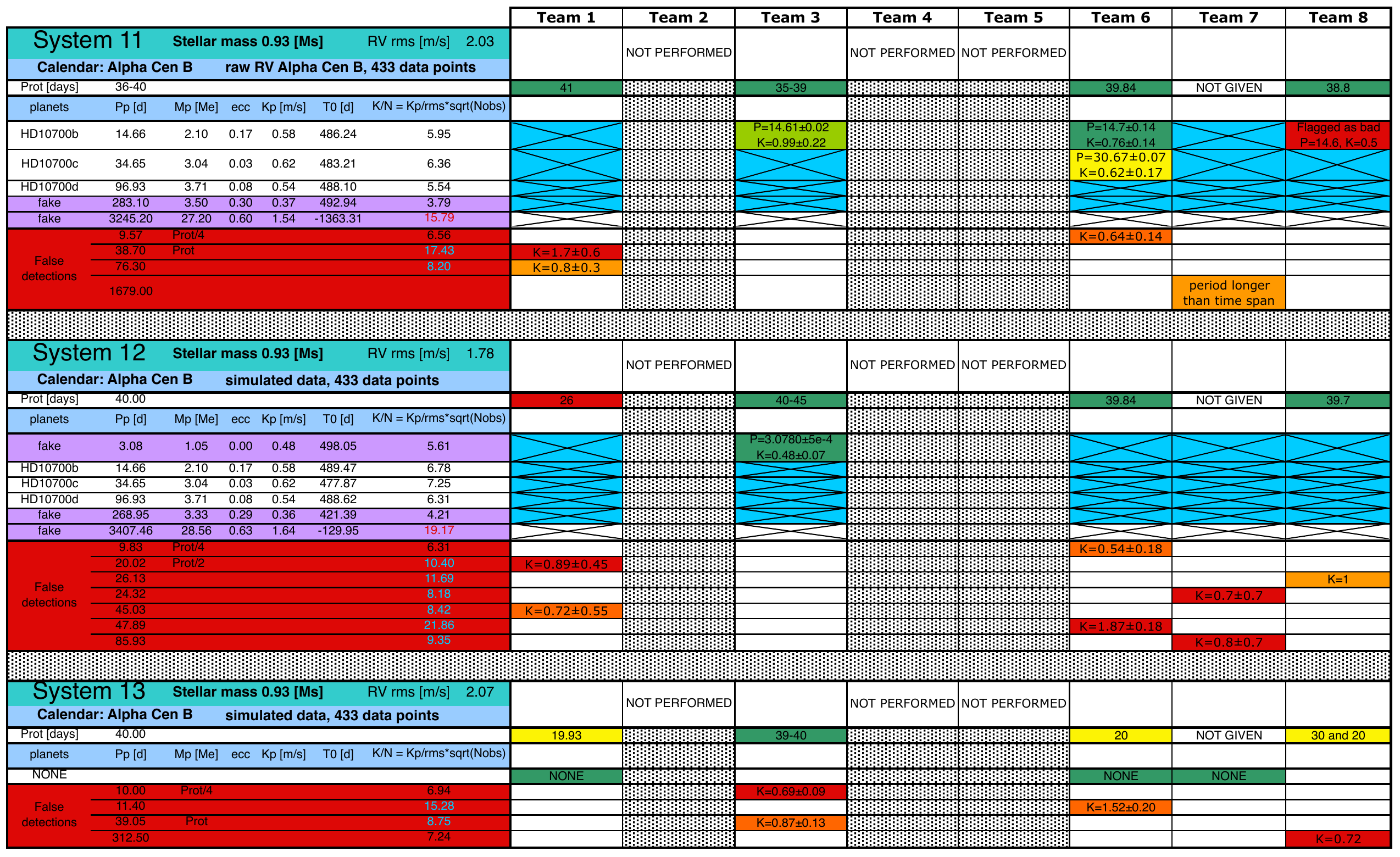}
\caption{Same as Fig. \ref{fig:summary_detection1} but for RV fitting challenge systems 11, 12 and 13.}
\label{fig:summary_detection5}
\end{center}
\end{figure*}

\begin{figure*}
\begin{center}
\includegraphics[width=24cm,angle=90]{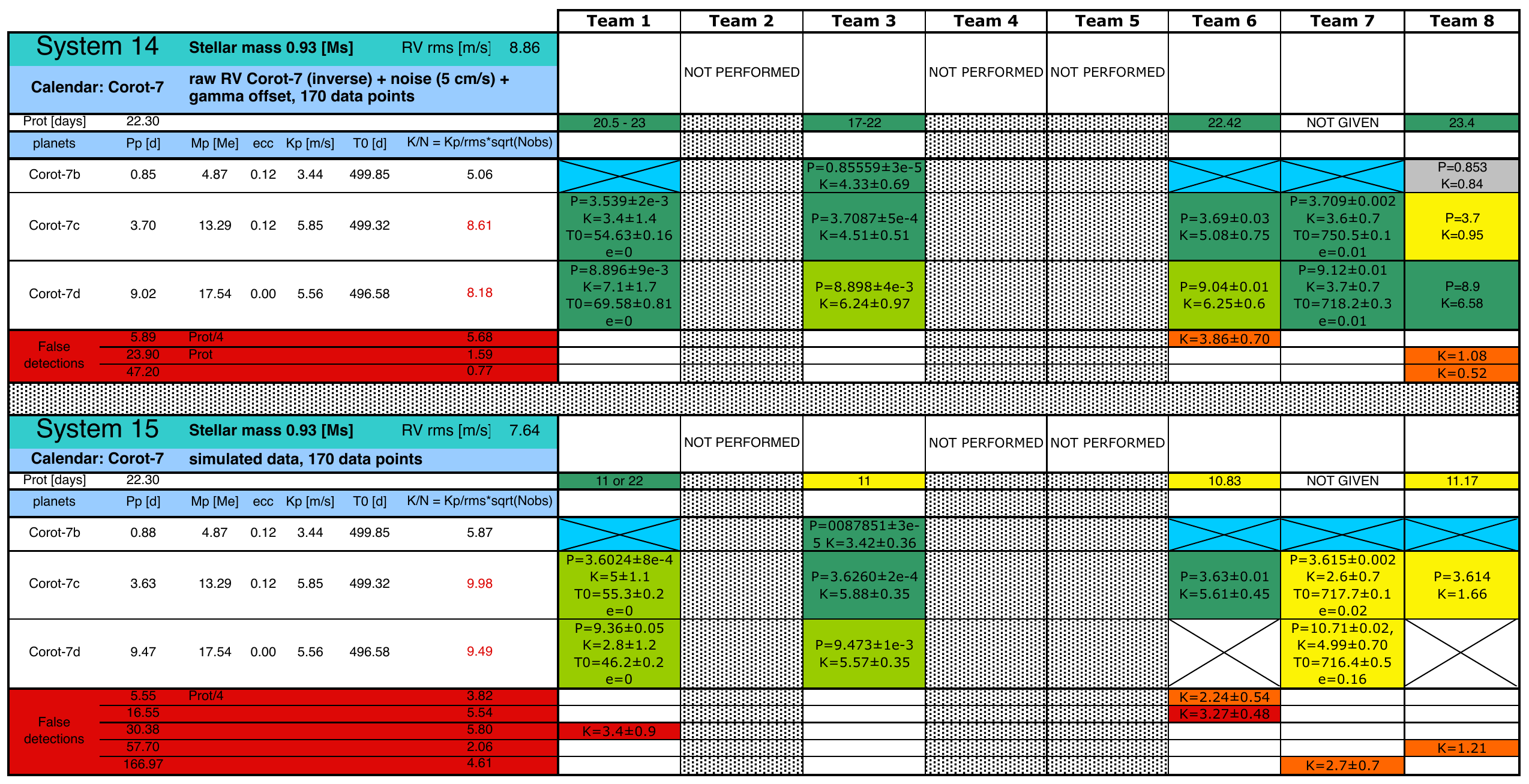}
\caption{Same as Fig. \ref{fig:summary_detection1} but for RV fitting challenge systems 14 and 15.}
\label{fig:summary_detection6}
\end{center}
\end{figure*}

Figs. \ref{fig:summary_detection1}, \ref{fig:summary_detection2}, \ref{fig:summary_detection3}, \ref{fig:summary_detection4}, \ref{fig:summary_detection5} and \ref{fig:summary_detection6} show a summary of the planet detection results obtained by the different teams.


\clearpage

\section{Details about the different algorithmes used by the different teams to analyze the data of the RV fitting challenge} \label{app:2}

%#######################################################################
% GROUP 1
%#######################################################################

\subsection{Team 1} \label{app:2-1}

\subsubsection{The step-by-step approach}

Here is a detailed summary of the different steps performed by team 1 to analyze the data of the RV fitting challenge.

\begin{enumerate}

\item \textit{Pre-treatment phase: removing long-term trends of stellar origin:}
When generating the data of the RV fitting challenge, \citet{Dumusque-2016a} considered magnetic cycles and their effect on the different observables, i.e., \logrhk, BIS SPAN and full width at half maximum (FWHM) of the CCF. In this case, strong correlation between \logrhk\,and FWHM, and between \logrhk\,and RV are expected \citep[][]{Lovis-2011b,Dumusque-2011,Lindegren-2003}. To test those correlations, the Torino team calculated Spearman's rank correlation coefficients and found, in most cases, a very strong correlation between these observables, i.e., $\rho>0.9$. In the case of significant correlation, i.e., $\rho>0.5$, team 1 detrended the RV and the \logrhk\,using linear fits between \logrhk\,and RV, and \logrhk\,and FWHM, respectively \citep[][]{Meunier-2013}. Detrending the RV and the \logrhk\,allows suppressing almost entirely the long-term activity effect induced by magnetic cycles, and therefore leaves only the short-term activity effect, that team 1 further modeled using a GP. In the case of systems 9, 10, and 11, RV were detrended with a linear fit as a function of time, as a significant long-term signal, probably due to a binary, was still visible after correcting for the magnetic cycle effect.

\vspace{0.2cm}
\item \textit{GP regression of the activity index \logrhk:}
To model \logrhk\,with a GP, team 1 used the combination of a \textit{rational quadratic} (RQ) and a \textit{quasi-periodic} (QP) covariance function \citep[][]{Pont-2013,Rasmussen-2006}:
\begin{eqnarray} \label{eq:2-0}
k_{RQ,QP}(t, t^{\prime}) = &A^2&\exp\left(-\frac{sin^{2}[\pi(t-t^{\prime})/\theta]}{2L^2}\right) \nonumber \\
&\times& \left(1 + \frac{(t-t^{\prime})^2}{2\alpha l^2}\right)^{-\alpha} +\sigma^{2}_{t}\delta_{tt^{\prime}},
\end{eqnarray}
where $t$ and $t^{\prime}$ represent epochs of observations, $\theta$ the stellar rotation period, $\sigma_{t}$ is the uncertainty of the measurement at time $t$, and  
$\delta_{tt^{\prime}}$ is the Kronecker's delta. When there is no a suitable guess about the timescale over which the data are varying, the RQ kernel can be assumed as a reasonable choice because it is intended to model the data by accounting for many different timescales. In fact, it is equivalent to an infinite sum of squared exponential (SE) kernels:
\begin{eqnarray} \label{eq:2-10}
k_{SE}(t, t^{\prime}) = {h ^2}\exp \left[ { - \frac{{{{(t - t')}^2}}}{{2{l ^2}}}} \right],
\end{eqnarray}
with different length-scales $l$ \citep{Rasmussen-2006}, with the inverse squared timescales $l^{ - 2}$ distributed according to a Gamma distribution with parameters $\alpha$ and $\beta=l^{ - 2}$. When $\alpha \rightarrow \infty$ the RQ kernel converges to the SE kernel. The function $k_{RQ,QP}(t, t^{\prime})$ describes the degree of correlation between each pair of measurements at times $t$ and $t^{\prime}$, reducing to uncorrelated noise, i.e., white noise, when $t$=$t^{\prime}$. This form of covariance function is suitable for data sets spanning a few years. For example, for the long-term photometry data set of HD189733, \citet{Pont-2013} discussed the choice of a $k_{RQ,QP}(t, t^{\prime}$) instead of a simpler exponential decay covariance function %$k_{QP,SE}(t, t^{\prime}$) 
to model the observed signal due to stellar activity.

The best-fit values of the covariance function hyper-parameters were obtained using an MCMC analysis. Initial guess for hyper-parameter $\theta$ was derived by performing a periodogram analysis with the Generalized Lomb-Scargle algorithm \citep[GLS,][]{Zechmeister-2009}. After a burn-in phase, typically consisting of 1500 steps per chain, team 1 maximized the following log-likelihood function:
\begin{equation} \label{eq:2-1}
\ln \mathcal{L} = -\frac{n}{2}\ln(2\pi) - \frac{1}{2}\ln(det\,\mathbf{K}) - \frac{1}{2}\underline{r}^{T}\cdot\mathbf{K}^{-1}\cdot\underline{r},
\end{equation}
where \textbf{K} is the covariance matrix built from the covariance function in Equation \ref{eq:2-0}, and $\underline{r}$ is the detrended \logrhk. The best-fit estimates of the hyper-parameters, inferred from their posterior distributions, were used as guess values for the subsequent modeling of the RVs, as explained below. Team 1 derived stellar rotation periods from the posterior distribution of $\theta$.

\vspace{0.2cm}
\item \textit{First identification of significant signals; GLS analysis of the RV time series:}
The Torino team applied the GLS algorithm to search for significant signals in the original RVs. Team 1 explored the frequency space below the Nyquist frequency and estimated peak significance using $p$-values determined through a bootstrap with replacement analysis consisting of 10'000 random shuffles of the data by keeping the time stamps fixed. Team 1 selected for further considerations only peaks with $p$-values$<$10$^{-3}$ (0.1$\%$), except for systems 14 and 15, because of the lower number of data points.

Team 1 iteratively removed sinusoidal fits from the data, with periodicity corresponding to the periodogram peaks, and obtained guess values for the orbital period of the candidate Keplerian signals. 
%The Torino team discarded signals if the periodicity found by the GLS was close to the stellar rotation period $\theta$ estimated with \logrhk, implying that the signal has likely a stellar activity origin. In addition, 
Team 1 looked at the window function to discard aliases.

As a general rule, only significant RV signals with period shorter than the data time span were considered, except for system 7, where a signal with a longer period than the data time span was modeled with a Keplerian in the global fit, despite the inability of characterizing reliably the potential orbit. Moreover, the approach followed by the team was conservative, i.e. aimed at avoiding as much false positives as possible, favoring the analysis of signals with the highest semi-amplitudes.% and not trusting those with orbital periods within $\sim1-2$ days from the identified stellar rotation period.

\vspace{0.2cm}
\item \textit{RV model and MCMC analysis:}
After the analysis of the GLS periodogram, and the identification of significant signals that could be due to planetary candidates, the Torino team performed a global fit of the RVs with a model consisting of Keplerian orbits and correlated noise, to account for short-term stellar activity signals. This correlated noise is modeled using the GP covariance function seen in Equation \ref{eq:2-0}. The training of the GP on the \logrhk\,gives initial guess for the GP hyper-parameters used when fitting the RVs. Doing so, team 1 assumes that short-term activity signals seen in RV and \logrhk\,have a similar covariance. 

The \textbf{general} Keplerian model fitted to the RVs is described by:
\textbf{\begin{eqnarray}\label{eq:2-2}
\Delta RV_{Kep}(t _{i}) &=&  \sum_{j=1}^{n_{planet}} \Delta RV_{Kep,j}(t _{i}) + \gamma \nonumber \\
				   &=& \sum_{j=1}^{n_{planet}} K_{i}\left[\cos(\nu(t_{i}, T_{0j,\,peri.}, P_{j}) + \omega_{j}) + e_{j}\cos(\omega_{j})\right]\nonumber \\
				   &+& \gamma.		
\end{eqnarray}}
Instead of fitting $e_{j}$ and $\omega_{j}$ separately, team 1 introduced:
\begin{equation}\label{eq:2-3}
C_{i} = \sqrt{e_{i}}\cdot \cos \omega_{i} \qquad S_{i} = \sqrt{e_{i}} \cdot \sin \omega_{i},
\end{equation}
to uniformly sample the eccentricity parameter space \citep[][]{Ford-2006b}. Short-term stellar activity is fitted simultaneously by the GP applied to the RV residuals obtained by subtracting the Keplerian model from the raw RV data. The best-fit is found by maximizing the log-likelihood seen in Equation \ref{eq:2-1}. Note however that in this case the array $\underline{r}$ represent the RV residuals.

The MCMC analysis used a number of random walkers, typically in the range 50-150, and was characterized by a burn-in phase, in general consisting of 1500 steps. For each fitted parameter the team adopted non-informative, uniform priors. The hyper-parameters of the covariance function were constrained within a range with reasonable finite lower and upper limits comprising the best-fit estimates found with the analysis of \logrhk, except for the semi-amplitude term $A$ of the covariance function, which for the RVs is necessarily different from that of \logrhk\,and was only imposed to be positive. No upper limits were fixed for $T_{0j,\,peri.}$ and semi-amplitude $K$, while the orbital periods were constrained over ranges of reasonable semi-amplitude centered on the guessed values obtained from the GLS periodogram analysis.
To test the convergence of the different chains, team 1 used the Gelman-Rubin statistics as described in \citet{Ford-2006b}. 
The best estimate of each parameter is derived using the median of its posterior distribution, with their asymmetric uncertainties derived from the 16$^{th}$ and 84$^{th}$ percentile (1$-\sigma$ uncertainty).

\vspace{0.2cm}
\item \textit{Model selection:}
The GP analysis requires a significant computational effort. Due to the relatively short timescale of the RV fitting challenge, team 1 could only test a limited number of different models for each system. Team 1 performed a Bayesian selection based on the truncated posterior mixture (TPM) method described in \citet{Tuomi-2012c}. In some cases, team 1 tested models with an equal number of planets, but fixing or not the eccentricities to zero. In few other cases, when signals could be of planetary or stellar nature, team 1 compared models with a different number of planets, limiting the analysis to circular orbits. Finally, when the Bayesian analysis showed to be inconclusive, team 1 selected the model with fewest parameters, following the ''Occam razor'' principle. Note however this was not the case for system 15, because the three candidate signals appear to be well modeled by a sinusoid, even if the true nature of one Keplerian was flagged as doubtful. 

%{EXPLAIN FURTHER An interesting outcome of our approach, i.e. when GP are involved in a global MCMC fit, should be that different results could be obtained applying different Bayesian selection criteria: how GPs affect the likelihood functions when a different number of planets are simultaneously fitted in a global noise model? Which are the best selection criteria to be used?  
%COMMENT: Actually, I cannot provide now a quantitative answer to your comment, which is indeed very relevant. I considered the opportunity of testing several Bayesian selection criteria other than TPM once I saw Rodrigo's poster in Yale, and then read his paper \cite{Diaz-2015} some months later. To this purpose, I first started to investigate the potential of the method based on MultiNest, developed by Feroz et al., but I've applied it to real data we are analyzing in Torino, not yet to your data sets, and I have to go deeper in the analysis when GPs are included in the model. I have not yet completed the development of the necessary software to compute the Bayes factor alternatively to TPM.
%In conclusion, it could be good to include your comment in the paper, following for example the footsteps of \cite{Diaz-2015}, but in a way to present our model selection analysis only as preliminary and necessarily incomplete. I usually do not like reading in a paper something like "this further analysis is left to a future work", but here this is the case. Let me know what do you think about it.}
\end{enumerate}

\subsubsection{Algorithms and Tools}
Here is a list of the different tools that team 1 used to perform the analysis:
\begin{itemize}
\item Spearman's rank correlation coefficients were evaluated with the \texttt{R$\_$CORRELATE} function, which is part of the IDL library.
\item Linear fits (\logrhk\,vs. FWHM and \logrhk\,(or time) vs. RV), and estimation of the GP hyper-parameters and Keplerian parameters were performed using the publicly available \texttt{EMCEE} Affine Invariant Markov Chain Monte Carlo Ensemble sampler, developed by \cite{Foreman-Mackey-2013} (see also \url{http://dan.iel.fm/emcee/current/}). 
\item The GP regression analysis was performed with the \texttt{George} Python library developed by \cite{Foreman-Mackey-2015} and \cite{	
Ambikasaran-2014}, and publicly available at \url{http://dan.iel.fm/george/current/}
\item The search for sinusoidal modulations in the \logrhk\,and RV data were performed with the Generalized Lomb-Scargle (GLS) algorithm developed by \citet{Zechmeister-2009}.
\end{itemize}
%

%
%%#######################################################################
%% GROUP 2
%%#######################################################################
%
%\subsection{Team 2} \label{app:2-2}
%
%\subsubsection{Algorithms and Tools}
%
%Here is a list of the different tools that team 2 used to perform the analysis:
%%
%\begin{itemize}
%\item MAP values for model hyper-parameters and parameters were found using multiple runs of a Nelder-Mead (``downhill simplex'') algorithm \citep{Lagarias-1998}
%\item Bayesian parameter inference was performed using the \textsc{MultiNest} nested-sampling algorithm \citep{Feroz-2013,Feroz-2008}
%\end{itemize}
%%

%#######################################################################
% GROUP 3
%#######################################################################

\subsection{Team 3} \label{app:2-3}

\subsubsection{The step-by-step approach}

Here is a detailed summary of the different steps performed by team 3 to analyze the data of the RV fitting challenge.

\begin{enumerate}

\item \textit{First identification of significant signals}
Team 3 first analyzed all time series of the RV fitting challenge using a likelihood-ratio periodogram including a first order moving average, i.e. a correlation dependence of each data point with their preceding neighbor (see last term of Equation \ref{eq:2-4}). Significant signals in activity observables, i.e. \logrhk, BIS SPAN and FWHM, were associated to stellar activity effect, and significant signal in the RVs were associated to potential planetary candidates if a similar signal was not seen in the activity observables.

\item \textit{RV model and MCMC analysis:}

To fit the RVs, team 3 used a model composed of:
\begin{itemize}
\item one or several Keplerians,
\item a polynomial function up to the 2nd order to fit any long-term trend due to distant companions, 
\item linear correlation with activity observables, to account for the effect of magnetic cycles, 
\item a Gaussian white noise $\epsilon_i$ with zero mean and variance $\sigma_{i}^{2} + \sigma^{2}$, where $\sigma_{i}$ is given by the data and $\sigma$ is a free parameter to account for additional instrumental white noise, 
\item and a first order moving average component with exponential smoothing accounting for the intrinsic correlations in the RVs.
\end{itemize}
In this case, the RV model that team 3 used can be described as:
\begin{eqnarray}\label{eq:2-4}
\Delta RV_{tot}(t _{i}) &=& \Delta RV_{Kep}(t _{i}) + \epsilon_i + Cte + \alpha\,t_i + \beta\,t_i^2 + \nonumber \\
		       &+& c_{01}\,\mathrm{BIS\,SPAN} + c_{02}\,\mathrm{FWHM} + c_{03}\,\log(R'_{HK})\nonumber \\
		       &+& \phi \left[\Delta RV_{tot}(t _{i-1}) - \Delta RV_{Kep}(t _{i-1})\right]exp^\frac{t_{i-1}-t_i}{\tau},
\end{eqnarray}
where $Cte$, $\alpha$, $\beta$, and $c_{01}$, $c_{02}$, $c_{03}$ are the free parameters of the polynomial fit and to account for correlation with activity observables, respectively. The parameter $\phi$ measures the strength of the correlation between consecutive measurements, $\tau$ is the correlation timescale, and $\Delta RV_{Kep}$ is the Keplerian model described in Equation \ref{eq:2-2}. Team 3 analyzed the data of the RV fitting challenge using adaptive-Metropolis Markov Chain Monte Carlo samplings \citep[][]{Haario-2001}, maximizing the following log-likelihood:
\begin{equation} \label{eq:2-5}
\ln \mathcal{L} = -\frac{1}{2}\ln(2\pi(\sigma_i^2+\sigma^2)) - \left[\frac{\left(RV(t_i)-\Delta RV_{tot}(t_i)\right)^2}{2 (\sigma_i^2+\sigma^2)}\right].
\end{equation}
Note that in their analysis, team 3 fixed the correlation timescale $\tau$ to 4 days, as this value seemed to give good results on previous analysis of HARPS high-cadence data.

\item \textit{Model selection:}
Team 3 used the MCMC samplings to calculate the integrated likelihoods and obtain Bayesian estimates for model probabilities when assuming equal prior probabilities. Team 3 applied the method based on the mixture of posterior and prior densities, described in \citet{Newton-1994}, to compare between models.

When detecting a Keplerian signal, team 3 interpreted it as existing if: 
\begin{enumerate}
\item including the signal in the model increased the model probability by a factor of 1000;
\item the corresponding signal was unique in the period space such that there were no other periods with posterior density (i.e. local maxima) in excess of 0.1\% of the global maximum;
\item and the period and the semi-amplitude of the signal were well constrained from above and below, and the semi-amplitude, in particular, statistically significantly different from zero. 
\end{enumerate}
In addition to these criteria \citep[][]{Tuomi-2014}, team 3 interpreted the signal to be related to activity-induced variations if any of the activity indices showed a significant signal at the same period.

\end{enumerate}

\subsubsection{Algorithms and Tools}

Here is a list of the different tools that team 3 used to perform the analysis:
\begin{itemize}
\item Moving average \citep[][]{Baluev-2013,Tuomi-2013a}
\item Adaptive Metropolis MCMC algorithm \citep[][]{Haario-2001}
\item Model selection using posterior and prior mixture \citep[][]{Newton-1994}
\end{itemize}
%

%#######################################################################
% GROUP 4
%#######################################################################

\subsection{Team 4} \label{app:2-4}

\subsubsection{The step-by-step approach}

Here is a detailed summary of the different steps performed by team 4 to analyze the data of the RV fitting challenge.

\begin{enumerate}

\item \textit{First identification of significant signals}
To look for significant signals in the RVs, P. Gregory corrected the RVs from the effect of the magnetic cycle using the \logrhk-RV correlation, and considered any signal in a GLS periodogram with $p$-values smaller than 0.01. Note that contrary to team 1 and 3, P. Gregory estimated $p$-values analytically.

\vspace{0.2cm}
\item \textit{RV model and MCMC analysis:}
Once the first peak is detected, P. Gregory runs a Bayesian Fusion MCMC \citep[][]{Gregory-2013} analysis to find the best parameters for the signal, and look for extra signals in the residuals using a GLS periodogram. To fit the RV data, P. Gregory used the following model:
\begin{eqnarray}\label{eq:2-6}
\Delta RV_{tot}(t _{i}) &=& \sum_{j=1}^{n_{signals}} \Delta RV_{Kep,j}(t _{i}) \times exp\left[-\frac{(t_i-t_{a,j})^2}{2\tau_j^2}\right] \nonumber \\
				 &+& \gamma + \epsilon_i + a\,\log(R'_{HK}),
\end{eqnarray}
where $n_{signals}$ is the number of significant signals in the data independent of their nature, i.e. planetary or stellar activity, $t_{a,j}$ and $\tau_j$ are the center and timescale of the apodized window of signal $j$, and $a$ is a free parameter to account for a possible correlation between RV and \logrhk. 

\vspace{0.2cm}
\item \textit{Distinguishing planetary from stellar short-term activity signals}:
If the signal $j$ is induced by a planet, the apodized term $exp\left[-\frac{(t_i-t_{a,j})^2}{2\tau_j^2}\right]$ will essentially be constant over the duration of the data because the semi-amplitude of the signal is constant. In this case $\tau$ will be greater than the time span of the data. On the other hand, the apodized term will strongly vary as a function of time in the case of stellar activity, due to appearance and disappearance of active regions on the stellar surface. In this case $\tau$ will be smaller than the time span of the data. 

To help distinguish between planetary and stellar activity signals, P. Gregory used a second approach based on the FWHM. The FWHM is first corrected for the effect of the magnetic cycle using the \logrhk-FWHM correlation. Then any significant signal found in either the initial corrected RVs or the later stage RV fit residuals, that coincides with a significant signal in the corrected FWHM, is associated with stellar activity. 

At each stage in the RV analysis, the number of Keplerian signals was extended to include the period with the highest peak in the periodogram of the residuals from the previous model as a starting point for a Bayesian Fusion MCMC exploration in parameter space.

\vspace{0.2cm}
\item \textit{Model selection:}

Model comparison was based on Bayes factors computed using the Nested Restricted Monte Carlo (NRMC) estimator (Section 2.2 in \citealt{Gregory-2016}, \citealt{Gregory-2013}, \citealt{Gregory-2010}, and in more details in Section 1.6 of the \emph{Supplement to Bayesian Logical Data Analysis for the Physical Sciences} available in the resources section of the Cambridge University Press website for P. Gregory's Textbook \emph{Bayesian Logical Data Analysis for the Physical Sciences: A Comparative Approach with Mathematica Support}).

\end{enumerate}

\subsubsection{Algorithms and Tools}

Here is a list of the different tools that P. Gregory used to perform the analysis:
\begin{itemize}
\item GLS periodogram \citep[][]{Zechmeister-2009} to look for significant signals,
\item Bayesian Fusion MCMC \citep[][]{Gregory-2013} to explore parameter space,
\item and Nested Restricted Monte Carlo estimator \citep{Gregory-2013,Gregory-2010} to compare between different models.
\end{itemize}
%

%%#######################################################################
%% GROUP 5
%%#######################################################################
%
%\subsection{Team 5} \label{app:2-5}
%
%\subsubsection{Algorithms and Tools}
%
%Here is a list of the different tools that team 5 used to perform the analysis:
%%
%\begin{itemize}
%\item The MCMC algorithm described in \citep[][]{Diaz-2014} was used to obtain the posterior sample. This sample was used for parameter inference and for the estimation of the Bayesian evidence.
%\item The comparison between models with different number of signals was carried out by estimating the Bayesian evidence log\,$\mathcal{Z}$ with the methods of  \cite{Chib-2001} and \cite{Perrakis-2014}. Two different estimates were used to control that the evidence estimation is correct, as it is well known that the estimation of this quantity is problematic in high-dimensional spaces \citep[e.g.][]{Diaz-2016,Ford-2007}.
%\end{itemize}
%%

%#######################################################################
% GROUP 7
%#######################################################################

\subsection{Team 7} \label{app:2-7}

\subsubsection{The step-by-step approach}

\begin{enumerate}

\item \textit{DFT of all observables and cleaning from the spectral window:}
To move from the time-domain to the frequency domain for the RV, BIS SPAN, FWHM and \logrhk, team 7 used, like team 6, a DFT. In the frequency domain, any uneven data as a function of time will be affected by the sampling of the signal. To reduce the effect of sampling, team 7 used the CLEAN algorithm \citep[][]{Roberts-1987}.  The team got as a result a cleaned DFT (CDFT) for all time series (see upper-left panel of Fig. \ref{fig:brera}). Particular attention is made to carefully select the \textit{gain} parameter, in order to remove as much spurious frequency peaks as possible without removing significant signals.

%The peaks obtained in the CLEANed power spectrum are dependent from the choice of the gain parameter. Only few peaks are always present. The other peaks, present only with one particular gain, but not for the others, are probably created by the window function used by the CLEAN algorithm and they have to be canceled. For this reason, team 7 applied the CLEAN algorithm to the data using three different gains (1.0, 0.8, and 0.5) and, using a function that compares the power spectra and takes the minimum of the three, team 7 keep just the g-independent peaks. 

\item \textit{Removing stellar signals:}
To remove stellar signals, team 7 subtracted the CDFT of all the activity observables (BIS SPAN, FWHM, \logrhk) from the CDFT of the RVs. The obtained CDFT, exempt of stellar signals, is used to create a \textit{pass-planet} filter in the frequency domain (see upper-right panel of Fig. \ref{fig:brera}). By applying this filter to the DFT of the RVs, team 7 obtained RVs in the frequency domain that are cleaned from any stellar signals. At this stage, any significant signal in those filtered RVs should be due to planets. Team 7 selected the highest peak and recorded its period, semi-amplitude and phase.

\item \textit{Fitting planets:}
To fit the planetary signal found at the previous step with a Keplerian, team 7 first transformed the filtered RVs back into the time-domain using an inverse DFT, and then used the RVLIN package \citep[][]{Wright-2009} to fit the planetary signal, fixing the initial parameters to what was previously found.

\item \textit{Iterative process:}
Once team 7 found the best-fit for the planet inducing the strongest RV signal, it removed the signal from the raw RVs. Team 7 applied the CLEAN algorithm on the residual RVs and restarted the whole process from the beginning. To be conservative and prevent the detection of false positives, team 7 stopped when the semi-amplitude of the signal found in the filtered residual RVs was smaller than the average uncertainty of the RV measurements.

\end{enumerate}

This method presents the advantage of being independent of any model to account for stellar signals, and it is computationally very fast compared to Bayesian methods. However this technique presents the disadvantage that planetary signals are only fitted one by one, thus it is difficult to constrain orbital parameters such as eccentricity and argument of periastron. In addition, the imperfect removal of a signal causes the introduction of spurious frequencies that can lead to false detections. In addition, lack of statistics forced team 7 to stop at a S/N level of 1 (S/N once the data have been filtered from stellar signals), preventing the detection of small S/N planetary signals. Finally significant signal around one day were not considered, to avoid strong residuals of the spectral window not fully cancelled by the cleaning process. 

\subsubsection{Algorithms and Tools}

Here is a list of the different tools that team 7 used to perform the analysis:
\begin{itemize}
\item Discrete Fourier Transform \citep[as described in ][]{Roberts-1987} 
\item CLEAN algorithm to remove sampling effects \citep[][]{Roberts-1987}
\item RVLIN package to fit Keplerians \citep[][]{Wright-2009}
\end{itemize}
%

%
%
%%#######################################################################
%% GROUP 8
%%#######################################################################
%
%\subsection{Team 8} \label{app:2-8}
%
%\subsubsection{Algorithm and Tools}
%
%Here is a list of the different tools that team 8 used to perform the analysis:
%\begin{itemize}
%\item Compressed Sensing for RV data \citep[][]{Hara-2016}, though in a less sophisticated way.
%\item The SPGL1 algorithm~\citep{vandenberg2008}.
%\end{itemize} 

\end{appendix}

\end{document}